\documentclass[prc,preprint,superscriptaddress,showkeys,showpacs,
   preprintnumbers,amsmath,amssymb,aps,floatfix]{revtex4-1}
\usepackage{graphicx}
\usepackage[sort&compress]{natbib}
\usepackage{dcolumn}
\usepackage{multirow}
\usepackage{bigstrut}
\usepackage{type1cm}
\usepackage{placeins}
\usepackage{rotating}

\begin{document}
\preprint{\today}

\title{Polarization observables in low-energy \\ antiproton-proton scattering}

\author{Daren Zhou}
\author{Rob G. E. Timmermans}
\affiliation{KVI, Theory Group, University of Groningen, Zernikelaan 25,
                   NL-9747 AA Groningen, The Netherlands}

\date{\today}
\vspace{3em}

\begin{abstract}
\noindent
We investigate the polarization parameters in low-energy antiproton-proton
elastic ($\overline{p}p\rightarrow\overline{p}p$) and charge-exchange
($\overline{p}p\rightarrow\overline{n}n$) scattering. The predictions for
unmeasured observables are based on our new energy-dependent
partial-wave analysis of all antiproton-proton scattering data below 925 MeV/$c$ 
antiproton laboratory momentum, which gives an optimal description of the existing
database. Sizable and remarkable spin effects are observed, in particular for
charge-exchange scattering. These result from the spin dependence of the long-
and medium-range one- and two-pion exchange antinucleon-nucleon potential
and the state dependence of the parametrized short-range interaction. We study the
possibility of polarizing a circulating antiproton beam with a polarized proton target by
exploiting the spin dependence of the $\overline{p}p$ total cross section. It appears
feasible to achieve a significant transverse polarization of an antiproton beam within
a reasonable time.
\end{abstract}
\maketitle

\section{Introduction}
The invention of stochastic cooling by van der Meer made it possible
to accumulate antiprotons in a high-quality beam~\cite{Mee85}. As a result,
the antinucleon-nucleon ($\overline{N}\!N$) interaction could be studied at the
Low Energy Antiproton Ring (LEAR) at CERN. Experimental data of good quality
could be obtained down to antiproton momenta of about 200 MeV/$c$. The
observables measured were mostly differential cross sections for antiproton-proton
elastic ($\overline{p}p\rightarrow\overline{p}p$) and charge-exchange
($\overline{p}p\rightarrow\overline{n}n$) scattering. With a polarized proton
target the analyzing power for elastic~\cite{Kun88,Kun89,Ber89,Per91}
and charge-exchange~\cite{Bir90,Bir91,Bir93b,Lam95} scattering was
measured for a number of antiproton momenta, and initial data for the
depolarization parameter in elastic~\cite{Kun91} and
charge-exchange~\cite{Bir93,Ahm95d} scattering and for
the spin-transfer parameter in charge-exchange~\cite{Ahm96} scattering were acquired.
It has always been a dream of the antiproton physics community to have
available a high-quality polarized antiproton beam. In recent years,
definite plans have been proposed for a physics program with polarized
antiprotons, for instance by the collaboration for Polarized Antiproton
eXperiments (PAX) \cite{Len05,PAX}. Experiments with a polarized
antiproton beam and a polarized proton target would give full access to
the complicated spin dependence of the $\overline{N}\!N$ interaction and
could help to unveil the spin structure of the (anti)proton and test predictions
of nonperturbative QCD.

In this article, we present theoretical predictions for spin observables
in $\overline{p}p$ elastic and charge-exchange scattering. 
Our results are based on the recent new energy-dependent partial-wave
analysis (PWA)~\cite{Zho12} that we performed of all $\overline{p}p$ scattering
data below 925 MeV/$c$. The method of PWA was originally developed for $pp$
and $np$ scattering~\cite{Ber88,Ber90,Tim93,Sto93,Ren99,Ren03,NNOnline} and
enabled major steps forward for the field of few-nucleon physics~\cite{Car98,Bet99}.
It was adapted to $\overline{p}p\rightarrow\overline{p}p$, $\overline{n}n$ scattering
in Refs.~\cite{Tim91,Tim94,Tim95}, and to the hyperon production reactions
$\overline{p}p\rightarrow\overline{Y}Y$ in Refs.~\cite{Tim88,Swa89,Tim90,Tim92}. 
This PWA exploits as much as possible
our knowledge of the long-range interaction to describe the energy dependence
of the (spin-dependent) scattering amplitude, while the unknown short-range
interaction that gives only slow energy variations to the amplitudes is
parametrized phenomenologically. In this way, a model-independent and
high-quality description of the scattering data is realized, and the predictive
power for thus-far unmeasured observables is optimal~\cite{Tim95}. 

In Ref.~\cite{Zho12} the long-range part of the strong $\overline{p}p$ interaction
is described by  the one-pion and two-pion exchange interactions derived from
the effective chiral Lagrangian of QCD. Notable features are a strong central
attraction in the elastic $\overline{p}p\rightarrow\overline{p}p$ channel and a
strong tensor force, in particular for charge-exchange
$\overline{p}p\rightarrow\overline{n}n$ scattering.
The short-range part of the interaction, including the coupling of the $\overline{p}p$,
$\overline{n}n$ system to the mesonic annihilation channels, is parametrized by a
complex, energy-dependent boundary condition at $r=1.2$ fm. The PWA describes
the existing $\overline{p}p$ database, which contains 3749 scattering data with
an excellent $\chi^{2}_{\text{min}}/N_{\text{df}}=1.048$, where $N_{\text{df}}=3578$
is the number of degrees of freedom.

The organization of our paper is as follows: In Sec.~\ref{sec:Spin} we explore
the various polarization observables in elastic and charge-exchange scattering
and the predictions of the PWA. We exhibit some sizable spin effects that could,
for instance, be exploited experimentally to produce polarized antineutrons. 
Our study is similar in spirit to that of Ref.~\cite{Dov82}, in which spin
observables were discussed qualitatively before LEAR came into operation
by using a simple optical-potential model. In Sec.~\ref{sec:Pol}, we discuss
how to exploit the spin dependence of the $\overline{p}p$ interaction to polarize
a low-energy antiproton beam. We summarize in Sec.~\ref{sec:Summary}.

\section{Spin obervables} \label{sec:Spin}
The scattering formalism that we use is standard (see for instance
Refs.~\cite{Wol52,Hos68,Bys78,LaF80,Bys84,LaF92}). Because the strong and
electromagnetic interactions obey the discrete symmetries  charge-conjugation,
parity, and time-reversal invariance, the $\overline{N}\!N$ scattering observables
can be calculated as a function of the scattering angle in the center-of-mass system
in terms of five (for the elastic case) or six (for the charge-exchange case) independent
scattering amplitudes $\langle s'm'|M(\theta)| s\,m\rangle$, where $s$ and $s'$ denote
the total initial and final spin, respectively, and $m$ and $m'$ are the corresponding
$z$ components. Neither for $\overline{N}\!N$ nor for the better known $N\!N$ case
are enough independent experimental quantities available to determine for a fixed
energy these five (or six) amplitudes for every angle $\theta$~\cite{Kun90,Puz57,Ric95}.
Therefore, so-called amplitude analyses are not feasible and energy-dependent partial-wave analyses
are the tool of choice~\cite{Tim95}.

We denote a general scattering observable by $X_{srbt}$, where the indices denote
the spin direction of the particles: $s=$ scattered, $r=$ recoil, $b=$ beam, and
$t=$ target. It is convenient to use two different coordinate systems for the spin
directions: (i) the vectors $\boldsymbol{\hat{x}}$, $\boldsymbol{\hat{y}}$, and
$\boldsymbol{\hat{z}}$ define a right-handed coordinate system wherein $\boldsymbol{\hat{z}}$
is the direction of the incoming beam and $\boldsymbol{\hat{y}}$ is normal to the
scattering plane; (ii) the spin directions $\boldsymbol{\hat{l}}$, $\boldsymbol{\hat{m}}$,
and $\boldsymbol{\hat{n}}$ are tied to the outgoing particles, where
$\boldsymbol{l}=\boldsymbol{p}_i+\boldsymbol{p}_f$
and $\boldsymbol{m}=\boldsymbol{p}_f-\boldsymbol{p}_i$
 lie in the scattering plane and
$\boldsymbol{\hat{n}}=\boldsymbol{\hat{l}}\times\boldsymbol{\hat{m}}=\boldsymbol{\hat{y}}$,
and $\boldsymbol{p}_i$ and $\boldsymbol{p}_f$ are the initial and final momenta in the
center-of-mass system.

The scattering observables can be expressed in terms of the scattering amplitude as
\begin{equation}
   I_0\,X_{srbt} = \frac{1}{4}{\rm Tr}
       \left[\sigma_{1s}\sigma_{2r}M\sigma_{1b}\sigma_{2t}M^\dagger\right] \ ,
\end{equation}
where $I_0=d\sigma/d\Omega={\rm Tr}\left[MM^\dagger\right]/4$ is the differential
cross section averaged over the spins of the initial-state particles and summed over
the spins of the final-state particles ($X_{0000}\equiv1$).  The indices take the values
$0$, $x$, $y$, and $z$ or $0$, $l$, $m$, and $n$; $\vec{\sigma}_{1,2}$ are the
Pauli spin matrices, and $(\sigma_{1,2})_0$ is the identity matrix used in case of an
unpolarized particle or unobserved spin direction. The simplest
spin observable is the analyzing power $A_n=A_{000n}$, defined by
\begin{equation}
  I_0\,A_n = \frac{1}{4}{\rm Tr}\left[M\sigma_{2n}M^\dagger\right]
\end{equation}
(where we replaced $X$ by the conventional symbol $A$), the measurement of which
requires only a polarized proton target. Differential cross sections and analyzing
powers of good quality were measured at LEAR for both elastic and charge-exchange
scattering.

In order to measure all possible spin observables one needs a polarized proton target,
a polarized antiproton beam, and ``analyzers'' of the polarization of the recoil proton
(or neutron) and the scattered antiproton (or antineutron). While there are standard
techniques to polarize a proton target, a high-intensity polarized antiproton beam does
not exist yet. The polarization of the recoil proton can be measured by secondary scattering
on a carbon target. However, carbon is not a good analyzer for the polarization of
antiprotons~\cite{Mar88} and a proton target gives only about 10\% analyzing power
for antiprotons. For these reasons, next to the analyzing power $A_n$, the depolarization
$D=D_{0n0n}$ is the easiest spin observable to measure. It was measured, with limited
statistical precision, at LEAR for both elastic and charge-exchange scattering for a few
antiproton momenta. For charge-exchange scattering also the spin transfer $D_t=K_{n00n}$
was measured. $D$ and $D_t$ are examples of depolarization and spin-transfer observables,
defined in general by
\begin{subequations}
\begin{eqnarray}
   I_0\,D_{ij} & = & \frac{1}{4}{\rm Tr}\left[\sigma_{2i}M\sigma_{2j}M^\dagger\right] \ , \\
   I_0\,K_{ij} & = & \frac{1}{4}{\rm Tr}\left[\sigma_{1i}M\sigma_{2j}M^\dagger\right] \ ,
\end{eqnarray}
\end{subequations}
where $D_{ij}=D_{0i0j}$ for depolarizations and $K_{ij}=K_{i00j}$ for spin transfers.
We also define the spin-correlation observables
\begin{subequations}
\begin{eqnarray}   
   I_0\,C_{ij} & = & \frac{1}{4}{\rm Tr}\left[\sigma_{1i}\sigma_{2j}MM^\dagger\right] \ ,  \\
   I_0\,A_{ij} & = & \frac{1}{4}{\rm Tr}\left[M\sigma_{1i}\sigma_{2j}M^\dagger\right] \ ,
\end{eqnarray}
\end{subequations}
where $C_{ij}=C_{ij00}$ and $A_{ij}=A_{00ij}$.

In a similar way, one can define higher-rank spin observables. In this paper, however,
we focus on the above observables, since even these rank-two spin observables are
difficult to measure. We point out that sometimes other notations are used in the literature,
for instance $A_{y} = A_{000n}$, $D_{yy} = D_{0n0n}$, and $K_{yy} = K_{n00n}$, etc. 
Actually, it is often convenient to mix the notations $l$, $m$, $n$ and $x$, $y$, $z$.
For instance, we can define $R_{t}=K_{lx}$, $R'_{t}=K_{mx}$,   $A_{t}=K_{\bar{l}\bar{z}}$,
and $A'_{t}=K_{\bar{m}\bar{z}}$, where the bar means that the spin of the particle points
in the opposite direction of the index. 

In order to calculate the spin observables, we use as input the scattering amplitudes
as predicted by our PWA. As discussed in detail in Refs.~\cite{Zho12,Tim94}, they
contain the Coulomb, magnetic-moment, and nuclear (hadronic) scattering amplitudes.
The spin-independent Coulomb amplitude for $\overline{p}p$ elastic scattering is given by 
\begin{equation}
   \langle s'm'|M_C(\theta)| s\,m\rangle = 
   -\delta_{ss'}\delta_{mm'}\, \frac{\eta}{2p}
   \frac{e^{2i\xi_{0}}}{(\sin^2\frac{1}{2}\theta)^{1+i\eta}} ~,
\label{Mcoul}
\end{equation}
where $\eta=-\alpha/v_{\rm lab}$ is the relativistic
Rutherford parameter, with $\alpha$ the fine-structure constant and $v_{\rm lab}$ the
velocity of the incoming antiproton in the laboratory frame; $p$ is its center-of-mass momentum;
and $\xi_{0}={\rm arg}\,\Gamma(1+i\eta)$ is the Coulomb phase shift for orbital angular momentum $\ell=0$.
The magnetic-moment interaction is treated in the Coulomb-distorted-wave Born approximation.
It contains a spin-orbit and a tensor force. The main effect is due to the spin-orbit force, which
for $\overline{p}p$ elastic scattering results in the spin-dependent amplitude
\begin{eqnarray}
 \hspace{-1.8em}\langle 1\,1|M^C_{C+M\!M}(\theta)| 1\,0\rangle & =&
    -\frac{e^{2i\xi_{0}}}{\sin\theta \sqrt{2}}\frac{(8\mu_{p}-2)\alpha}{4M_{p}}
    \left[e^{-i\eta\ln\frac{1}{2}(1-\cos\theta)}-
        \frac{1}{2}(1-\cos\theta)\right] \nonumber \\
 & = &
            -\langle 1\,0|M^C_{C+M\!M}(\theta)| 1\,1\rangle ~,
\label{Mmagn}            
\end{eqnarray}
where $M_{p}$ is the mass of the proton and $\mu_p$ its magnetic moment. In practice, $\eta$
in the square brackets of Eq.~(\ref{Mmagn}) is set to zero; the difference from the case when
$\eta$ is nonzero is very small. The nuclear elastic and charge-exchange amplitudes
are given by
\begin{eqnarray}
   \langle s'm'| M^{C}_{C+N}(\theta) | s\,m \rangle = 
   \sum_{\ell\,\ell' J}\, && \sqrt{4\pi(2\ell+1)} \: i^{\ell-\ell'} \:
   C^{\ell}_{0}\,^{s}_{m}\,^{J}_{m} \:
   C^{\ell'}_{m-m'}\,^{s'}_{m'}\,^{J}_{m} \:
   Y^{\ell'}_{m-m'}(\theta) \nonumber \\
  \hspace{-1em}&&\,\times\langle\ell's'|S_C^{1/2}
   \left(S^{C}_{C+N}-1\right)
   S_C^{1/2}|\ell\,s\rangle /(2ip)~,
\label{Mnucl}
\end{eqnarray}
where the $C$'s are Clebsch-Gordan coefficients, $Y$ is a spherical harmonic,
and $\left|\ell'-\ell\right|=0, \,2$. $S^{C}_{C+N}$ is the nuclear $S$ matrix in the presence
of the Coulomb force, i.e., matched outside the range of the nuclear forces to Coulomb
wave functions. We use matrix notation because of the presence of a tensor force
which couples the partial waves with $\left|\ell'-\ell\right|=2$. $S_C$ is the Coulomb
$S$ matrix, which is diagonal in spin and orbital angular momentum, with entries
$\langle \ell's'|S_C|\ell\,s\rangle = \delta_{\ell\ell'}\delta_{ss'} \exp(2i\xi_{\ell})$,
where $\xi_{\ell} \: = \: \arg\,\Gamma(\ell+1+i\eta)$ for elastic scattering.  $S_C$ is
equal to the identity matrix for charge-exchange scattering.

\begin{figure}[bhtp]
   \centering
   \includegraphics[width=0.45\textwidth]{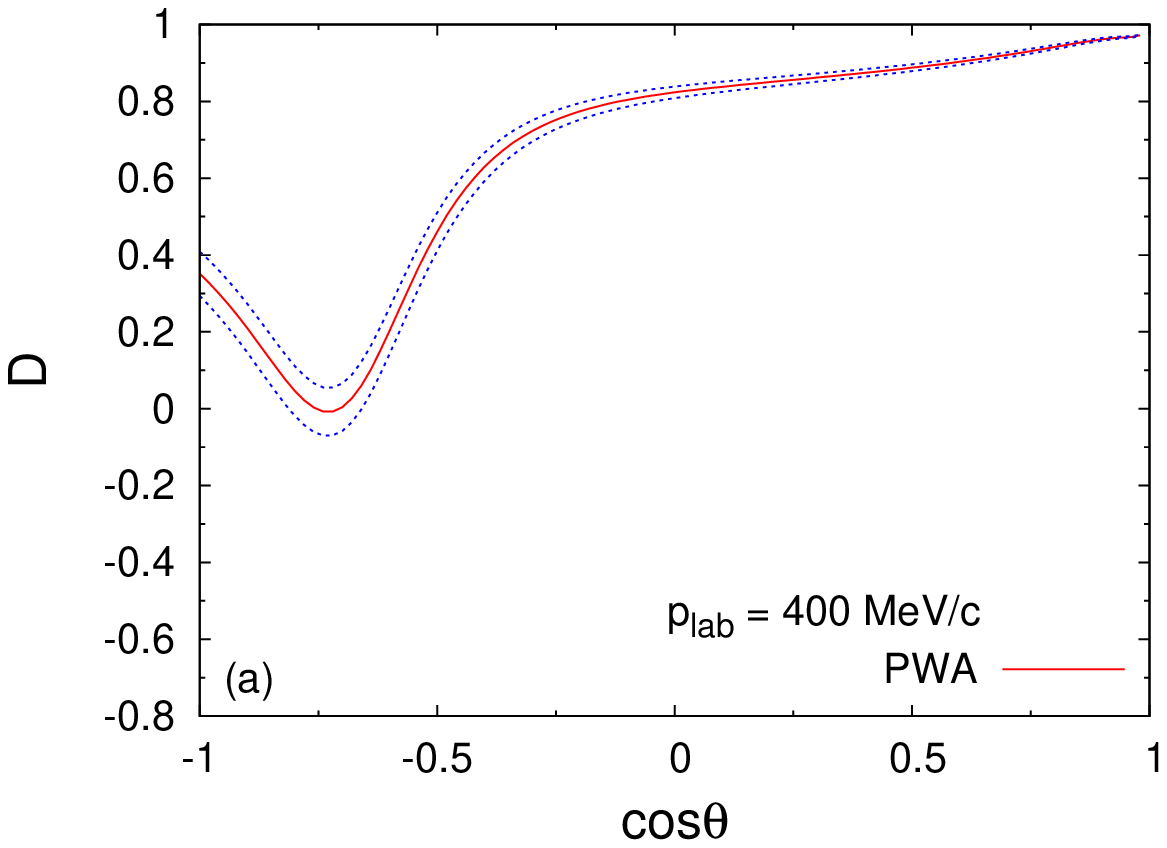} \hspace{2em}
   \includegraphics[width=0.45\textwidth]{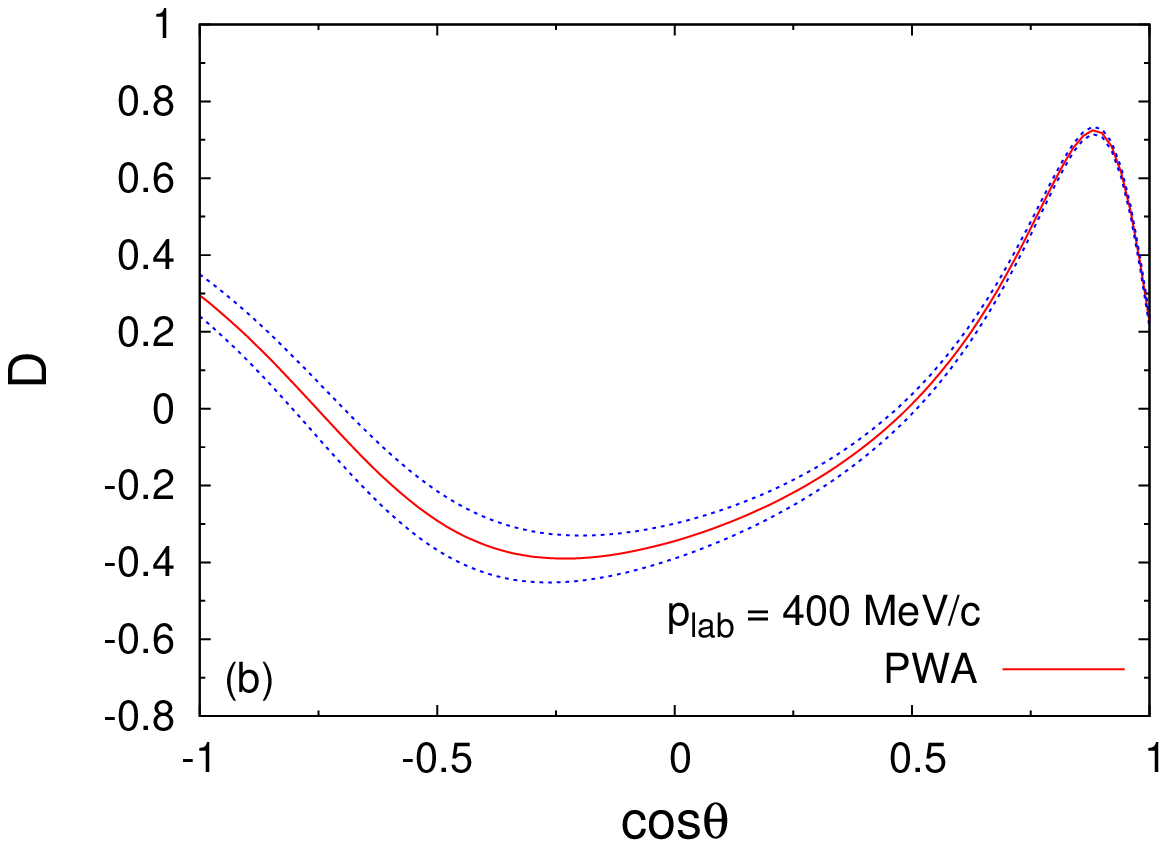}\\
   \includegraphics[width=0.45\textwidth]{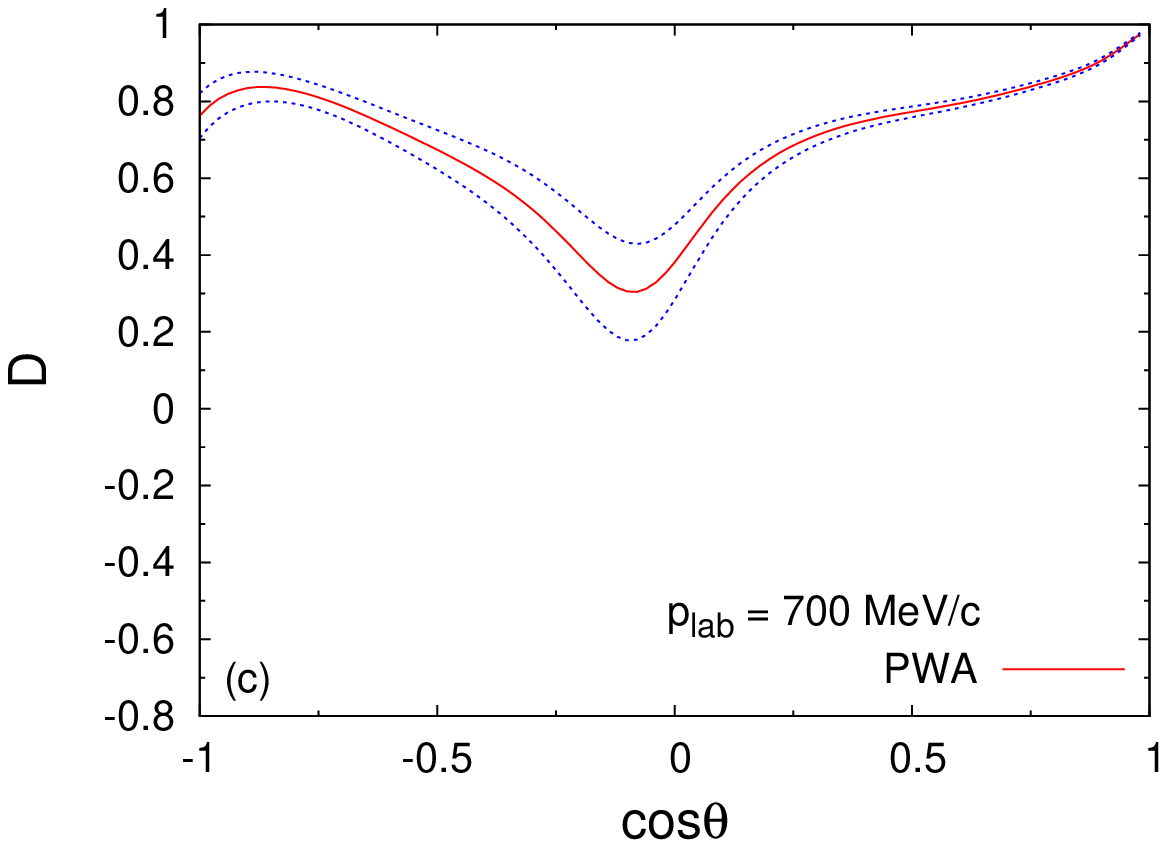} \hspace{2em}
   \includegraphics[width=0.45\textwidth]{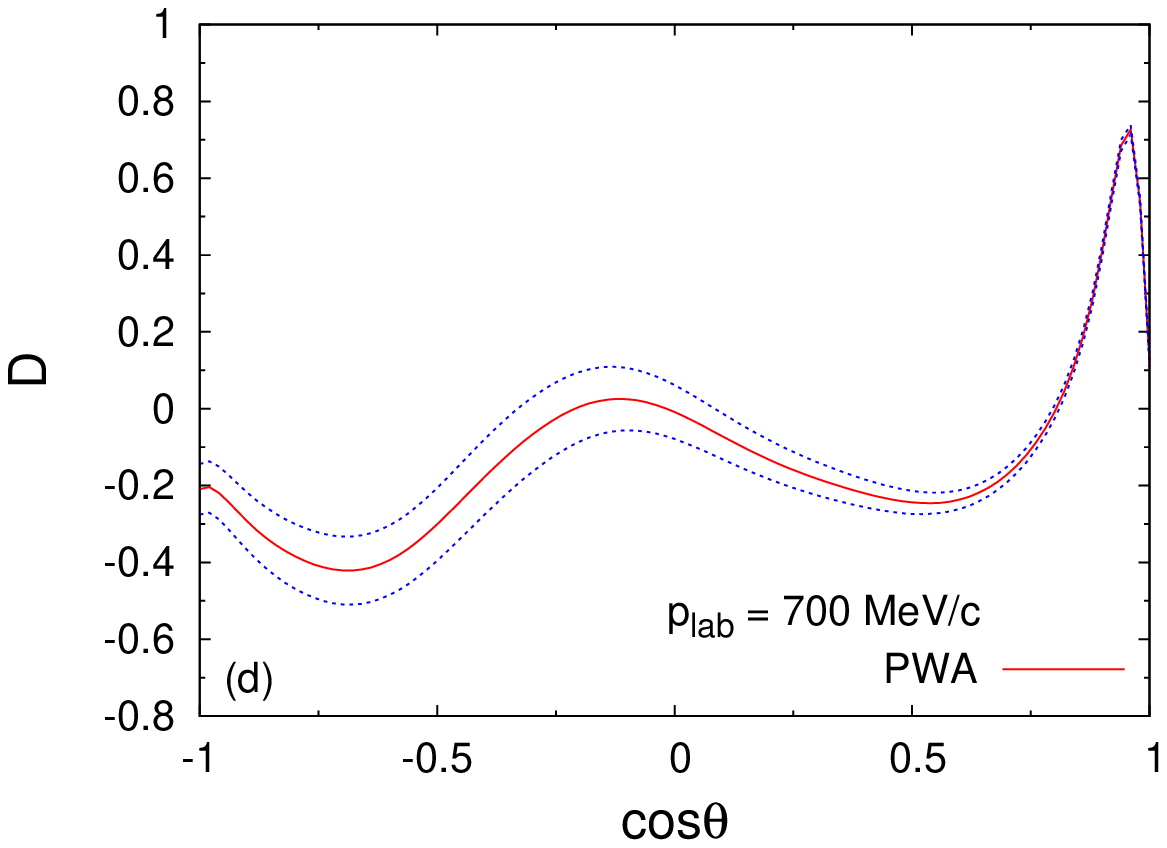}\\   
   \caption{\label{D}{(Color online) The depolarization $D$ for $\overline{p}p$ elastic (left) and charge-exchange (right)
   scattering at 400 and 700 MeV/$c$ laboratory momentum.
   The PWA result is given by the solid red line and the dotted blue lines indicate the $1\sigma$ uncertainty region.}}
\end{figure}

\begin{figure}[htbp]
   \centering
   \includegraphics[width=0.45\textwidth]{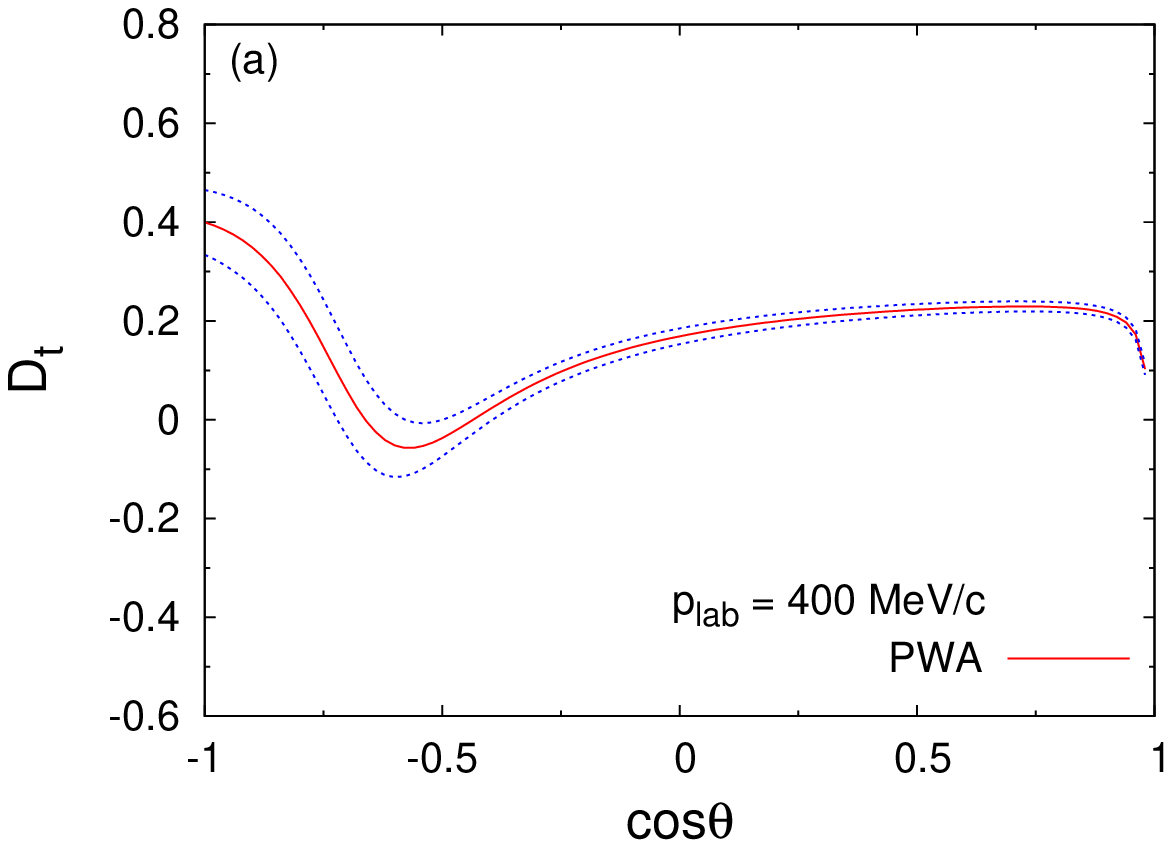} \hspace{2em}
   \includegraphics[width=0.45\textwidth]{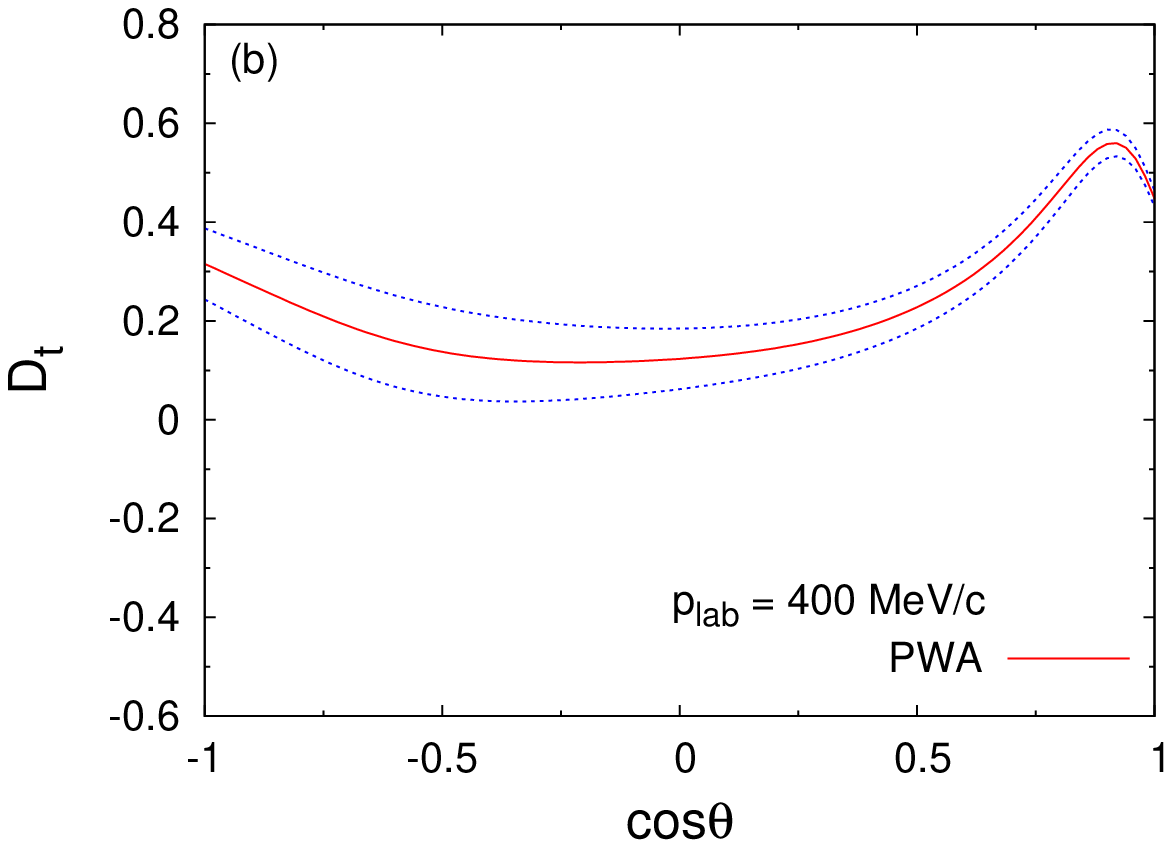}\\
   \includegraphics[width=0.45\textwidth]{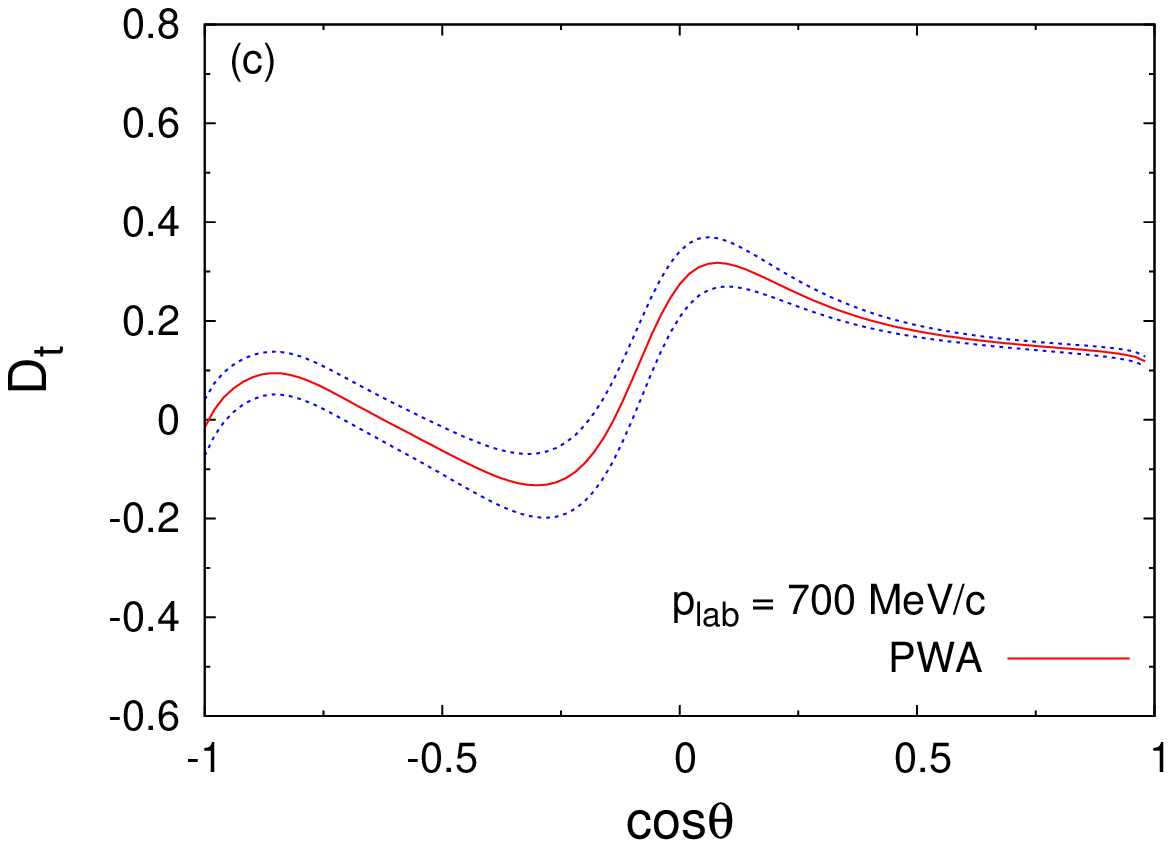} \hspace{2em}
   \includegraphics[width=0.45\textwidth]{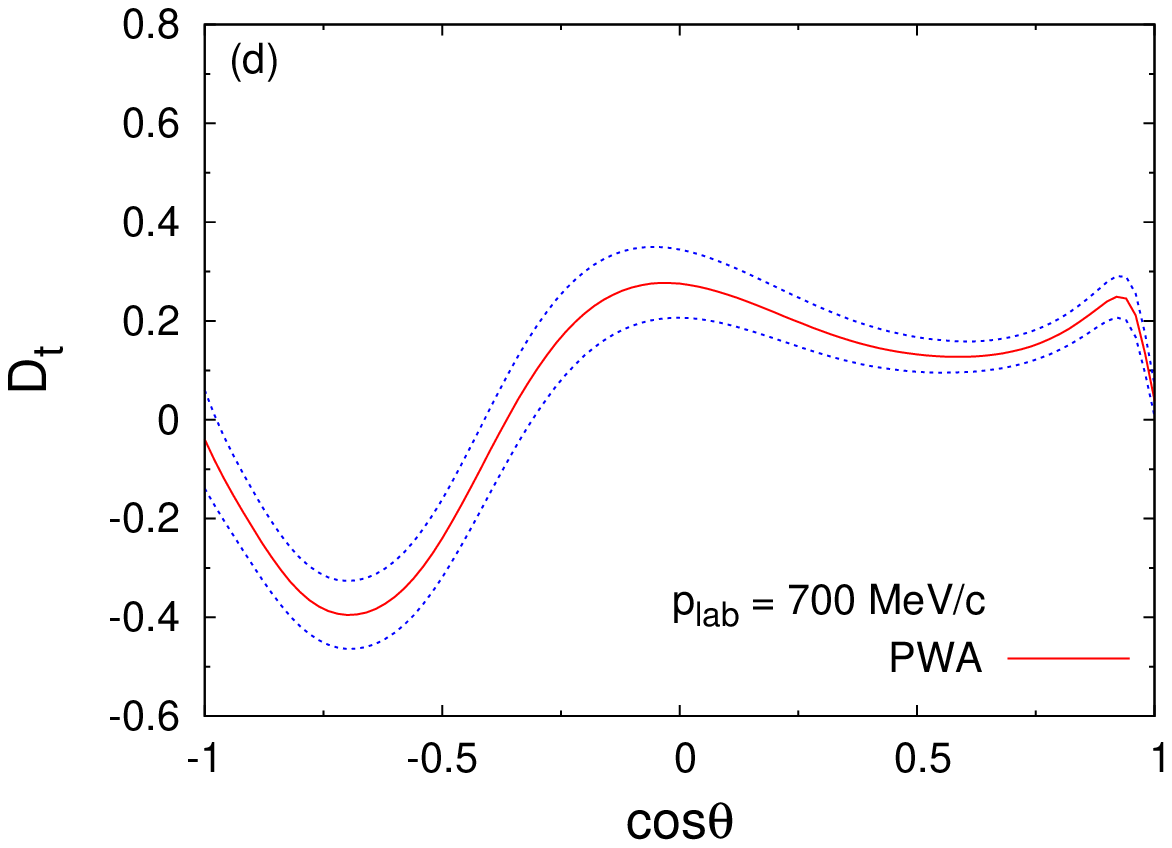}\\   
   \caption{\label{Dt}{(Color online) The polarization transfer $D_{t}$ for $\overline{p}p$ elastic (left) and charge-exchange (right)
   scattering at 400 and 700 MeV/$c$ laboratory momentum.
   The PWA result is given by the solid red line and the dotted blue lines indicate the $1\sigma$ uncertainty region.}}
\end{figure}

\begin{figure}[htbp]
   \centering
   \includegraphics[width=0.45\textwidth]{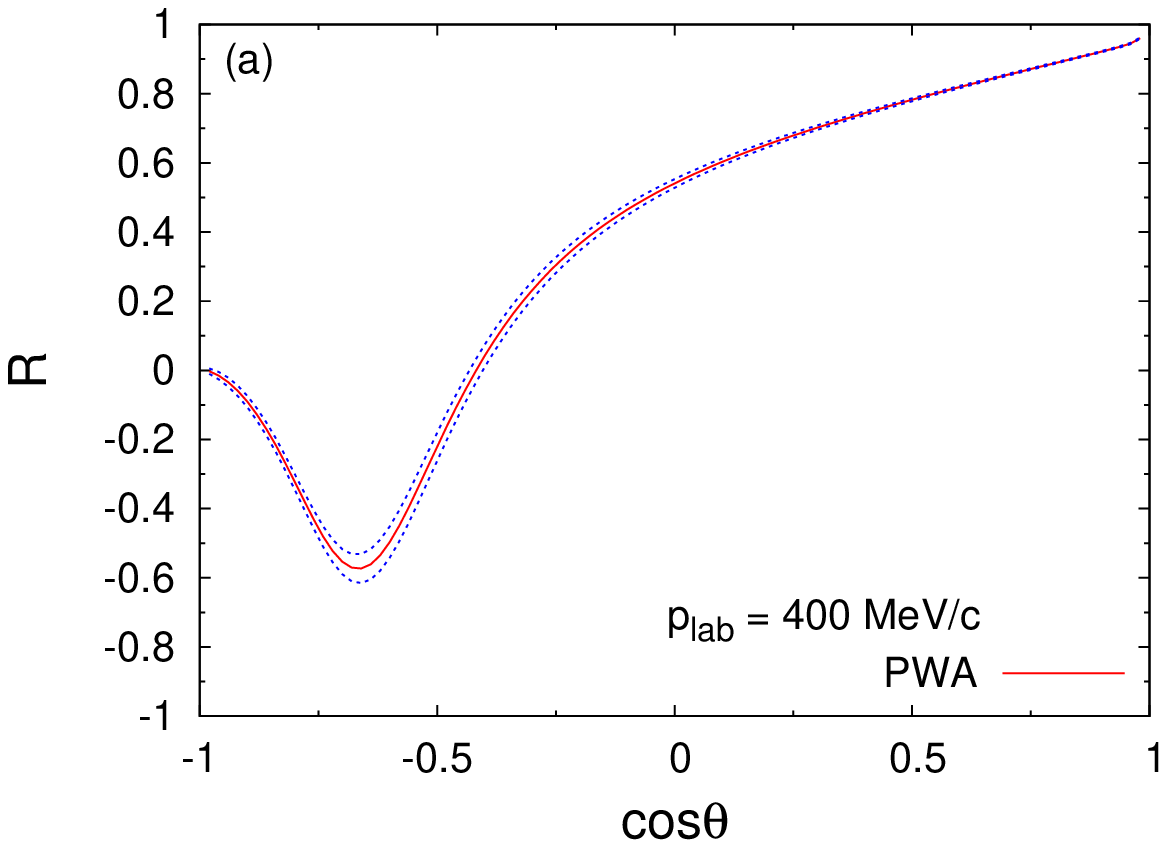} \hspace{2em}
   \includegraphics[width=0.45\textwidth]{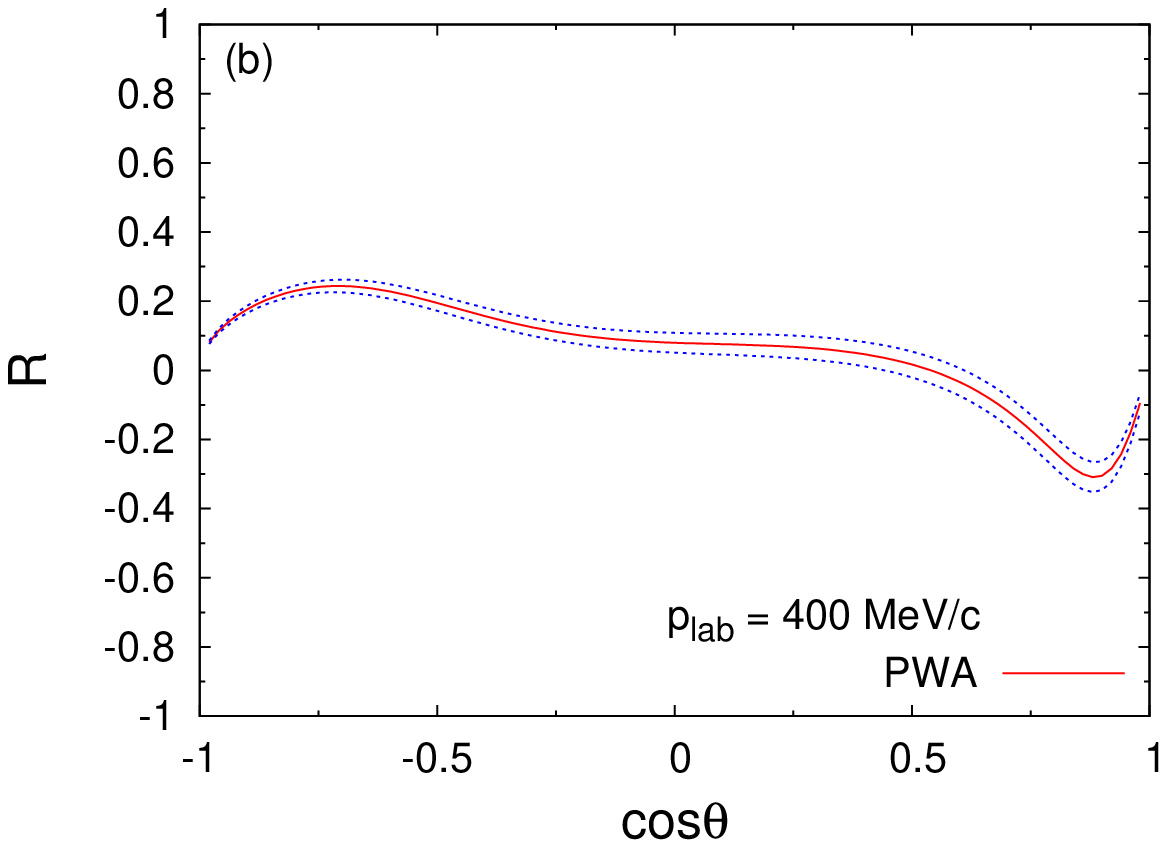}\\
   \includegraphics[width=0.45\textwidth]{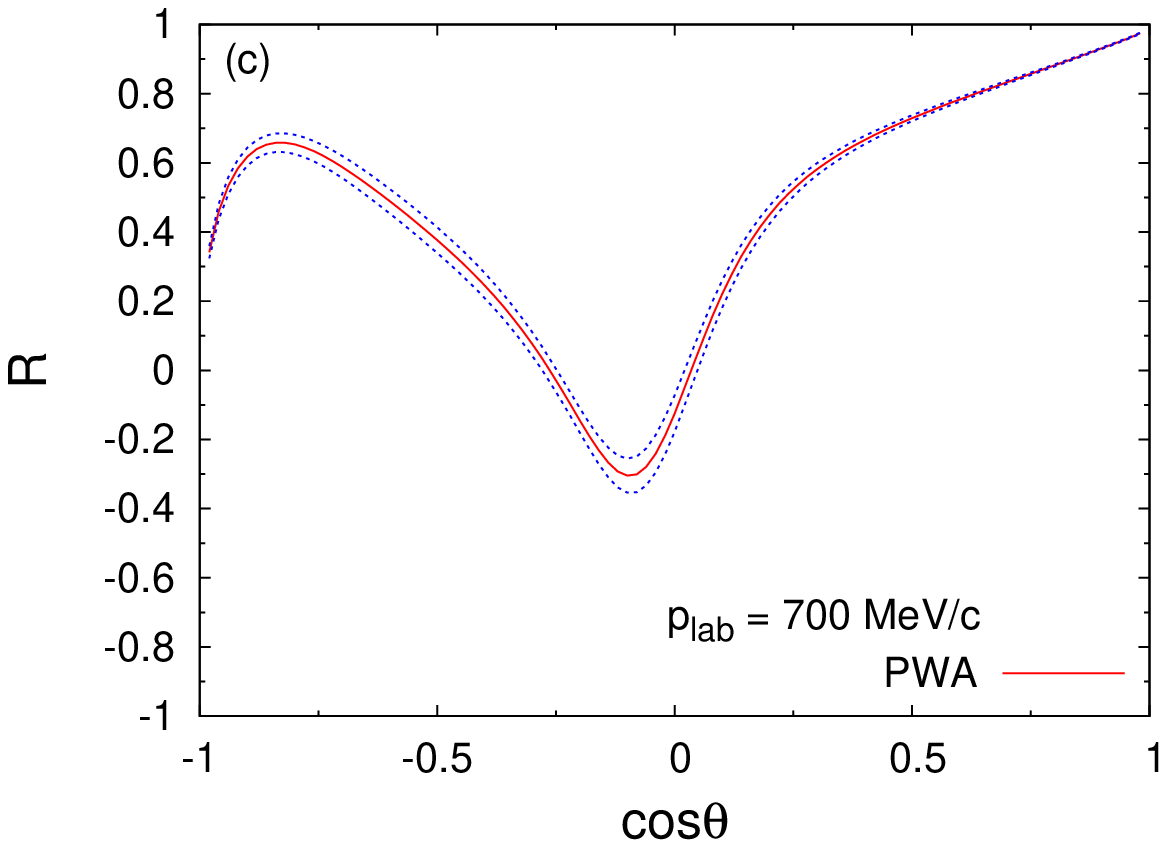} \hspace{2em}
   \includegraphics[width=0.45\textwidth]{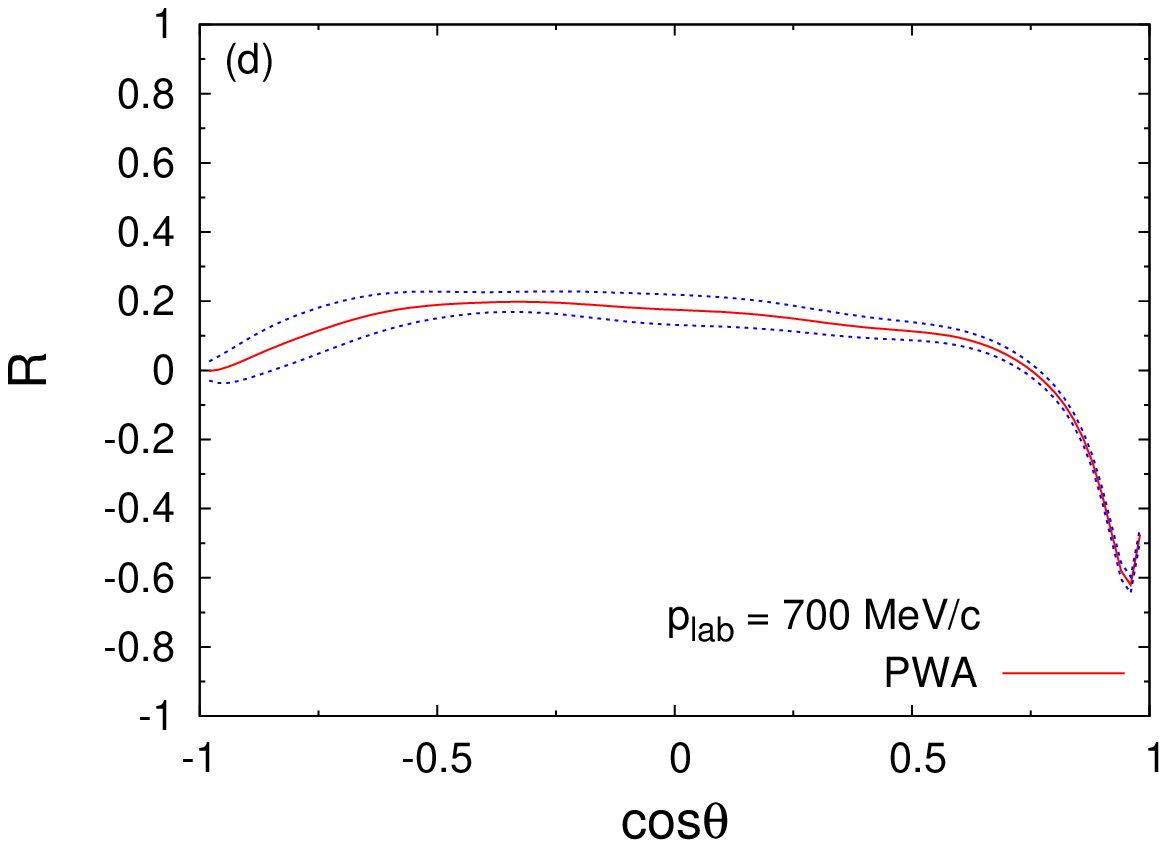}\\   
   \caption{\label{R}{(Color online) The rotation parameter $R$ for $\overline{p}p$ elastic (left) and charge-exchange (right)
   scattering at 400 and 700 MeV/$c$ laboratory momentum.
   The PWA result is given by the solid red line and the dotted blue lines indicate the $1\sigma$ uncertainty region.}}
\end{figure}

\begin{figure}[htbp]
   \centering
   \includegraphics[width=0.45\textwidth]{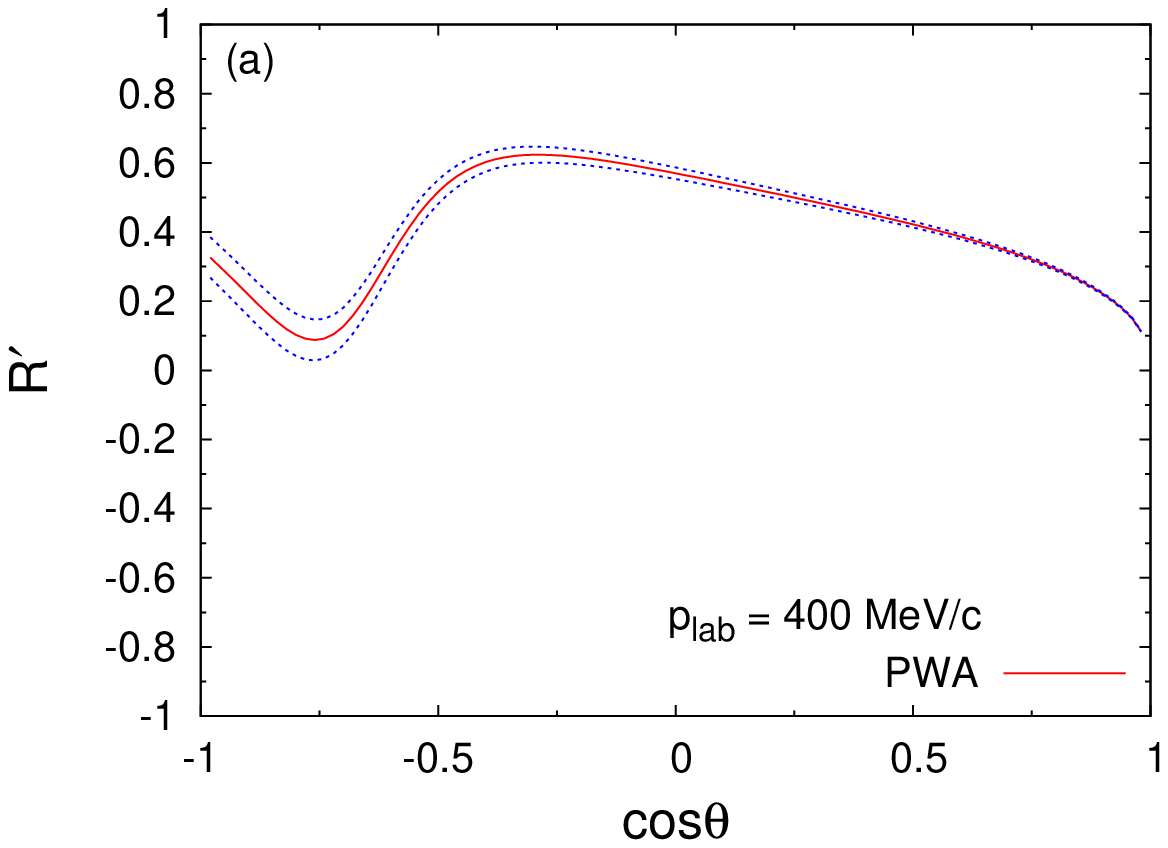} \hspace{2em}
   \includegraphics[width=0.45\textwidth]{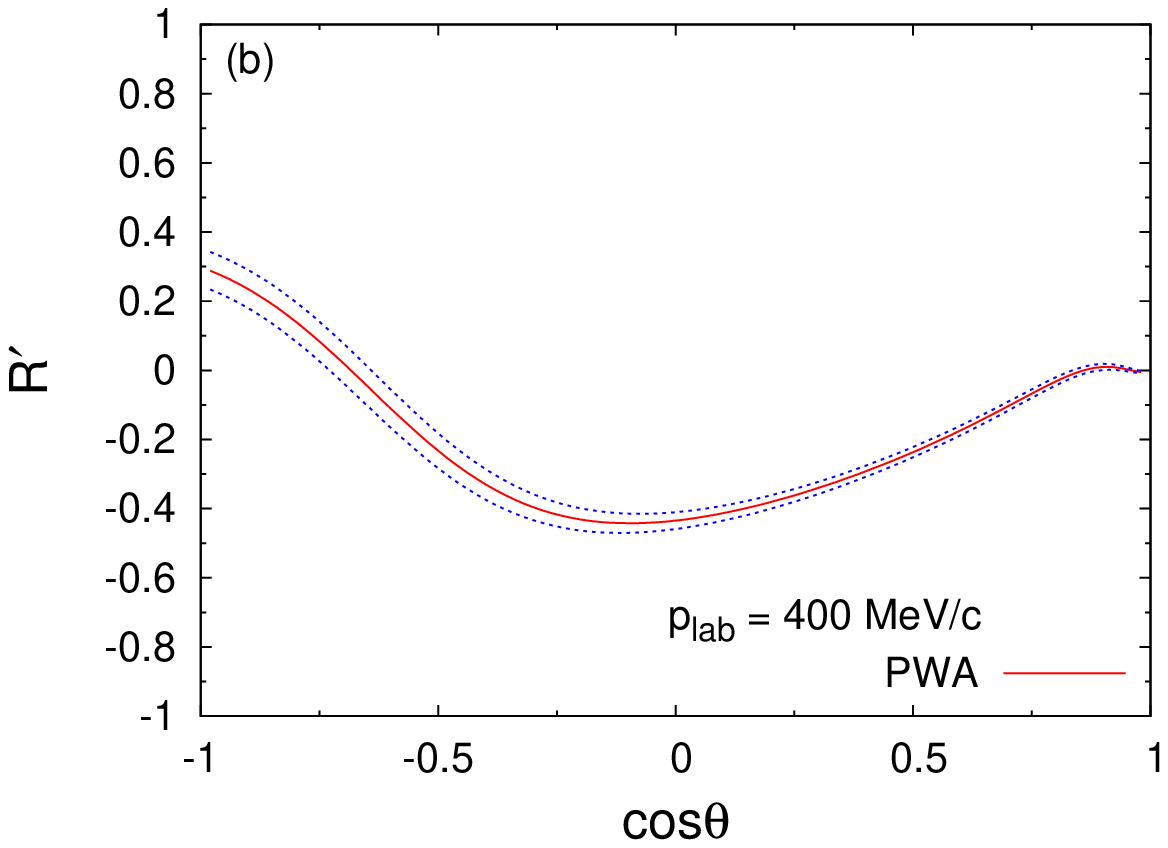}\\
   \includegraphics[width=0.45\textwidth]{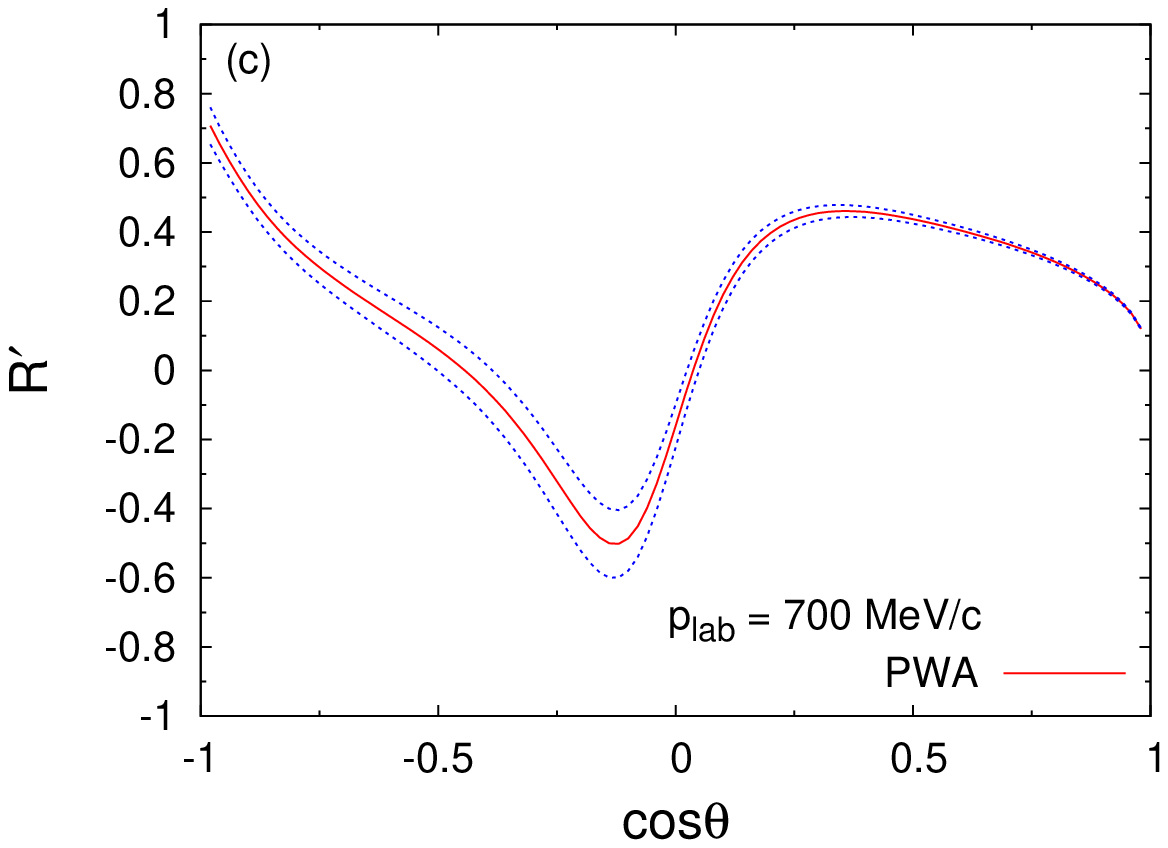} \hspace{2em}
   \includegraphics[width=0.45\textwidth]{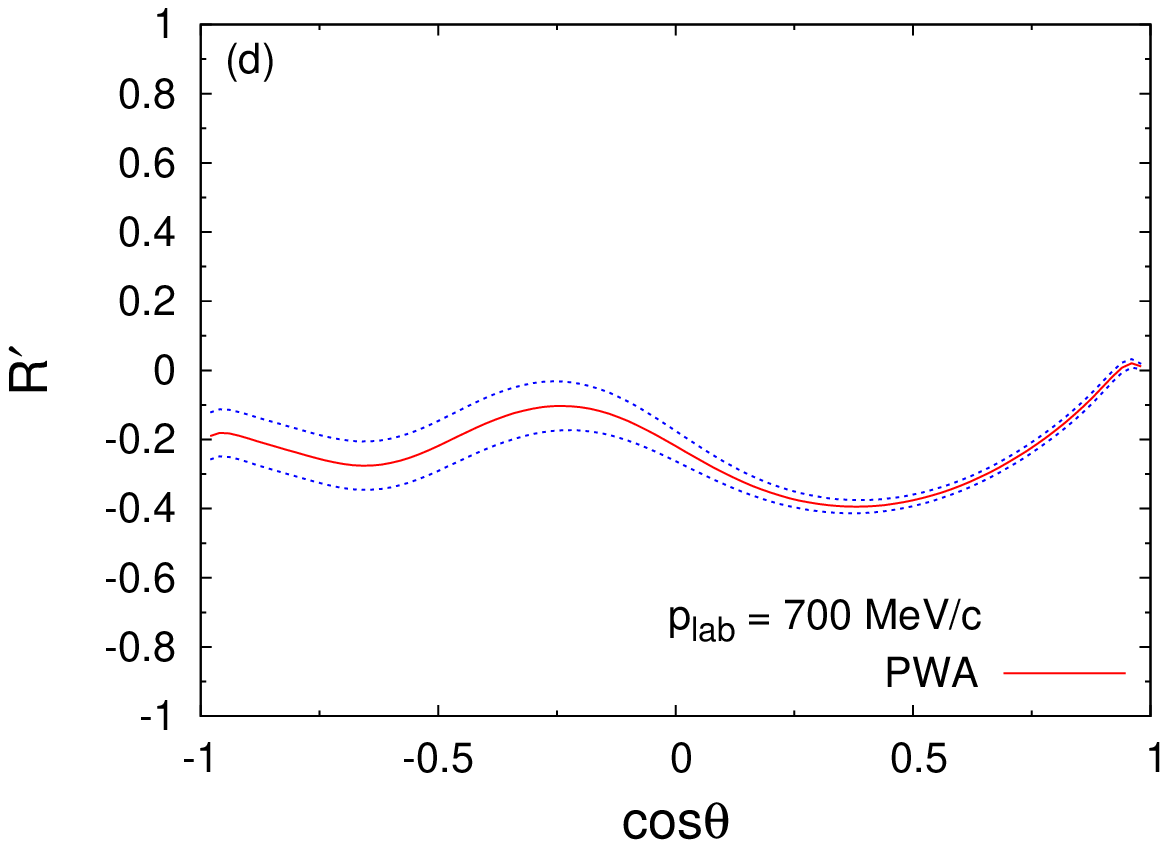}\\   
   \caption{\label{Rp}{(Color online) The rotation parameter $R'$ for $\overline{p}p$ elastic (left) and charge-exchange (right)
   scattering at 400 and 700 MeV/$c$ laboratory momentum.
   The PWA result is given by the solid red line and the dotted blue lines indicate the $1\sigma$ uncertainty region.}}
\end{figure}

\begin{figure}[htbp]
   \centering
   \includegraphics[width=0.45\textwidth]{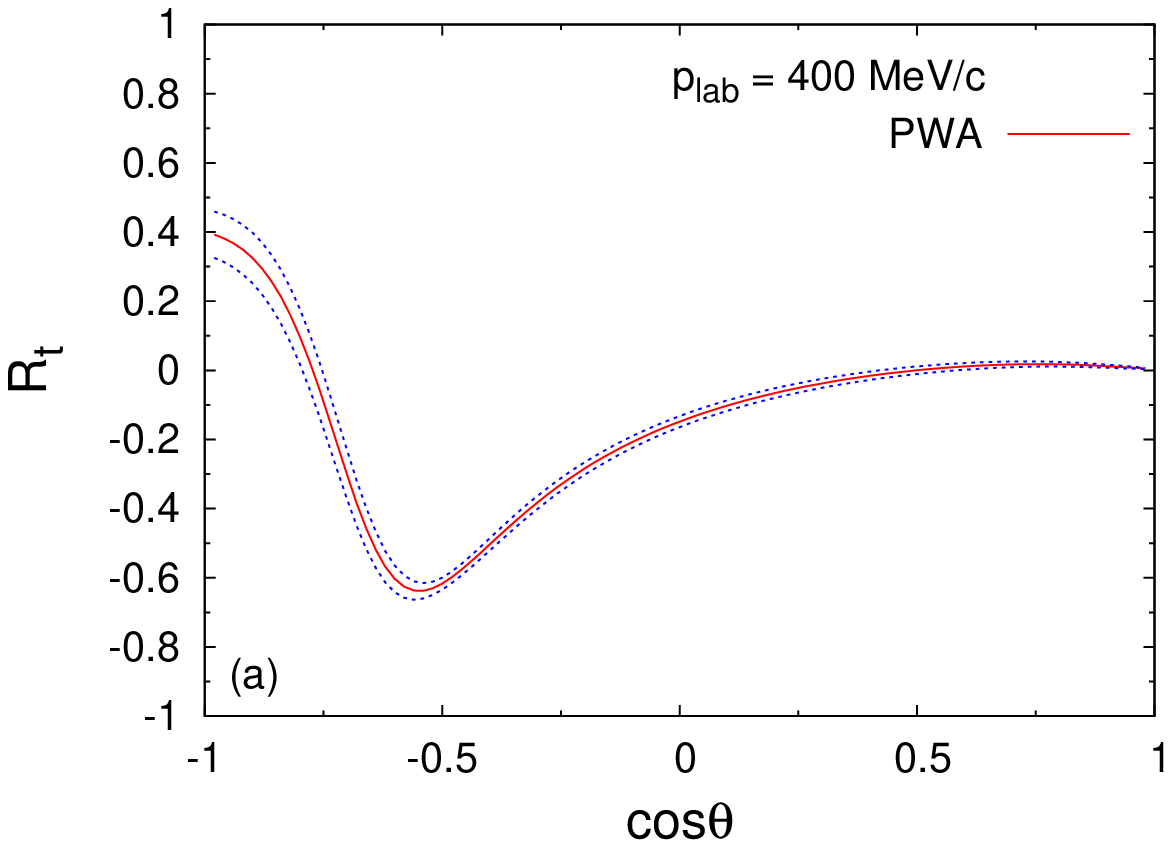} \hspace{2em}
   \includegraphics[width=0.45\textwidth]{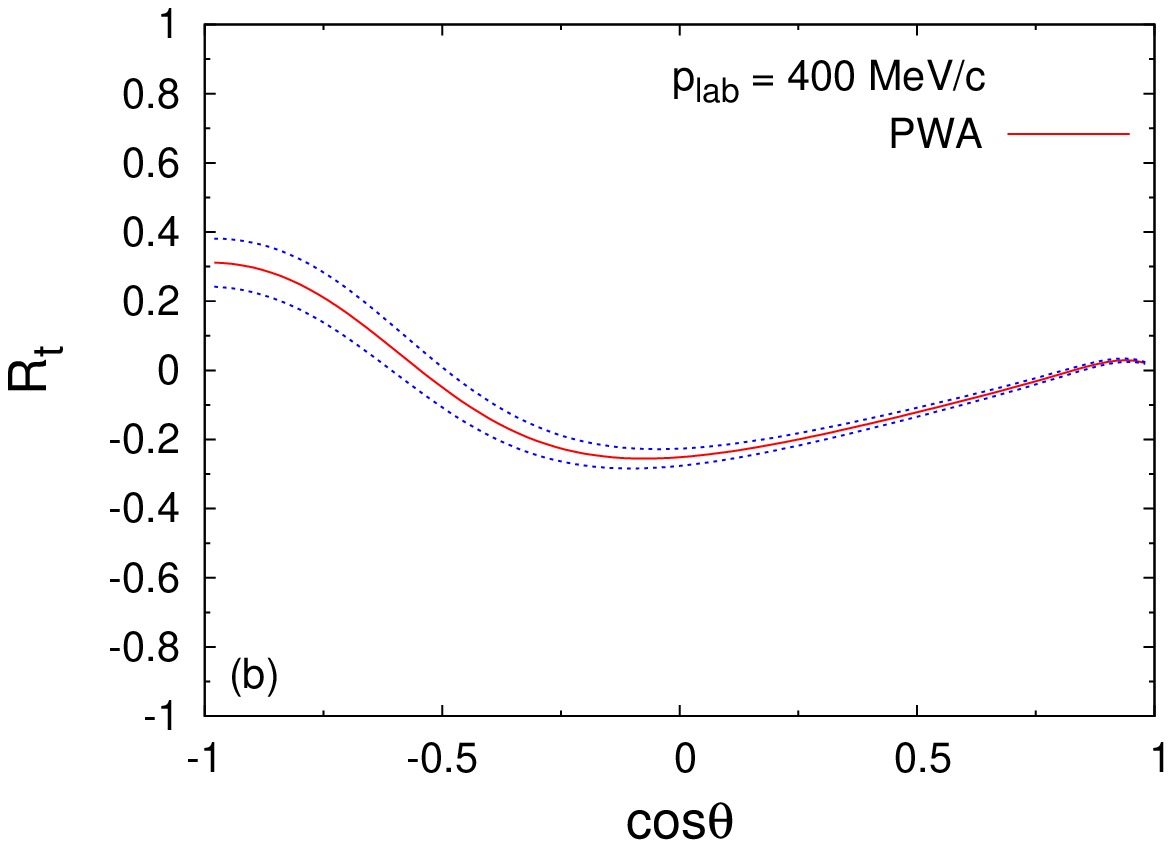}\\
   \includegraphics[width=0.45\textwidth]{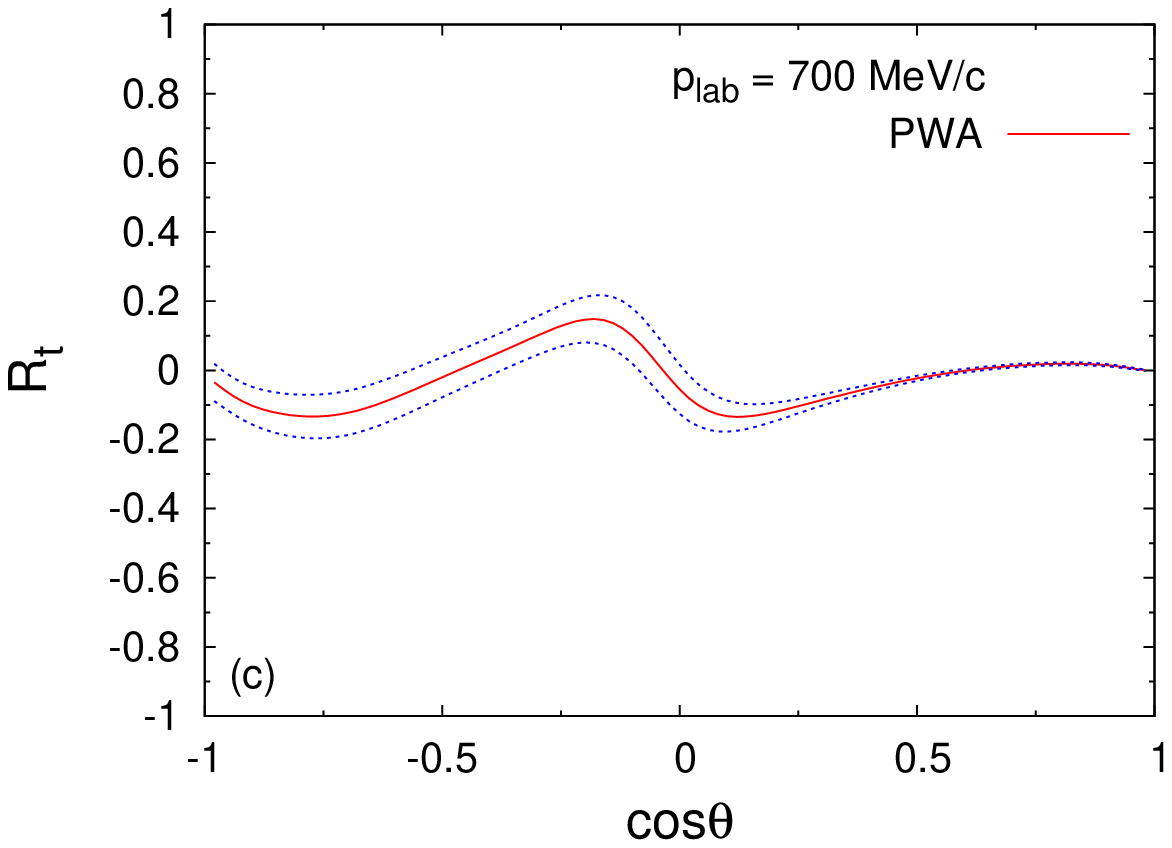} \hspace{2em}
   \includegraphics[width=0.45\textwidth]{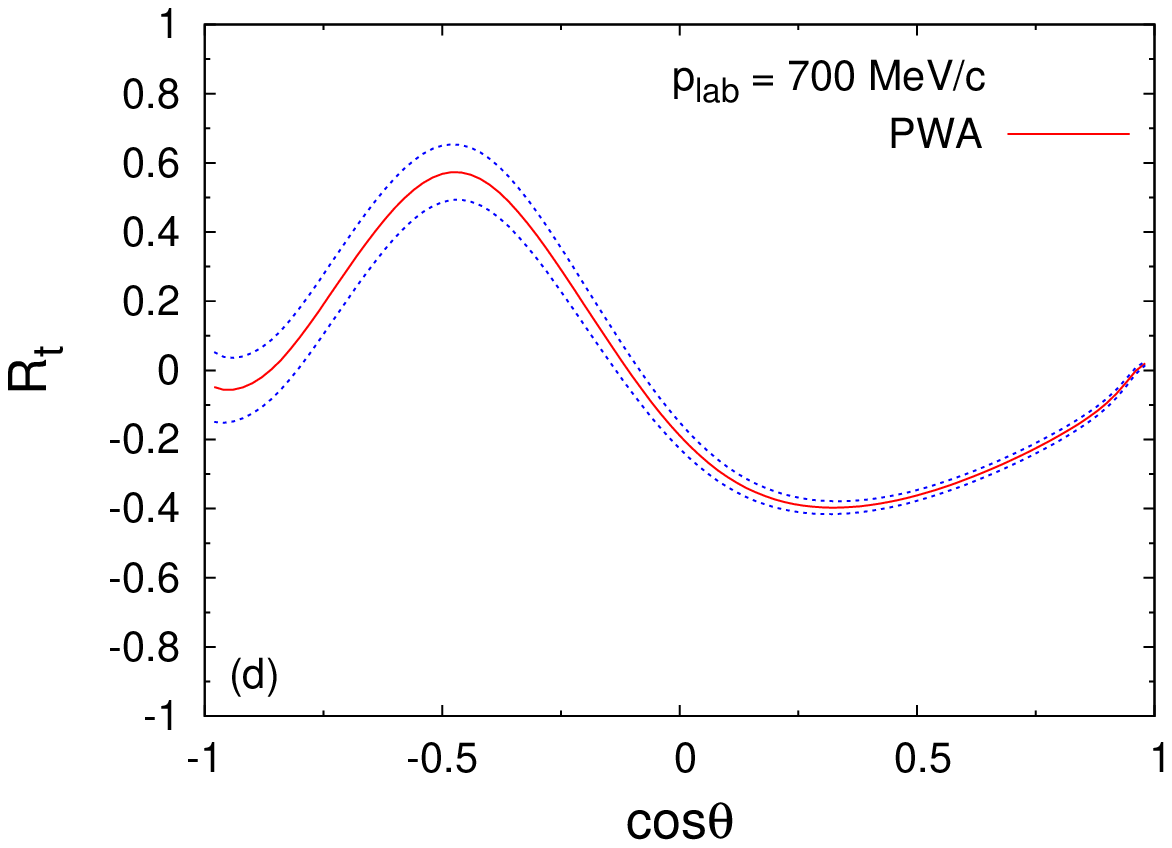}\\   
   \caption{\label{Rt}{(Color online) The polarization transfer $R_{t}$ for $\overline{p}p$ elastic (left) and charge-exchange (right)
   scattering at 400 and 700 MeV/$c$ laboratory momentum.
   The PWA result is given by the solid red line and the dotted blue lines indicate the $1\sigma$ uncertainty region.}}
\end{figure}

\begin{figure}[htbp]
   \centering
   \includegraphics[width=0.45\textwidth]{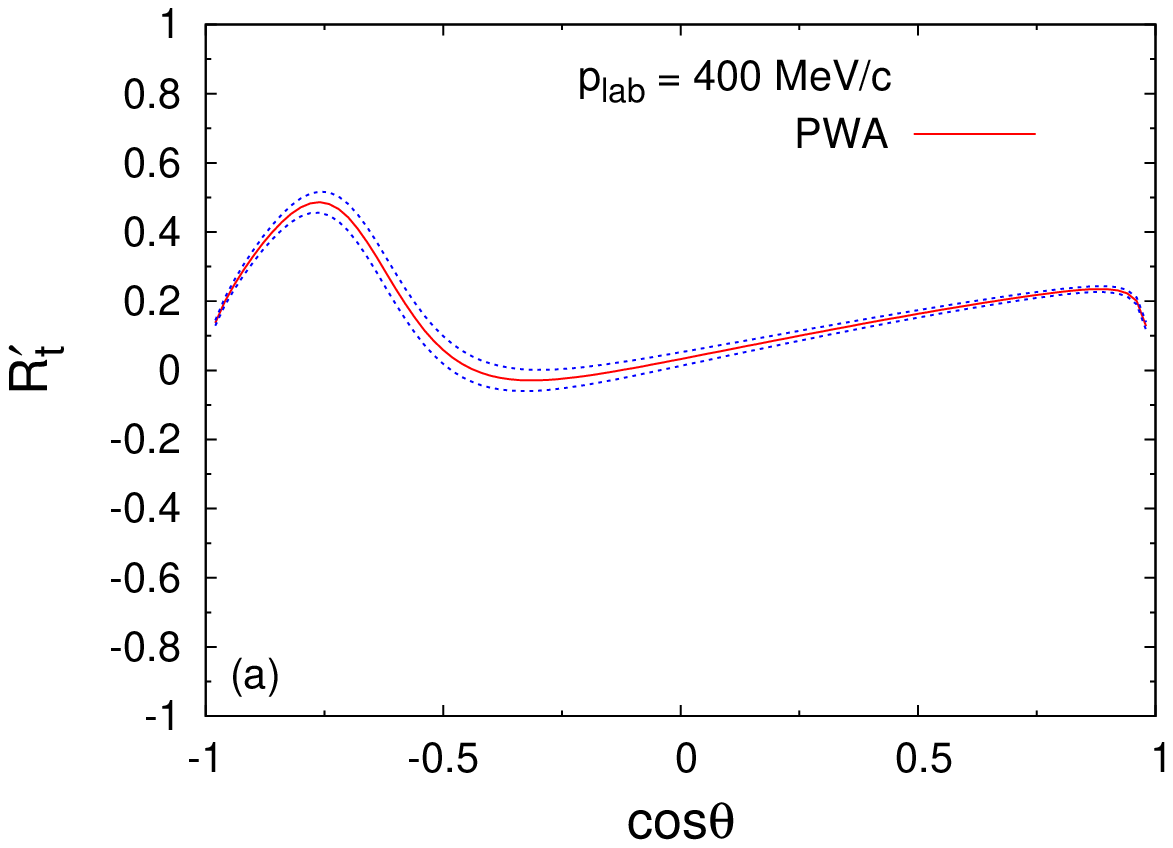} \hspace{2em}
   \includegraphics[width=0.45\textwidth]{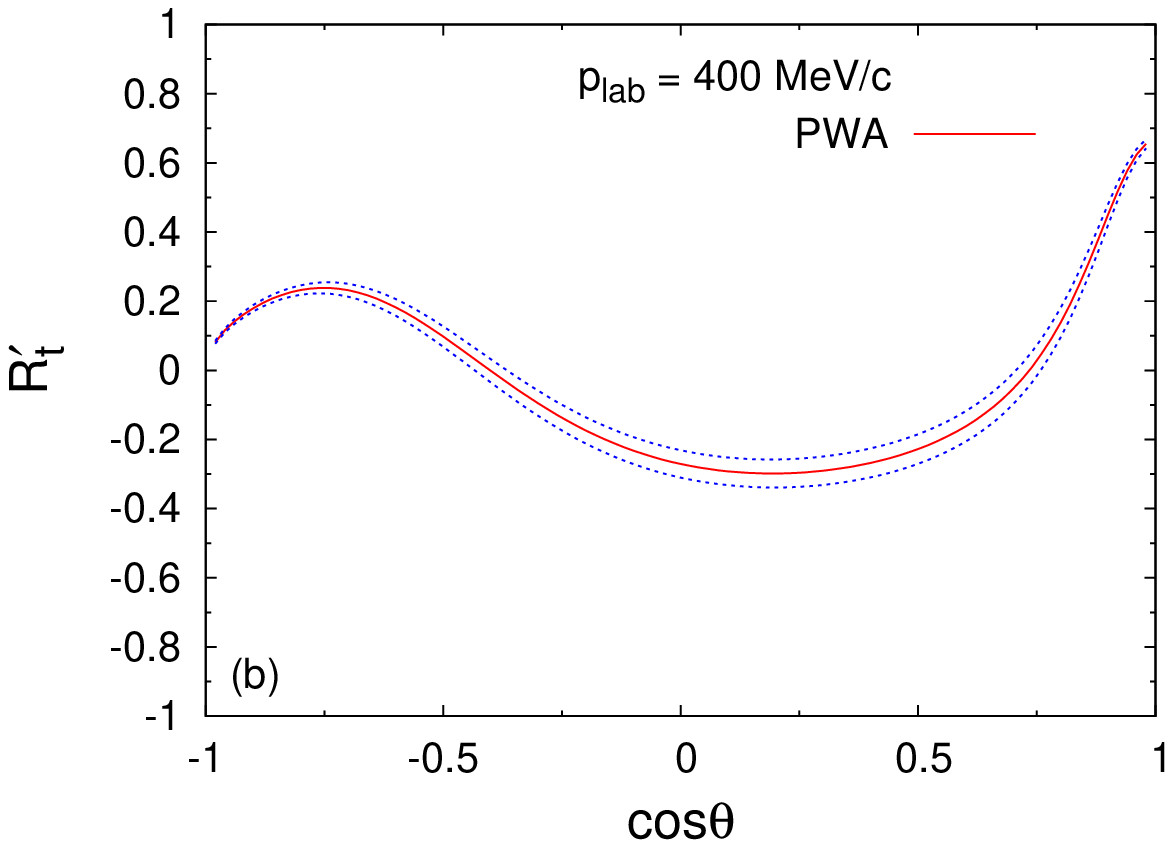}\\
   \includegraphics[width=0.45\textwidth]{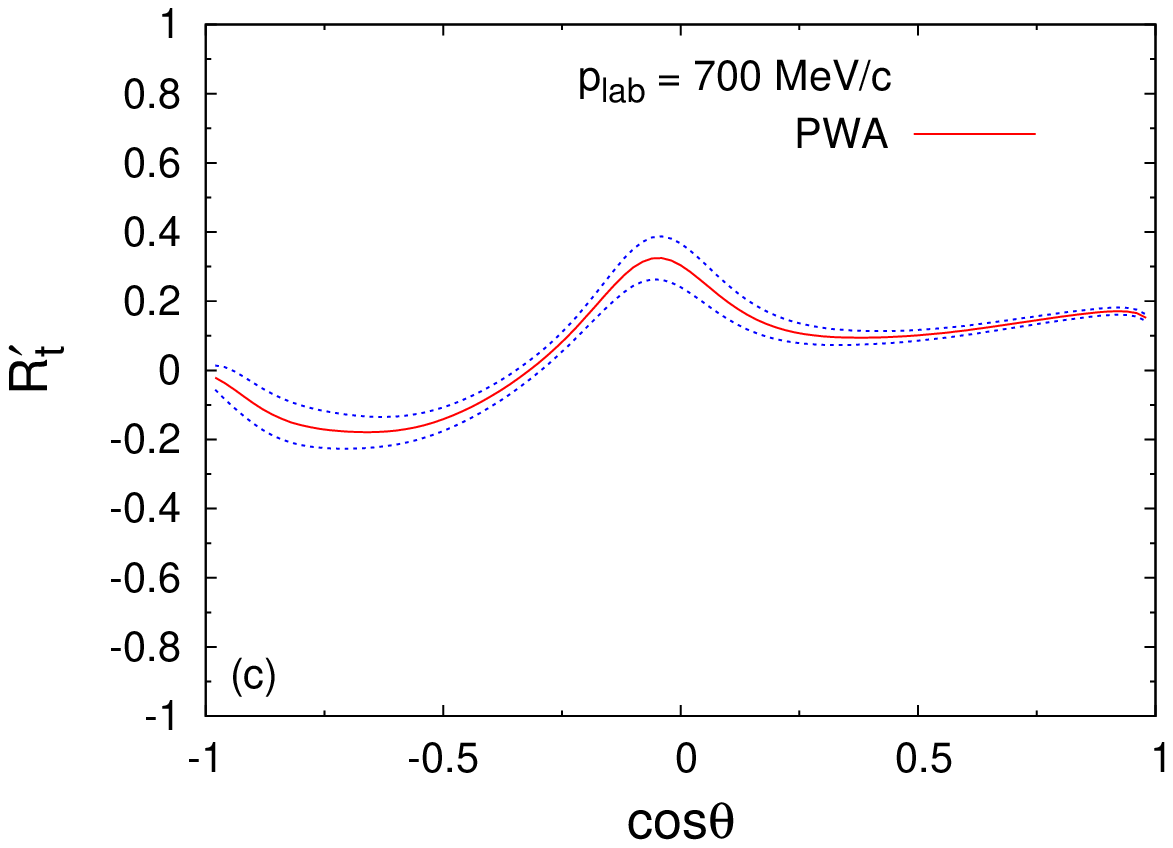} \hspace{2em}
   \includegraphics[width=0.45\textwidth]{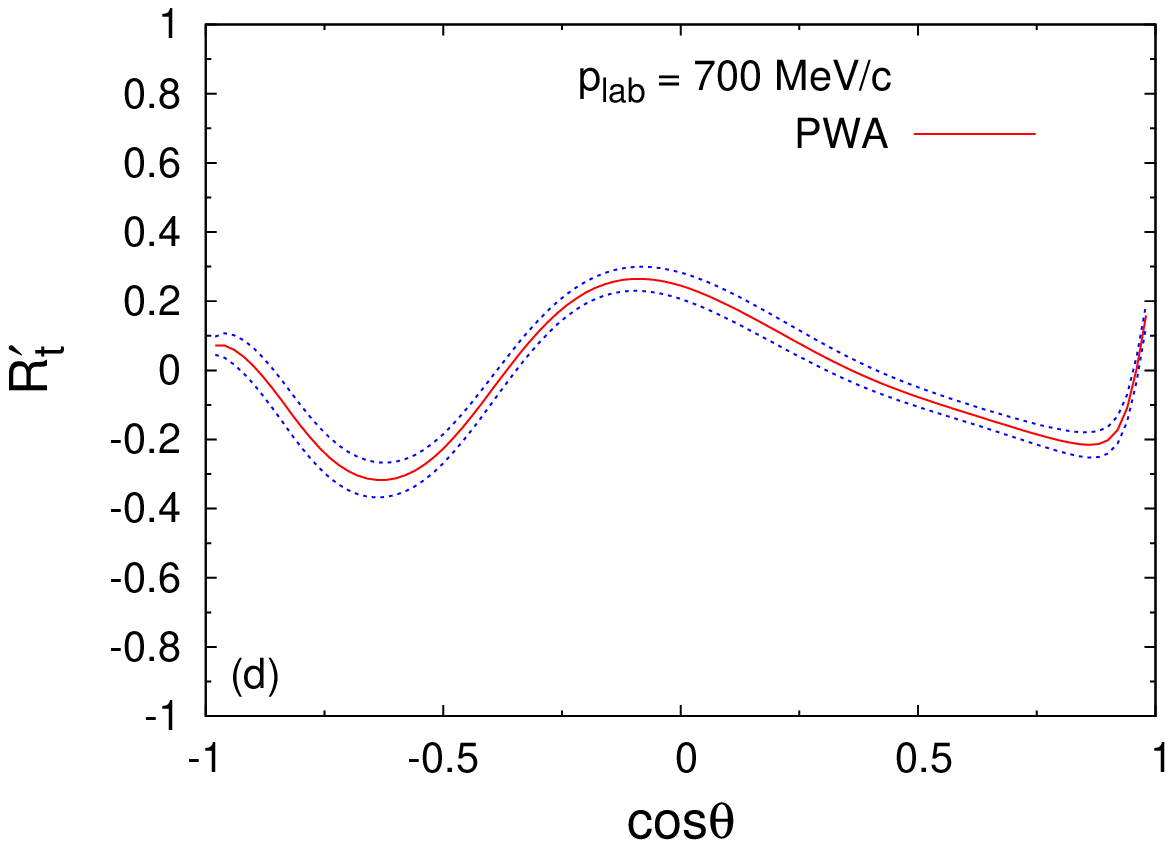}\\   
   \caption{\label{Rtp}{(Color online) The polarization transfer $R'_{t}$ for $\overline{p}p$ elastic (left) and charge-exchange (right)
   scattering at 400 and 700 MeV/$c$ laboratory momentum.
   The PWA result is given by the solid red line and the dotted blue lines indicate the $1\sigma$ uncertainty region.}}
\end{figure}

\begin{figure}[htbp]
   \centering
   \includegraphics[width=0.45\textwidth]{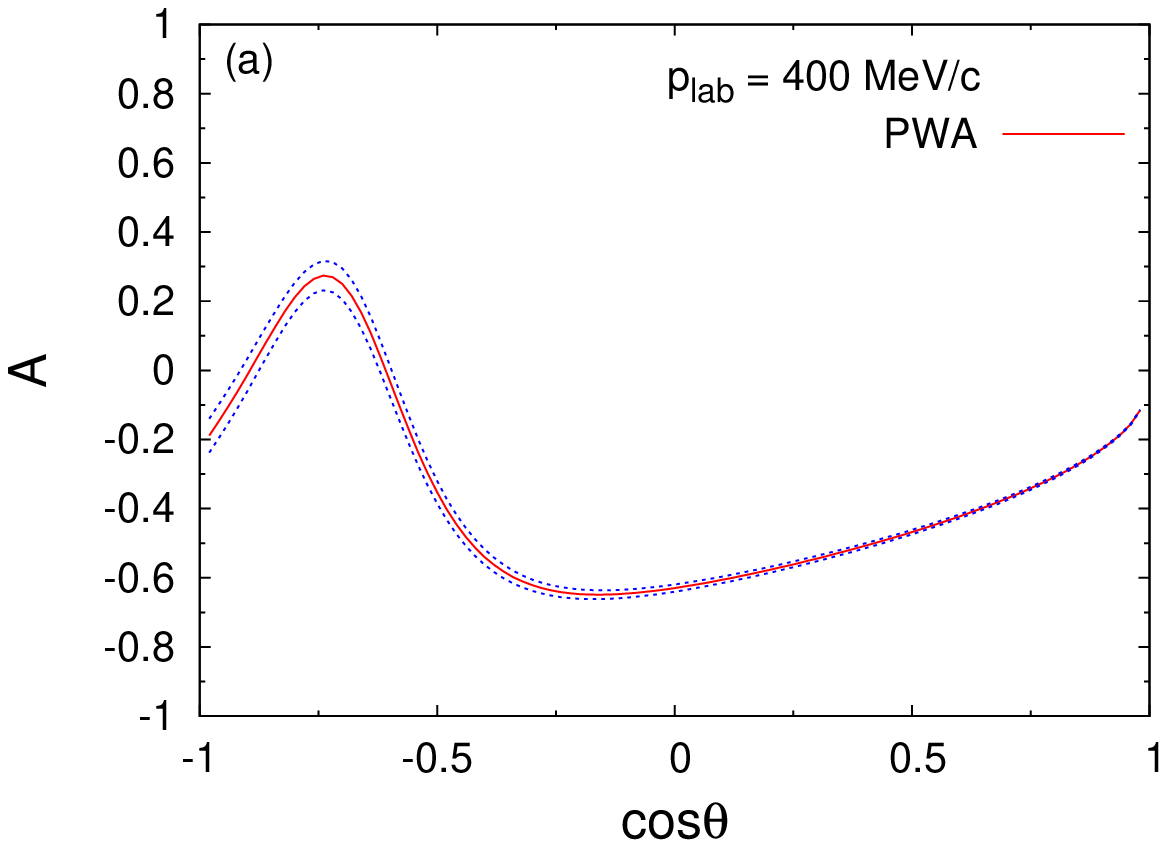} \hspace{2em}
   \includegraphics[width=0.45\textwidth]{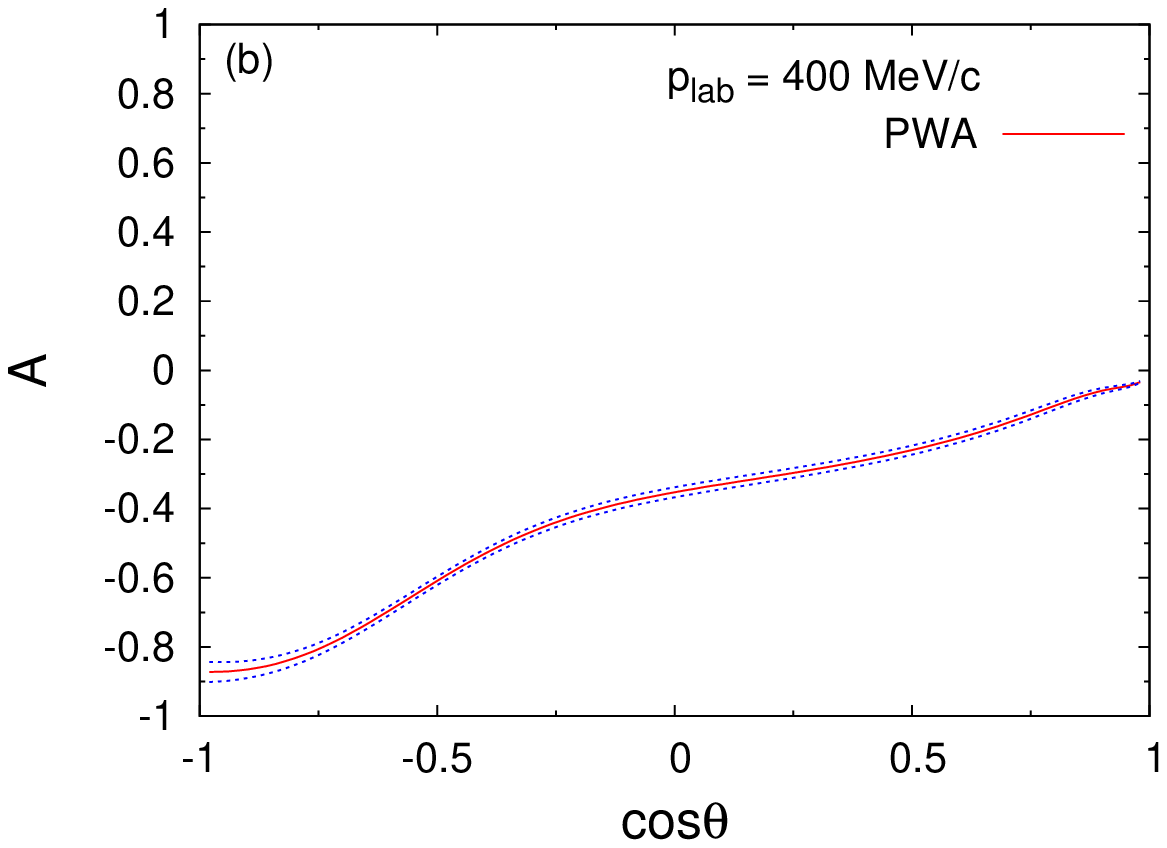}\\
   \includegraphics[width=0.45\textwidth]{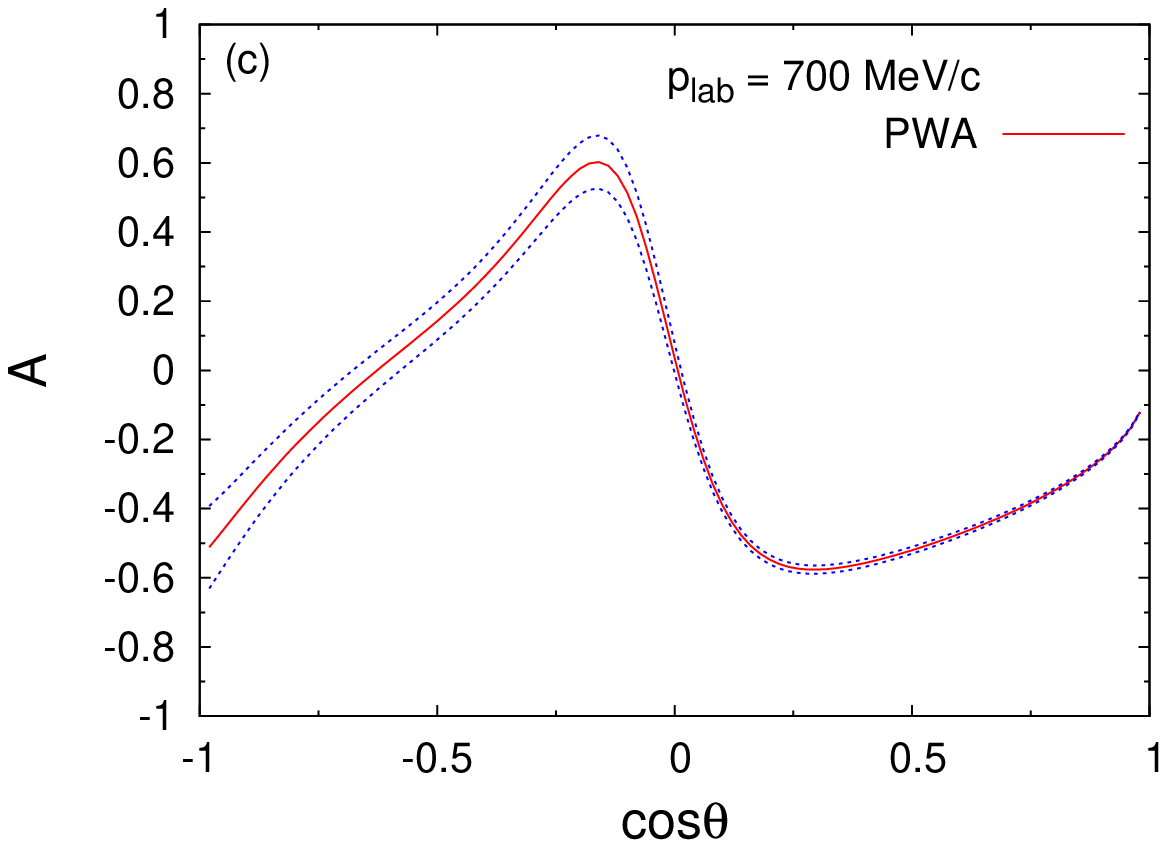} \hspace{2em}
   \includegraphics[width=0.45\textwidth]{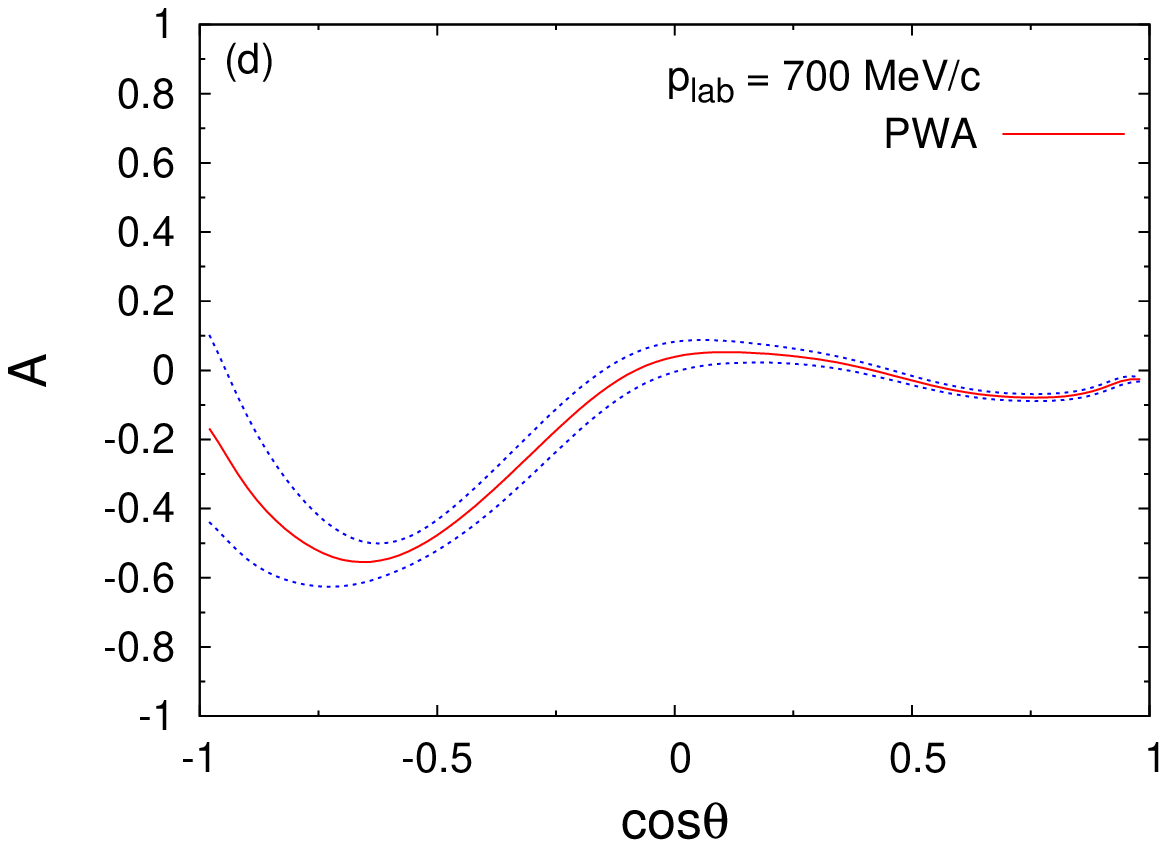}\\
    \caption{\label{A}{(Color online) The rotation parameter $A$ for $\overline{p}p$ elastic (left) and charge-exchange (right) scattering at
   400 and 700 MeV/$c$ laboratory momentum.
   The PWA result is given by the solid red line and the dotted blue lines indicate the $1\sigma$ uncertainty region.}}
\end{figure}

\begin{figure}[htbp]
   \centering
   \includegraphics[width=0.45\textwidth]{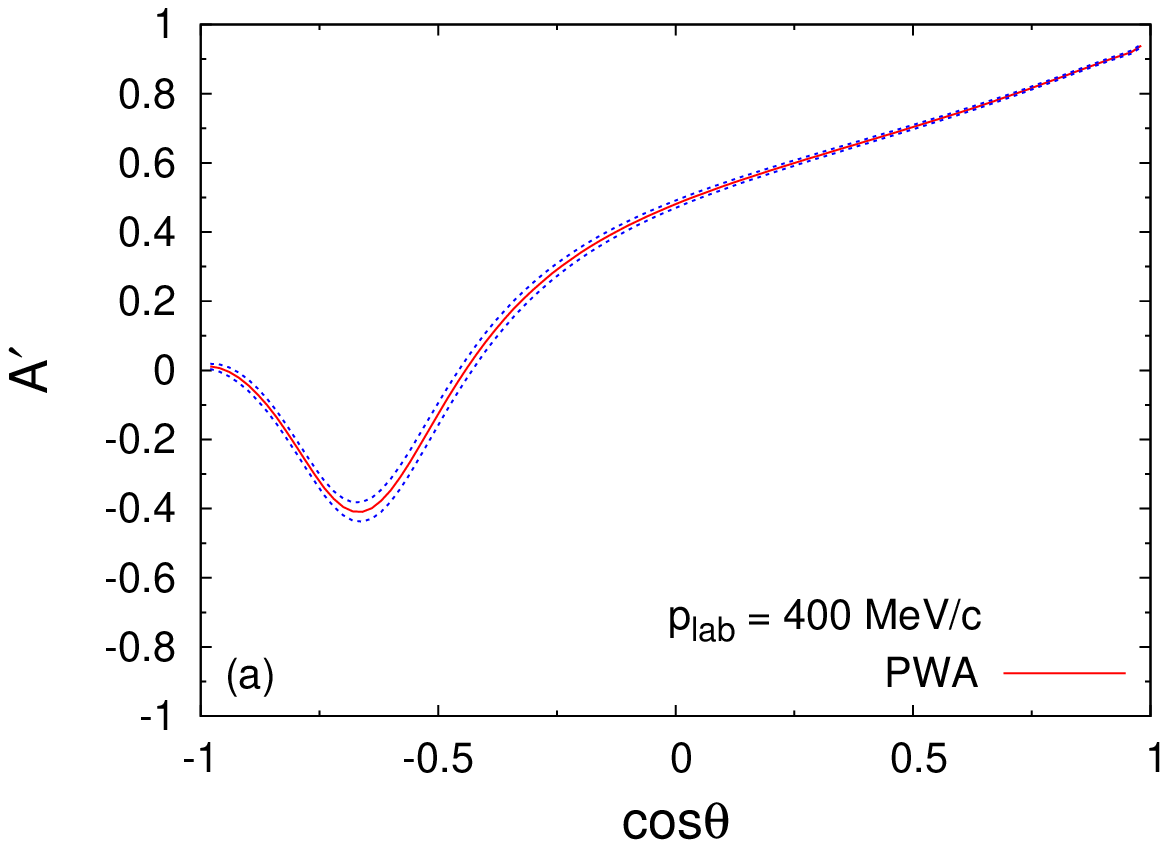} \hspace{2em}
   \includegraphics[width=0.45\textwidth]{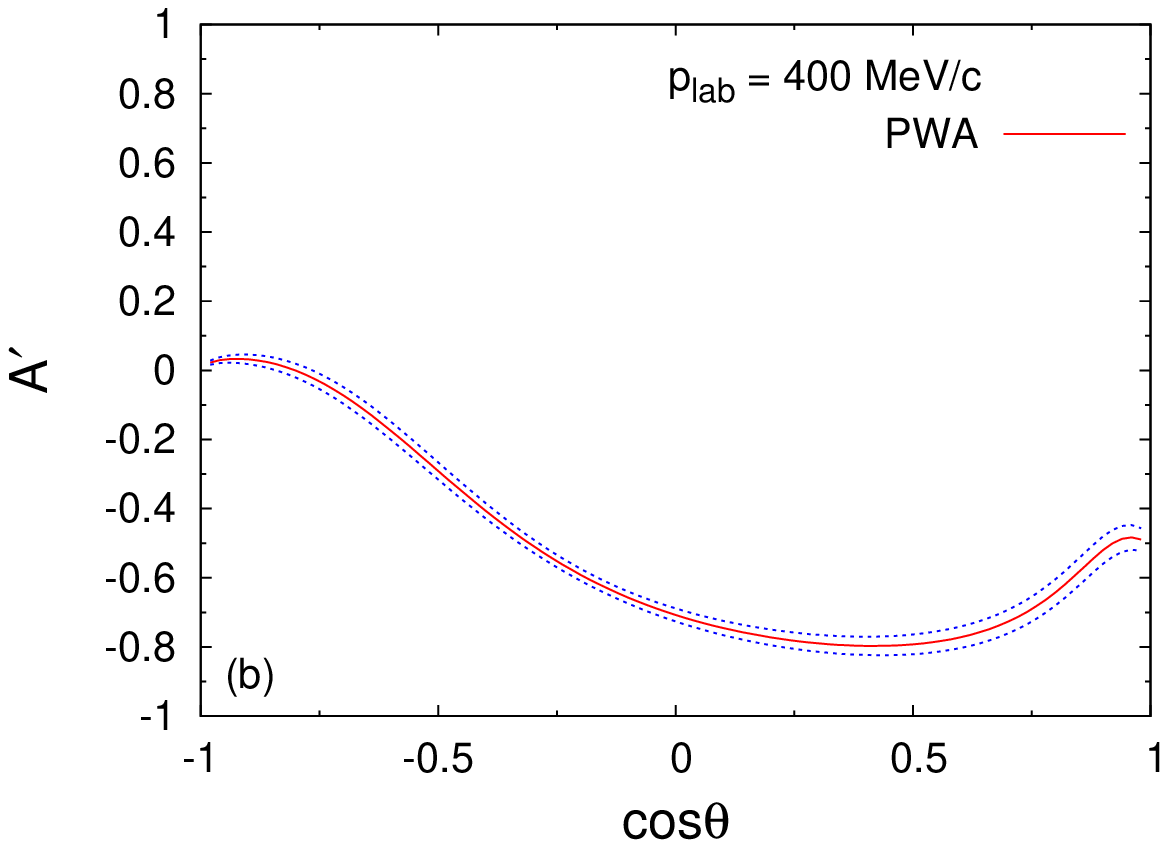}\\
   \includegraphics[width=0.45\textwidth]{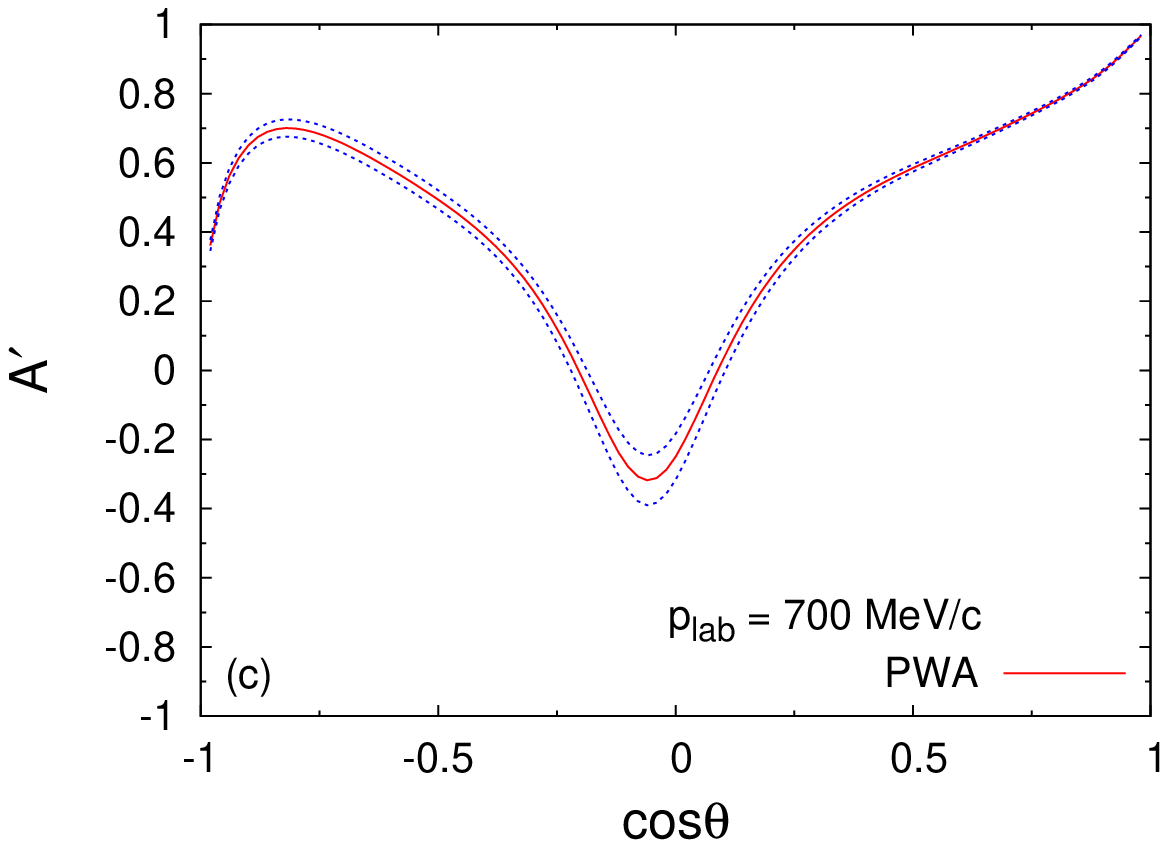} \hspace{2em}
   \includegraphics[width=0.45\textwidth]{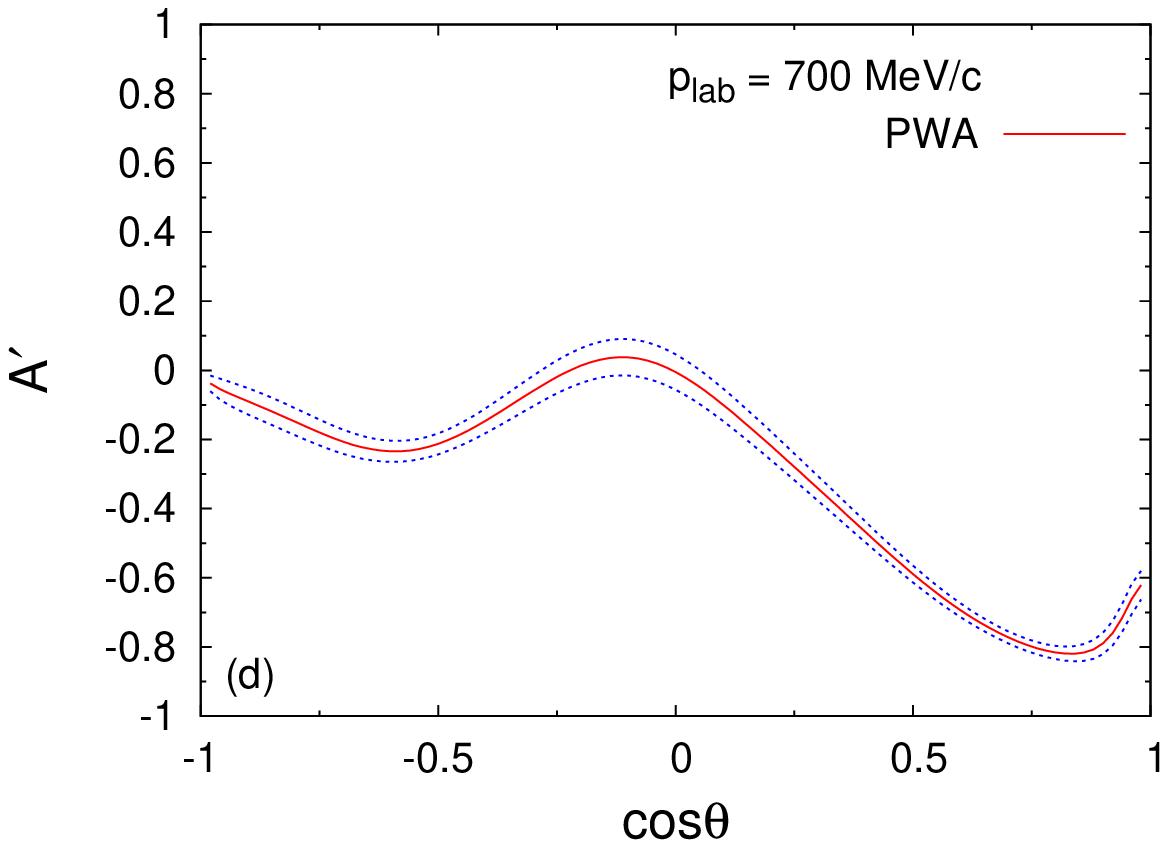}\\
    \caption{\label{Ap}{(Color online) The rotation parameter $A'$ for $\overline{p}p$ elastic (left) and charge-exchange (right) scattering at
   400 and 700 MeV/$c$ laboratory momentum.
   The PWA result is given by the solid red line and the dotted blue lines indicate the $1\sigma$ uncertainty region.}}
\end{figure}

\begin{figure}[htbp]
   \centering
   \includegraphics[width=0.45\textwidth]{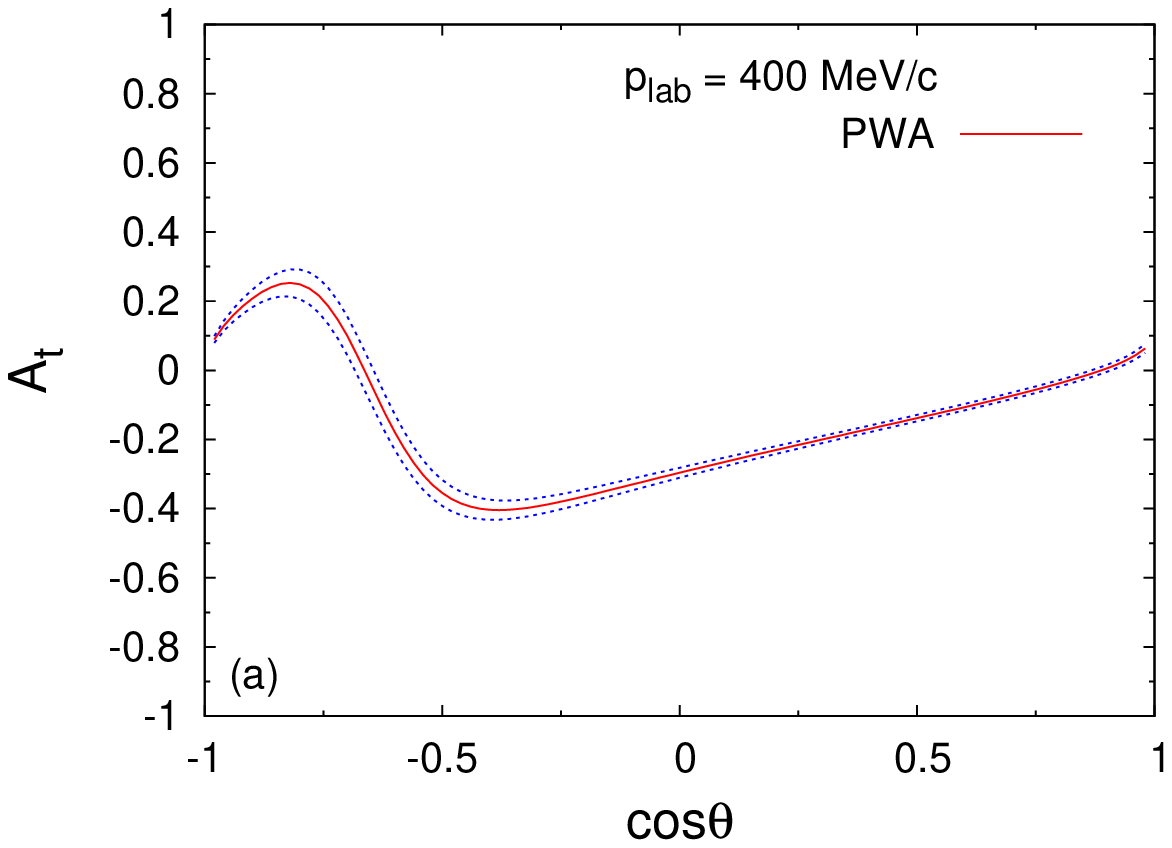} \hspace{2em}
   \includegraphics[width=0.45\textwidth]{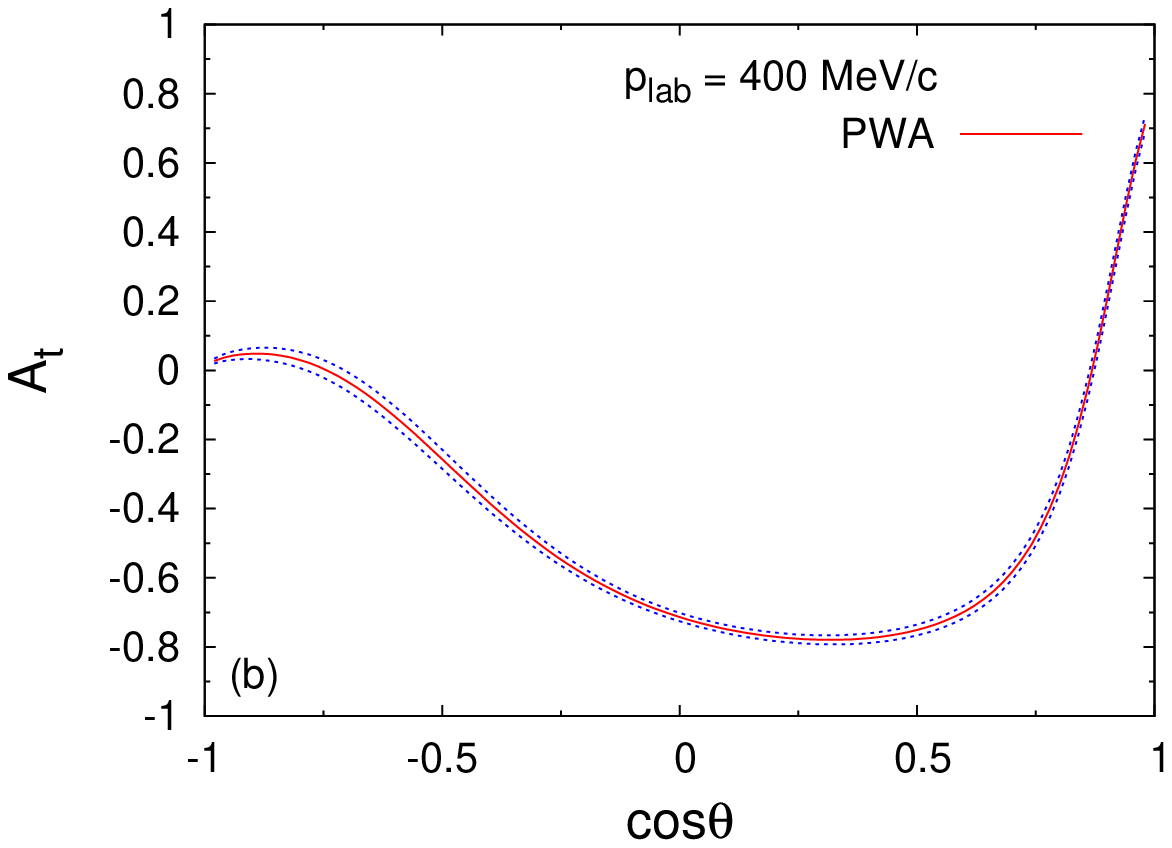}\\
   \includegraphics[width=0.45\textwidth]{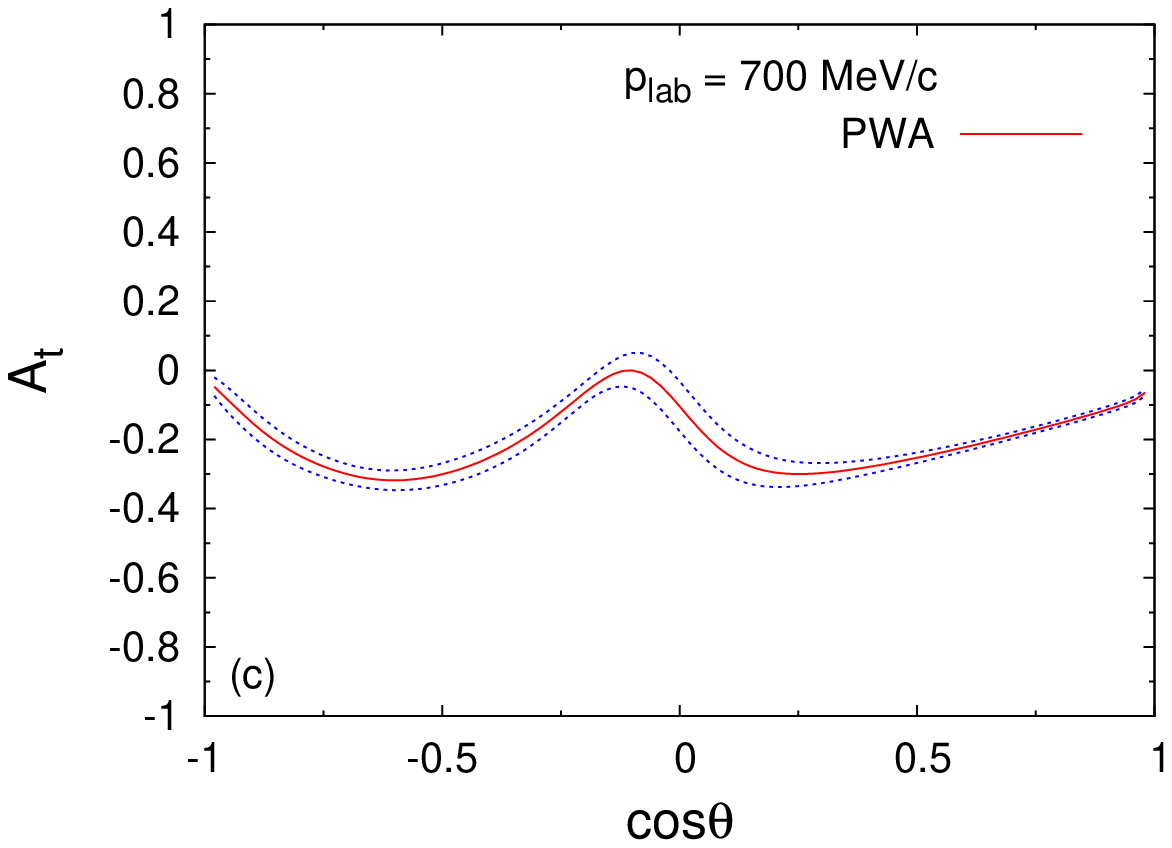} \hspace{2em}
   \includegraphics[width=0.45\textwidth]{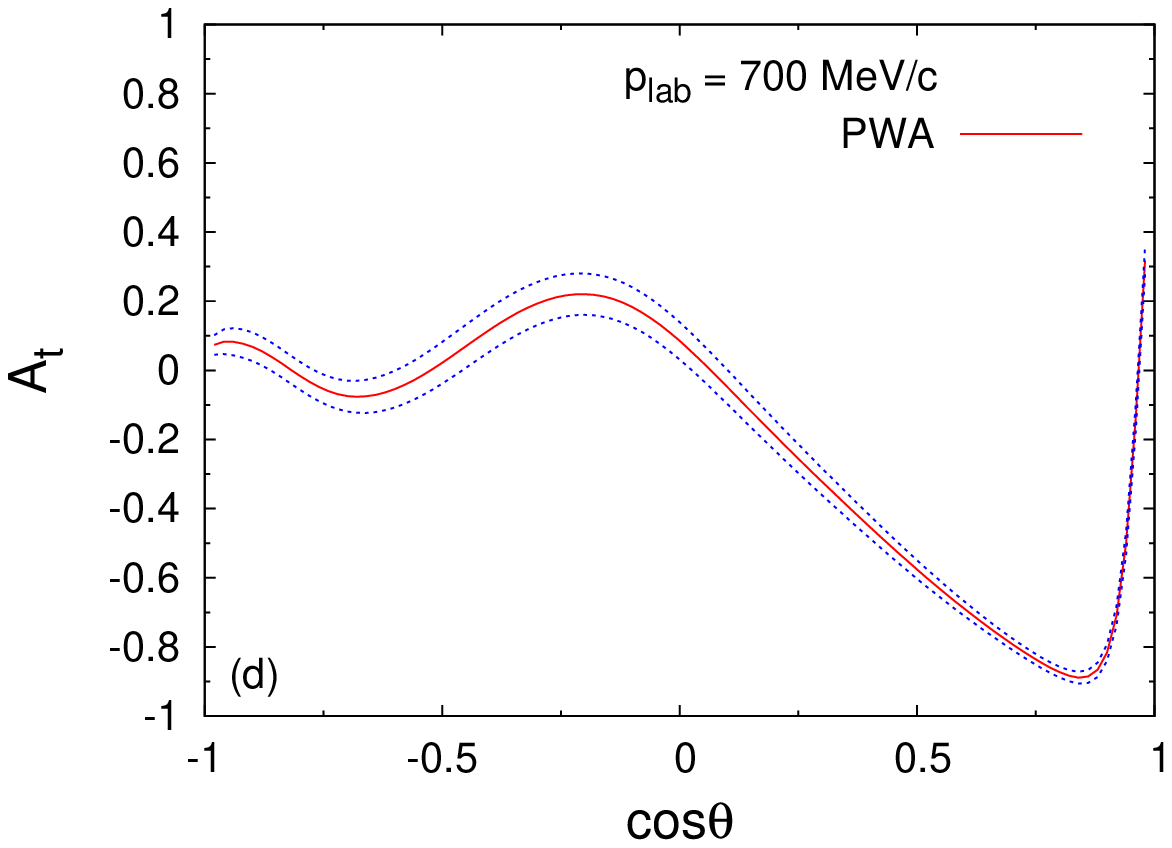}\\
    \caption{\label{At}{(Color online) The polarization transfer $A_{t}$ for $\overline{p}p$ elastic (left) and charge-exchange (right) scattering at
   400 and 700 MeV/$c$ laboratory momentum.
   The PWA result is given by the solid red line and the dotted blue lines indicate the $1\sigma$ uncertainty region.}}
\end{figure}

\begin{figure}[htbp]
   \centering
   \includegraphics[width=0.45\textwidth]{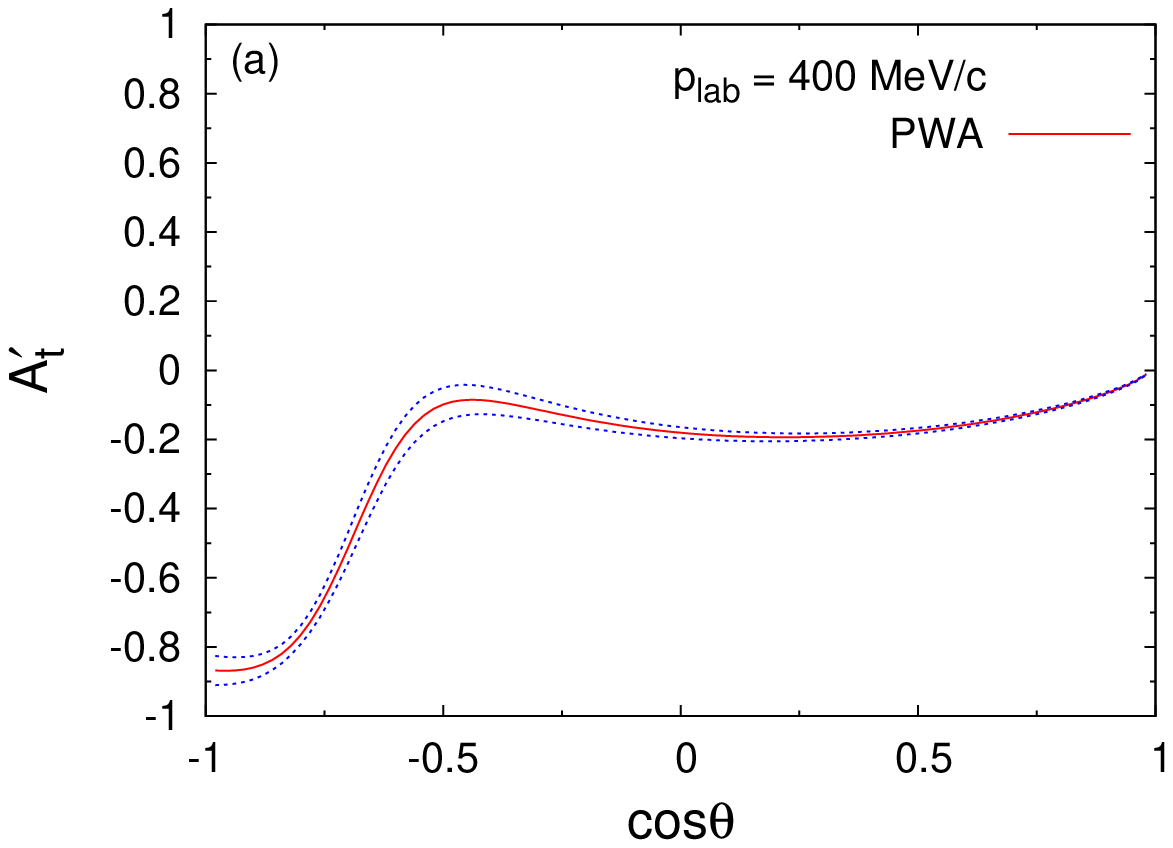} \hspace{2em}
   \includegraphics[width=0.45\textwidth]{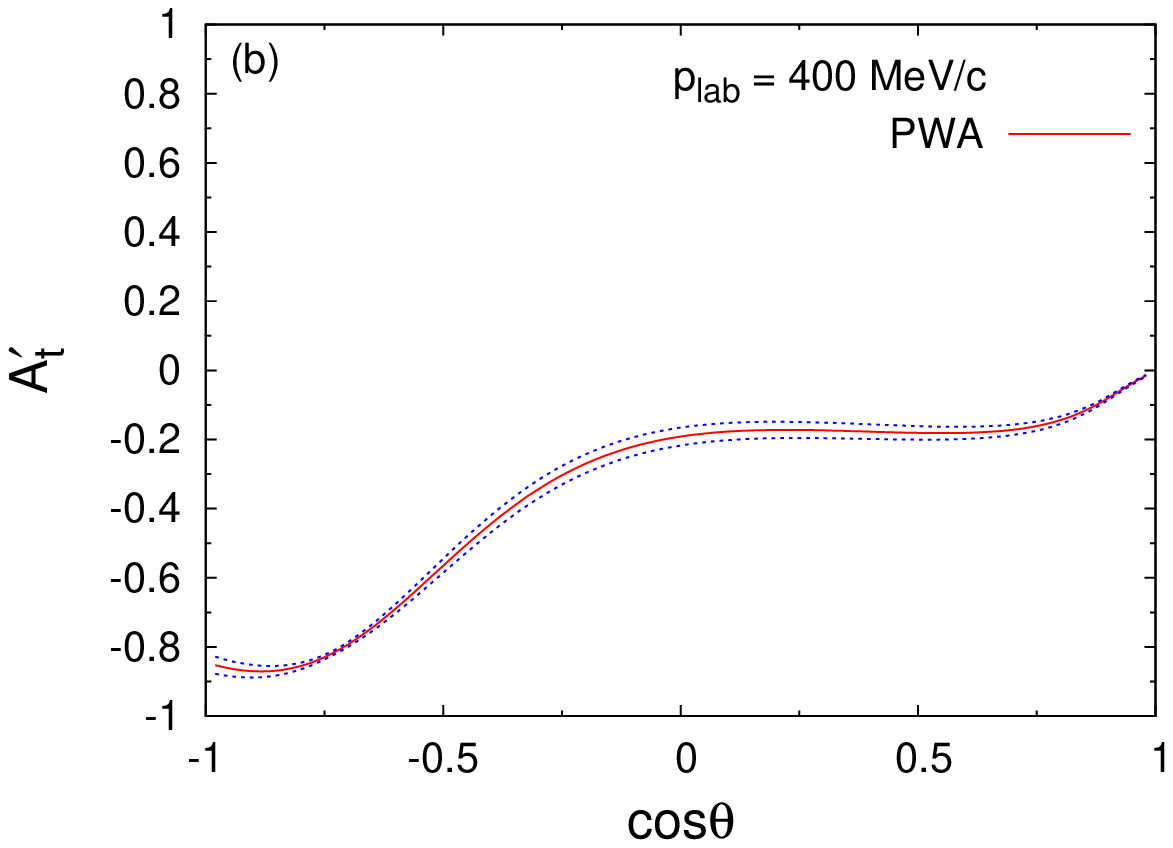}\\
   \includegraphics[width=0.45\textwidth]{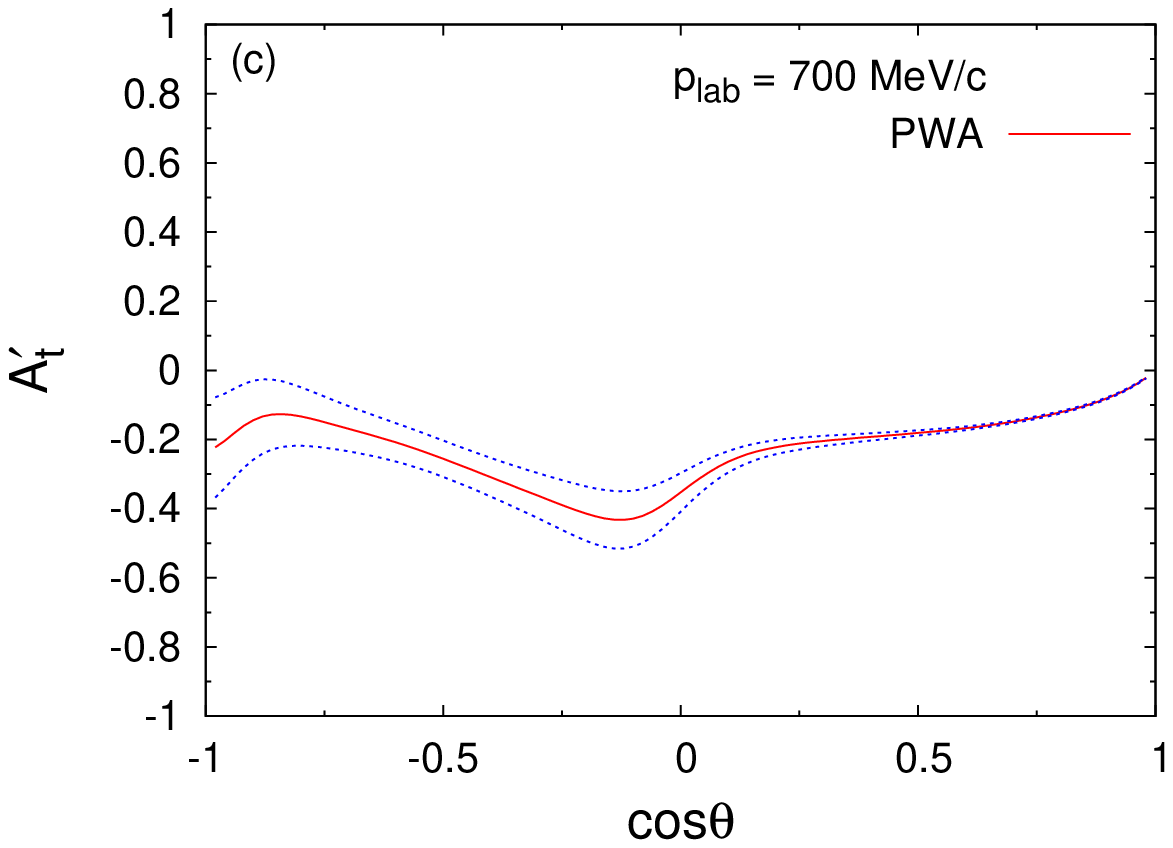} \hspace{2em}
   \includegraphics[width=0.45\textwidth]{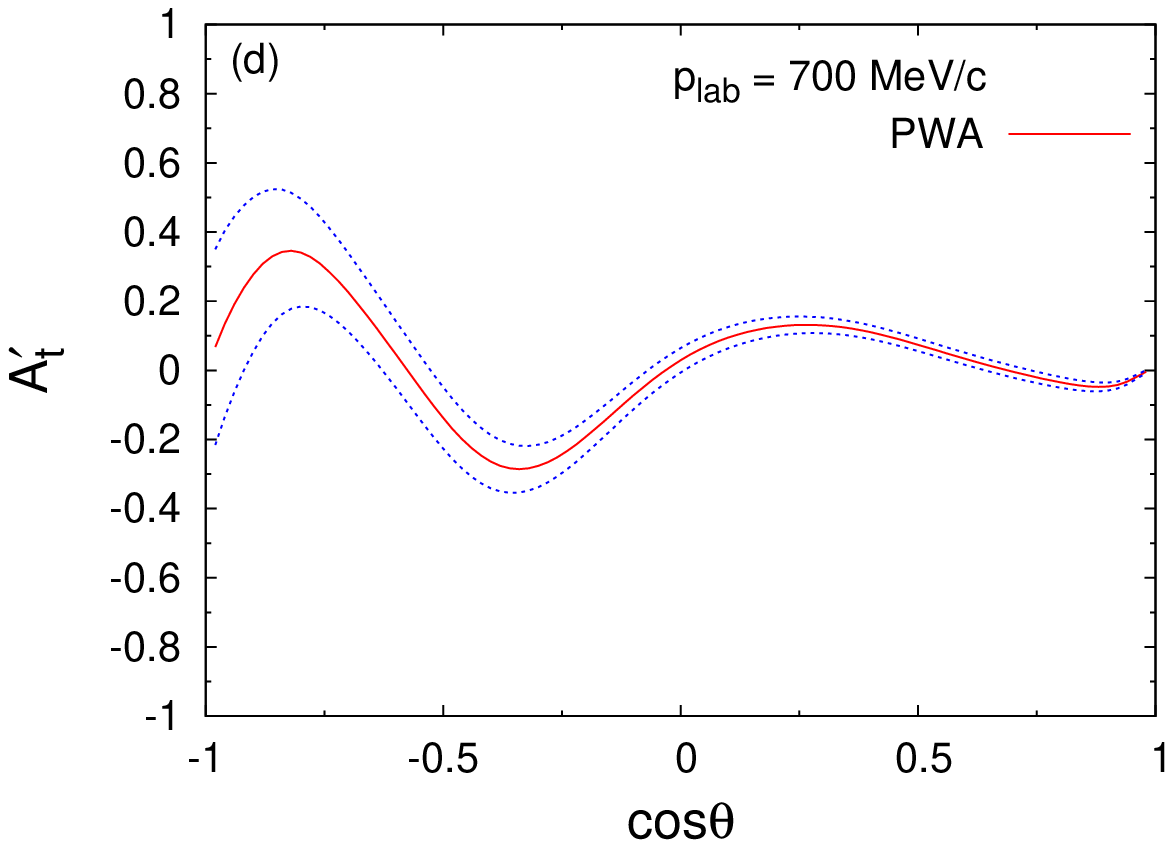}\\
   \caption{\label{Atp}{(Color online) The polarization transfer $A'_{t}$ for $\overline{p}p$ elastic (left) and charge-exchange (right) scattering
   at 400 and 700 MeV/$c$ laboratory momentum.
   The PWA result is given by the solid red line and the dotted blue lines indicate the $1\sigma$ uncertainty region.}}
\end{figure}

\begin{figure}[htbp]
   \centering
   \includegraphics[width=0.45\textwidth]{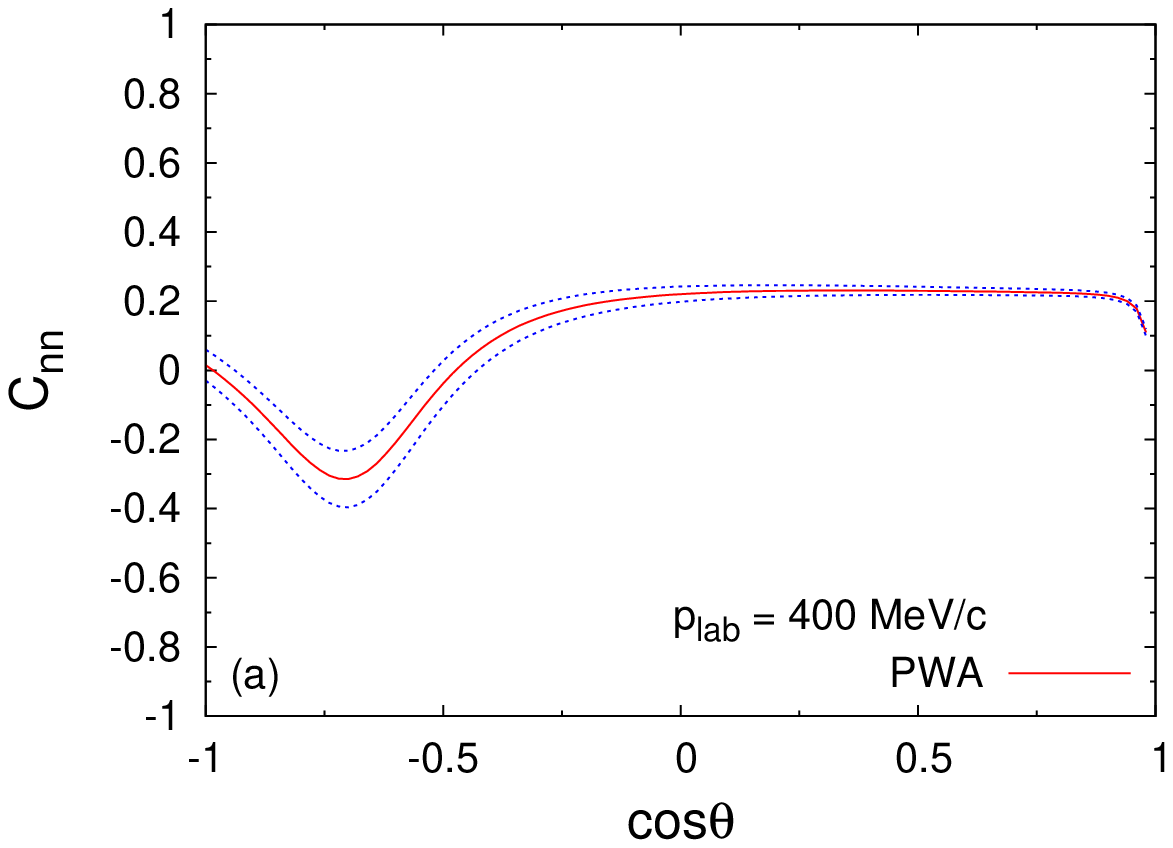} \hspace{2em}
   \includegraphics[width=0.45\textwidth]{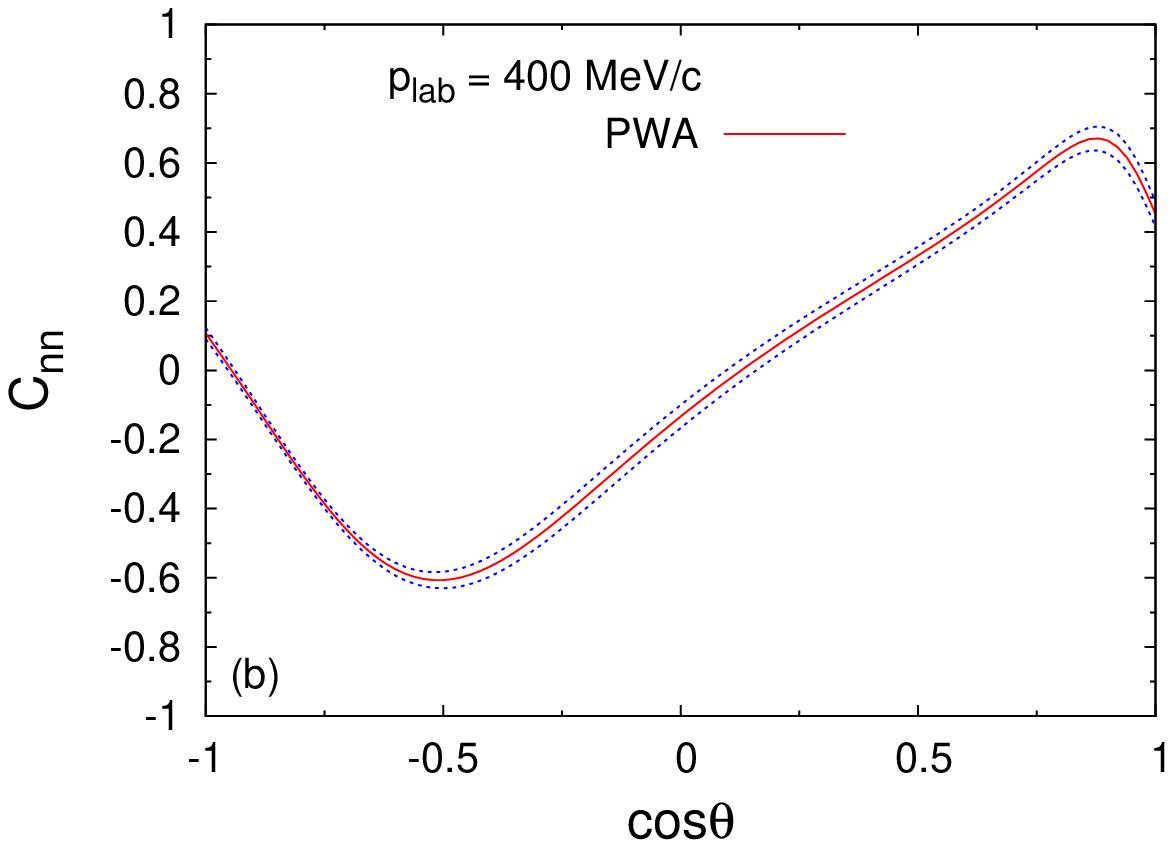}\\
   \includegraphics[width=0.45\textwidth]{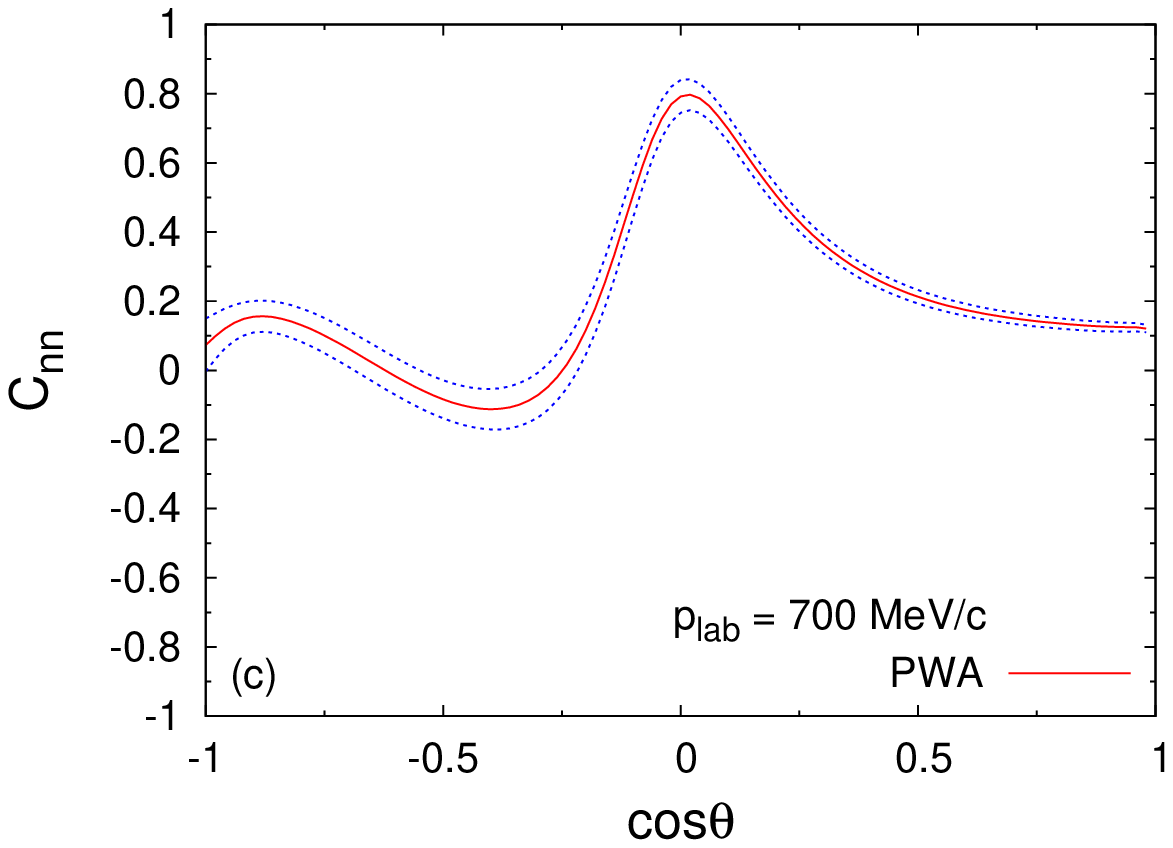} \hspace{2em}
   \includegraphics[width=0.45\textwidth]{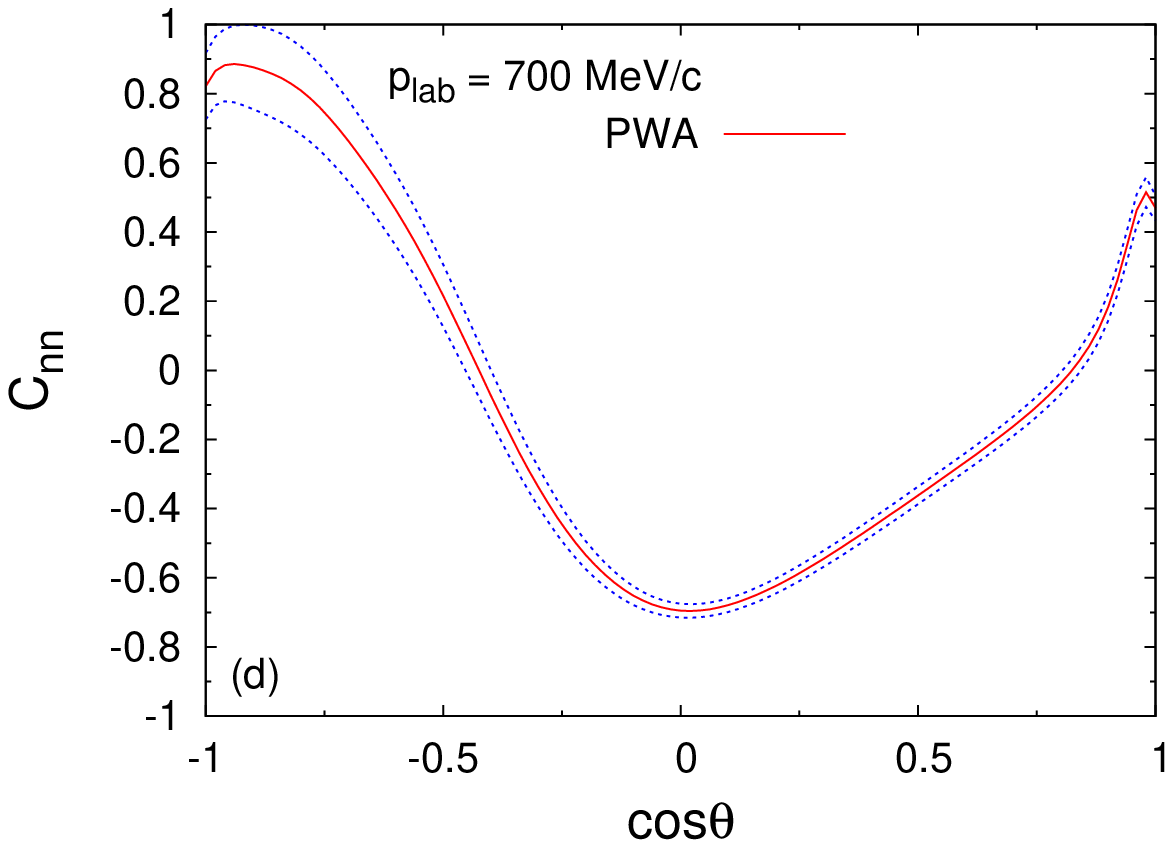}\\
   \caption{\label{Cnn}{(Color online) The spin correlation $C_{nn}$ for $\overline{p}p$ elastic (left) and charge-exchange (right) scattering
   at 400 and 700 MeV/$c$ laboratory momentum.
   The PWA result is given by the solid red line and the dotted blue lines indicate the $1\sigma$ uncertainty region.}}
\end{figure}

\begin{figure}[htbp]
   \centering
   \includegraphics[width=0.45\textwidth]{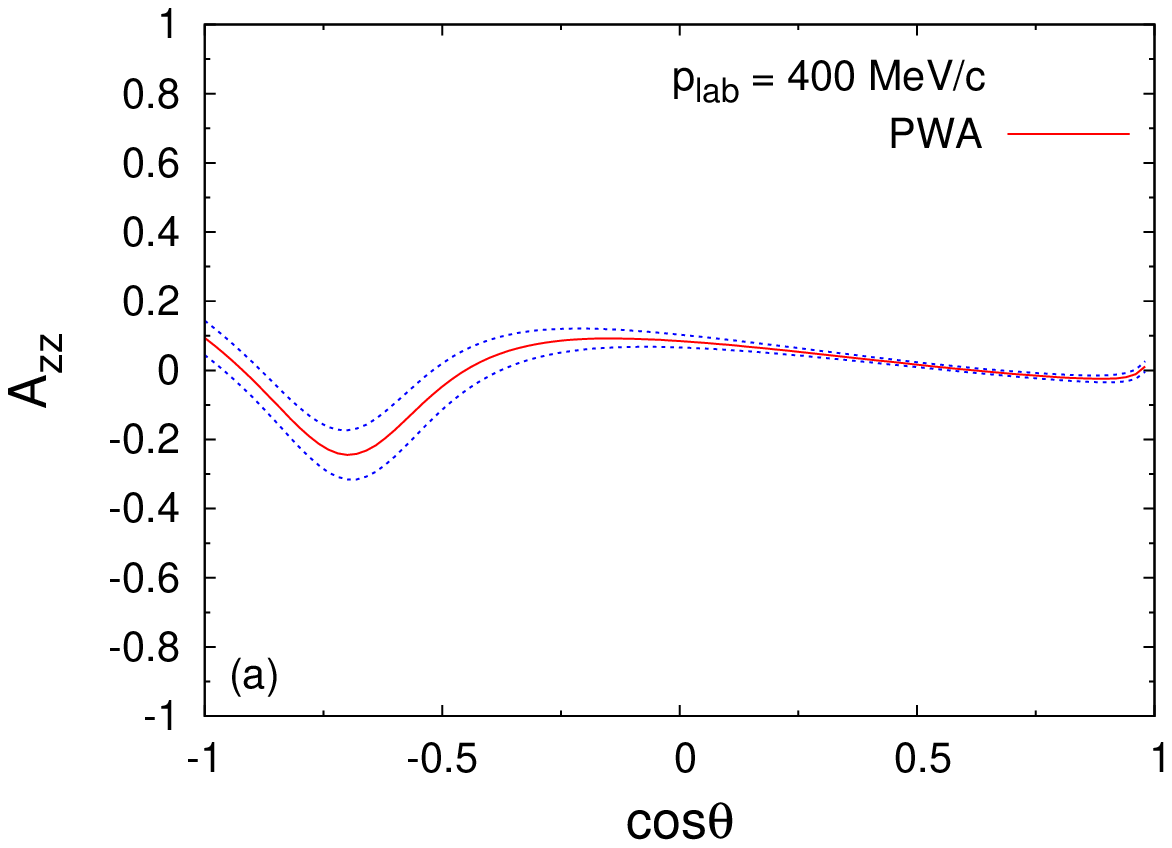} \hspace{2em}
   \includegraphics[width=0.45\textwidth]{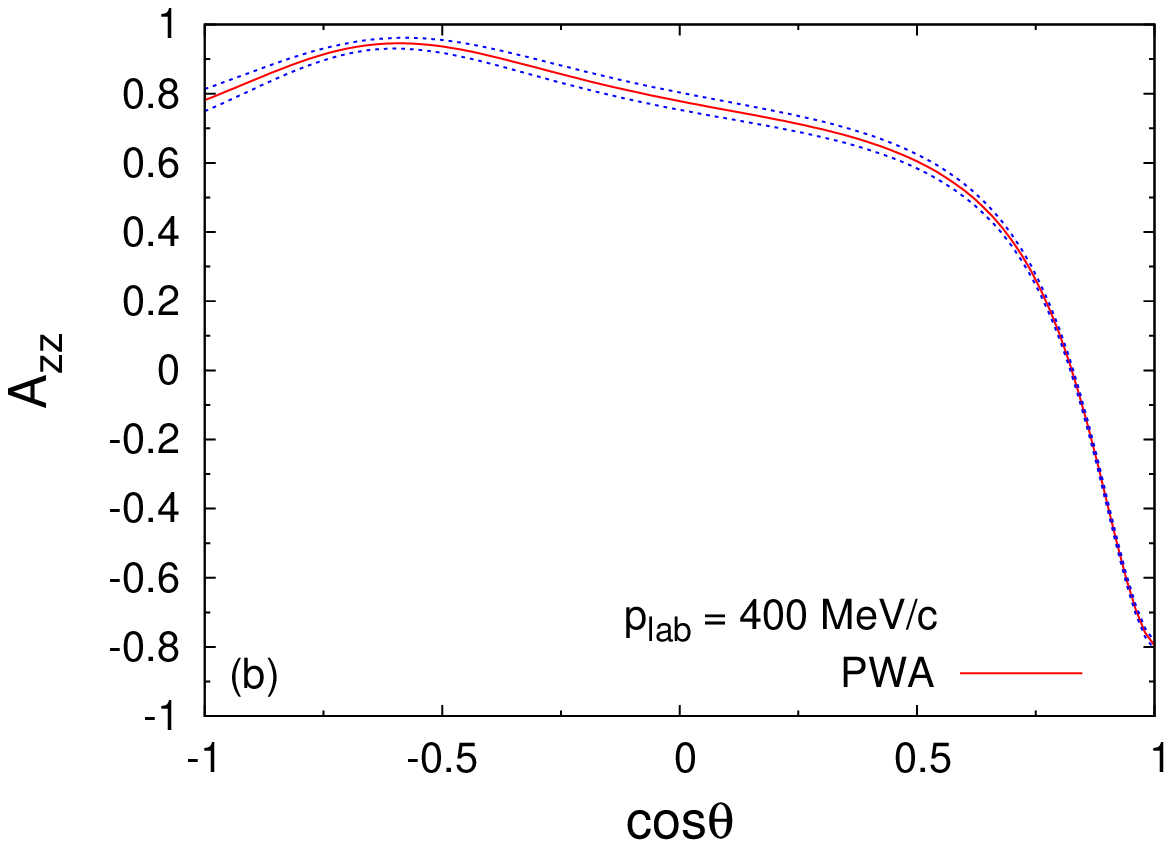}\\
   \includegraphics[width=0.45\textwidth]{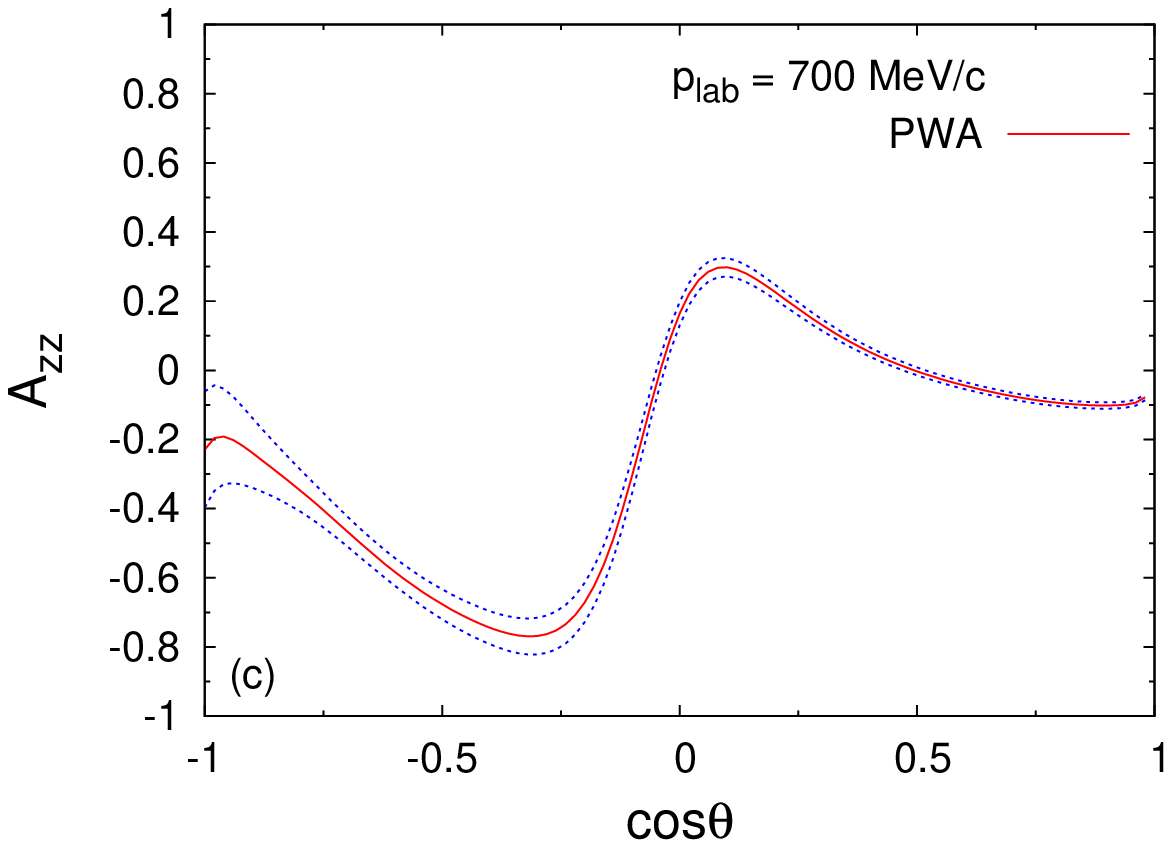} \hspace{2em}
   \includegraphics[width=0.45\textwidth]{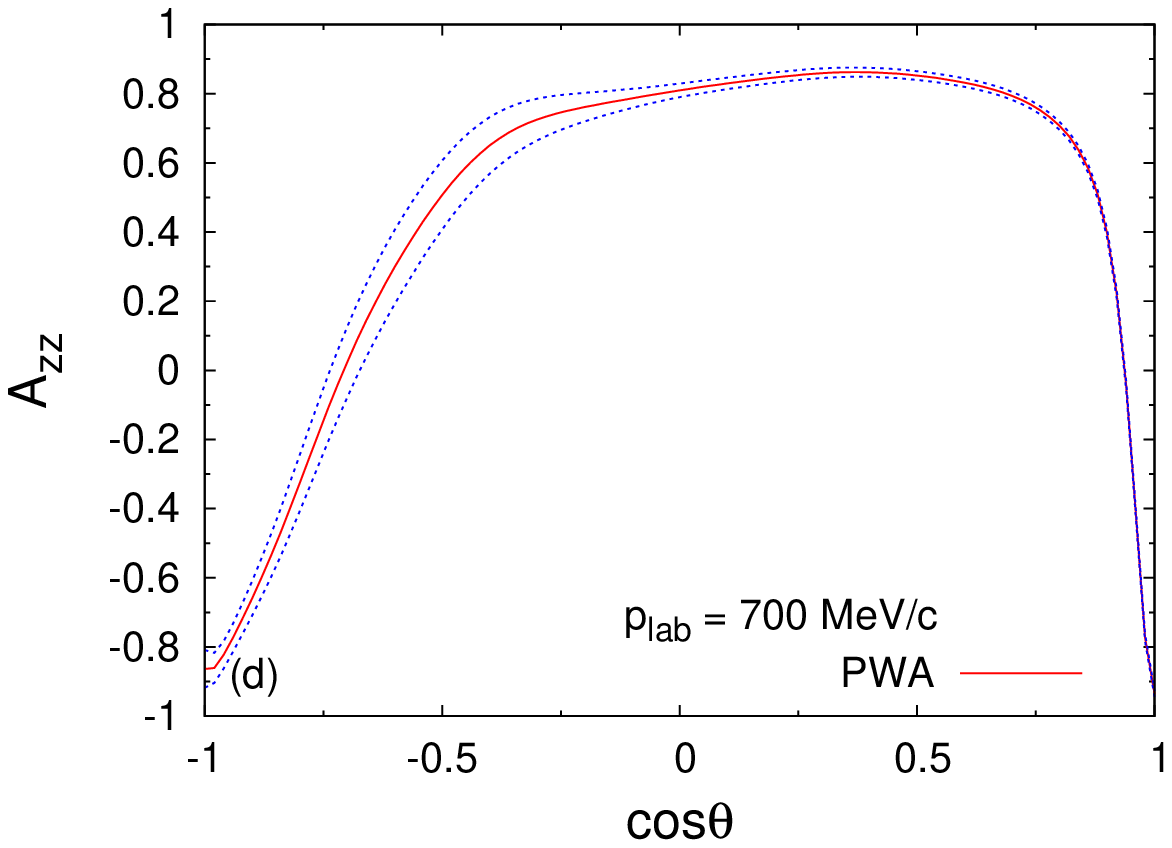}\\
   \caption{\label{Azz}{(Color online) The spin correlation $A_{zz}$ for $\overline{p}p$ elastic (left) and charge-exchange (right) scattering
   at 400 and 700 MeV/$c$ laboratory momentum.
   The PWA result is given by the solid red line and the dotted blue lines indicate the $1\sigma$ uncertainty region.}}
\end{figure}

In Figs.~\ref{D}--\ref{Azz}, we plot the predictions of the PWA~\cite{Zho12} for the
spin observables $D$, $D_{t}$, $R$, $R'$, $R_{t}$, $R'_{t}$, $A$, $A'$, $A_{t}$, $A'_{t}$,
$C_{nn}$, and $A_{zz}$ for both elastic and charge-exchange scattering at two typical
laboratory momenta $p_{\rm lab} = 400$ and 700 MeV/$c$. In all cases, the solid
(red) line is the PWA prediction and the two dashed (blue) lines border the
one-standard-deviation theoretical uncertainty region. Our results for these observables
at other momenta, or predictions for other, higher-rank spin observables can be calculated
straightforwardly and are available upon request. We discuss some of the salient features
that are apparent from the results plotted in Figs.~\ref{D}--\ref{Azz}.

In Figs.~\ref{D}--\ref{R} we show the results for the depolarization $D=D_{0n0n}$, the polarization
transfer $D_{t}=K_{n00n}$, and the rotation parameter $R=D_{0m0x}$. The depolarization
$D$ for elastic scattering is close to the value 1 for forward angles, especially for low momenta.
In the charge-exchange case, $D$ rises rapidly for very forward angles, after which it decreases
to negative values for larger angles. The rise for the forward angles is due to one-pion
exchange. For elastic scattering, the spin rotation $R$ approaches the value 1 for forward angles.

In Figs.~\ref{Rp}--\ref{Rtp} we show the results for the rotation parameter $R'=D_{0\ell0x}$ and
the polarization transfers $R_{t}=K_{\ell00x}$ and $R'_{t}=K_{m00x}$. For the charge-exchange
case, the value of the spin transfer $R_{t}'$ is large for the very forward angles, especially
at low momenta, which implies that the outgoing antineutron is strongly polarized. It may be
interesting to study whether it is feasible to produce a polarized antineutron beam in this way.

In Figs.~\ref{A}--\ref{At} we show the results for the rotation parameters $A=D_{0\bar{m}0\bar{z}}$
and $A'=D_{0\bar{\ell}0\bar{z}}$ and the polarization transfer $A_{t}=K_{\bar{\ell}00\bar{z}}$.
The spin rotation $A'$ is large and positive for forward angles for the elastic case, while for the
charge-exchange case it is large and negative. For charge-exchange scattering, the spin
transfer $A_{t}$ is forward peaked (due to one-pion exchange), especially at low momenta,
resulting in strongly polarized antineutrons. This characteristic could again be exploited to
produce a polarized antineutron beam~\cite{Dov82}.

In Figs.~\ref{Atp}--\ref{Azz} we show the results for the polarization transfer $A'_{t}=K_{\bar{m}00\bar{z}}$
and the spin correlations $C_{nn}=C_{nn00}$ and $A_{zz}=A_{00zz}$. For the charge-exchange
case, the spin correlations $C_{nn}$ and $A_{zz}$ vary strongly as a function of angle and reach
large values, both positive and negative.
A measurement of $C_{ij}$ for the elastic or charge-exchange reactions requires secondary scattering
of both the recoil nucleon and the scattering antinucleon. In the strangeness-exchange reactions
$\overline{p}p\rightarrow\overline{Y}Y$ the ``self-analyzing'' weak decays of the (anti)hyperons
give access to the spin correlations $C_{ij}$ by measuring the angular distributions of the decay
products~\cite{Dur64,Pas00,Bas02,Pas06}. A measurement of $A_{zz}$ requires a polarized antiproton
beam and a polarized proton target.

We also study the transverse and longitudinal spin-dependent total cross sections
for elastic and charge-exchange scattering. They require a polarized antiproton beam.
The spin-dependent cross sections can be written as~\cite{Hos68,Bys78,LaF80,Bys84,LaF92,Bil63}
\begin{equation}
\sigma = \sigma_{\textrm{tot}}
         - \frac{1}{2}\:\boldsymbol{\zeta}_{B}\!\cdot\boldsymbol{\zeta}_{T}\:\Delta\sigma_{\!_\perp}
         - \frac{1}{2}\:\boldsymbol{\zeta}_{B}\!\cdot\boldsymbol{\hat{p}}\:
           \boldsymbol{\zeta}_{T}\!\cdot\boldsymbol{\hat{p}}\:
           (\Delta\sigma_{\!_\parallel} - \Delta\sigma_{\!_\perp})~,
\label{Eq:total_xs}
\end{equation}
where $\boldsymbol{\zeta}_{B}$  and $\boldsymbol{\zeta}_{T}$ 
are the unit polarization vectors of the beam and target, respectively, 
$\boldsymbol{\hat{p}}$ is the unit vector in the direction of the beam momentum,
and $\sigma_{\textrm{tot}}$ is the integrated spin-independent cross section.
The differences in the cross sections for antiparallel versus parallel spins,
transversely and longitudinally oriented with respect to the beam direction, are
\begin{subequations}
\begin{eqnarray}
\Delta\sigma_{\!_\perp} &=& \sigma_{\!_{\uparrow\downarrow}}-\sigma_{\!_{\uparrow\uparrow}} ~,\\
\Delta\sigma_{\!_\parallel} &=& \sigma_{\rightleftarrows}-\sigma_{\rightrightarrows} ~,
\end{eqnarray}
\end{subequations}
where the double arrows mean that the spins of the beam and target particles are antiparallel
or parallel, respectively. In terms of $\sigma_{S\mu}$, the integrated cross section with total
spin of beam and target particles equal to $S$ with a $z$ component of $\mu$, one has 
\begin{subequations}
\begin{eqnarray}
\sigma_{\textrm{tot}} &=& \frac{1}{2}\sigma_{11} + \frac{1}{4}(\sigma_{10} + \sigma_{00}) ~, \\
\Delta\sigma_{\!_\perp} &=& -\frac{1}{2}(\sigma_{10} - \sigma_{00}) ~, \\
\Delta\sigma_{\!_\parallel} &=& -\sigma_{11} + \frac{1}{2}(\sigma_{10} + \sigma_{00}) ~.
\label{Eq:xs012}
\end{eqnarray}
\end{subequations}

\begin{figure}[t]
\centering
\includegraphics[width=0.9\textwidth]{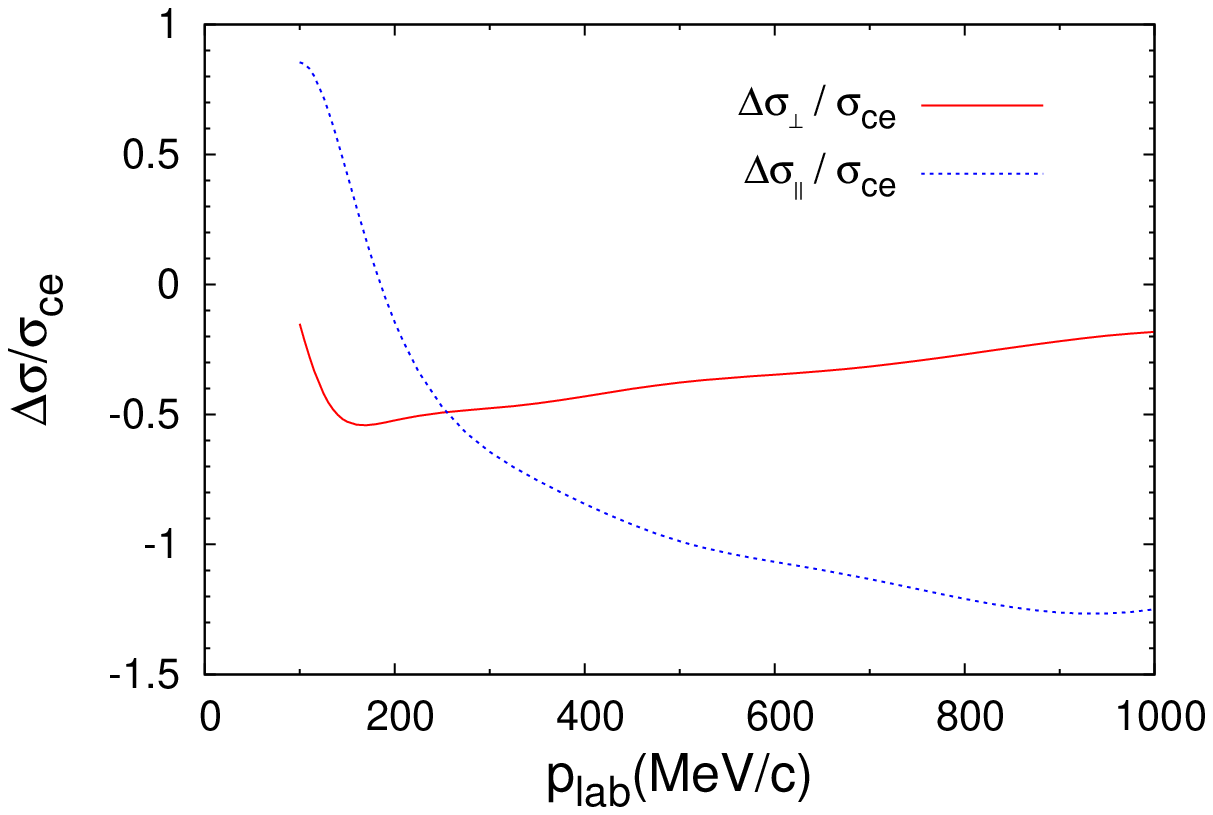}
\caption{\label{Deltace}(Color online) 
The differences of the transverse  
and longitudinal cross sections, $\Delta\sigma_{\!_\perp}$ and $\Delta\sigma_{\!_\parallel}$,
relative to the total cross section $\sigma_{\textrm{ce}}$ for charge-exchange scattering
$\overline{p}p\rightarrow\overline{n}n$, as function of the antiproton laboratory momentum.
The solid red line shows the result for the transverse case and the dotted blue line shows the
result for the longitudinal case.}
\end{figure}

In Fig.~\ref{Deltace} we plot $\Delta\sigma_{\!_\perp}$ and $\Delta\sigma_{\!_\parallel}$ for
the charge-exchange reaction $\overline{p}p\rightarrow\overline{n}n$ for momenta between
100 and 1000 MeV/$c$. The longitudinal difference $\Delta\sigma_{\!_\parallel}$ is very
large, because of the coherent tensor force from one- and two-pion exchange, and it varies
strongly as a function of momentum.

\section{Towards polarized antiprotons} \label{sec:Pol}
In this section we investigate the buildup of polarization of a circulating antiproton
beam by a polarized proton target. The basic idea is that, since the antiproton-proton total
cross section is spin dependent, the number of antiprotons that remain in the beam depends
on the spin state of the antiprotons relative to that of the protons. The antiprotons remain in
the beam when they are elastically scattered within a certain, very small scattering angle,
called the ``acceptance angle.'' These antiprotons can be scattered again in the next
revolution of the beam. The antiprotons that interact elastically with larger scattering
angle, or undergo charge exchange, or get annihilated are removed from the beam.
When the beam is circulated many times in a ring with the target, the remaining beam
loses intensity but acquires a net polarization. This mechanism is sometimes called
``spin filtering.'' The feasibility of polarizing a proton beam in this way was demonstrated
at the Test Storage Ring in Heidelberg by the experiment FILTEX \cite{Rat93}. Recently, the PAX
Collaboration confirmed these results in an experiment at the COSY ring \cite{Aug12}. 
This raises hope that it could also work experimentally for an antiproton beam.

We follow the methods developed in Refs.~\cite{Cso68,Mil05,Dmi08,Dmi10,Hai11,Uzi11}
to calculate the buildup of polarization of an antiproton beam. We present results
for antiproton laboratory momenta from 100 to 1000 MeV/$c$ and for typical
acceptance angles $\theta_{\rm{acc}}^{\rm{lab}} = 5$, 10, 20, and 30 mrad in the
laboratory frame. As input we use the scattering amplitudes predicted by our PWA.
Our results  can be compared to the model calculations of Refs.~\cite{Dmi08,Dmi10,Hai11}. 
We assume that the polarization buildup due to the filtering mechanism dominates
and that spin-flip mechanisms in the beam can be neglected~\cite{Mil05}. This was
verified by an experiment in the COSY ring~\cite{Oel09}. 

The scattering amplitude in spin space for a certain momentum and angle is the sum
of the electromagnetic and nuclear contributions. Because the filtering mechanism
involves forward scattering of a circulating beam, care has to be taken with the treatment
of electromagnetic effects~\cite{Swa97}. The electromagnetic amplitude, in particular
the standard Coulomb scattering amplitude [Eq.~(\ref{Mcoul})], is infinite for $\theta=0$
(where, in this section, we neglect the small effects of the magnetic-moment interaction).
In reality, of course, the Coulomb interaction is screened at very large distances.
Since it is not known how to treat this long-range screening, the overall phase of
the Coulomb amplitude is unknown. The same holds for the nuclear amplitude,
Eq. (\ref{Mnucl}). This implies that electromagnetic effects cannot be separated
completely from nuclear effects. The nuclear amplitude contains remnants of the
electromagnetic interaction, and one cannot properly define, e.g., the concept
of a total hadronic cross section.

The use of a nonzero acceptance angle alleviates the problems associated with extreme
forward scattering. Partially integrated elastic and charge-exchange cross sections can be
calculated by integrating the differential cross sections for angles
$\theta^{\rm{lab}}>\theta_{\rm{acc}}^{\rm{lab}}$.
The annihilation amplitude, however, cannot be calculated theoretically. Instead, the
annihilation cross section has to be obtained by using the optical theorem for the total
$\overline{p}p$ cross section and subtracting the elastic and charge-exchange cross sections.
However, since the optical theorem again involves the forward scattering amplitude, it is
strictly not valid for scattering of charged particles~\cite{Swa97}. With these
caveats in mind, we first review briefly the formalism for the filtering mechanism and then
present our results.

Suppose $N_{+}(t)$ and $N_{-}(t)$ are the number of antiprotons with spin
``up'' and spin ``down,'' respectively, at time $t$; $N_{+}(0)=N_{-}(0)$, since
the beam is initially unpolarized.  The number of antiprotons in the beam as
a function of time is given by
\begin{equation}
   N(t) = \frac{1}{2}N(0)\left[e^{-\Omega_{-}^{\rm{out}}t}+e^{-\Omega_{+}^{\rm{out}}t}\right]~,
\end{equation}
where $\Omega_{\pm}^{\rm{out}}$ characterize how many antiprotons with spin up
or down are scattered out of the acceptance angle~\cite{Mil05}. 
The polarization of the beam is then
\begin{equation}
   P_{B}(t) = \frac{N_{+}(t)-N_{-}(t)}{N_{+}(t)+N_{-}(t)}
                  = \tanh\left[\frac{t}{2}(\Omega_{-}^{\rm{out}}-\Omega_{+}^{\rm{out}})\right]~.
\end{equation}
The relation of $\Omega_{\pm}^{\rm{out}}$ to the spin-dependent cross sections is given by
\begin{equation}
   \Omega_{\pm}^{\rm{out}} = nf\left\{\sigma_{\textrm{tot}}\mp \frac{1}{2} P_{T}
      \left[\Delta\sigma_{\!_\perp}+(\boldsymbol{\zeta}_{T}\cdot\boldsymbol{\hat{p}})^{2}
      (\Delta\sigma_{\!_\parallel}-\Delta\sigma_{\!_\perp})\right]\right\}~,
\end{equation}
where $n$ is the areal density of the target, $f$ is the revolution frequency of the beam,
and $P_{T}$ is the polarization of the target.
If $\left|\Omega_{-}^{\rm{out}}-\Omega_{+}^{\rm{out}}\right|\ll\left|\Omega_{-}^{\rm{out}}
+\Omega_{+}^{\rm{out}}\right|$, as in the case discussed in Refs.~\cite{Mil05,Dmi08,Dmi10},
the beam lifetime is given by
\begin{equation}
   \tau_{B} = \frac{2}{\Omega_{-}^{\rm{out}}+\Omega_{+}^{\rm{out}}}=\frac{1}{nf\sigma_{\textrm{tot}}}~.
\end{equation}
One can define a figure of merit by ${\cal F}(t)=P_{B}^{2}(t)N(t)$, which is maximal
at $t=t_{0}=2\tau_{B}$. The polarizations at $t_{0}$ are
\begin{subequations}
\begin{eqnarray}
   P_{B}(t_{0}) = P_{\!_\perp}(t_{0})=P_{T}\frac{\Delta\sigma_{\!_\perp}}{\sigma_{\textrm{tot}}} ~, 
     &\;\;\;& \textrm{when} \;\;\boldsymbol{\zeta}_{T}\cdot\boldsymbol{\hat{p}} = 0 ~, \\
   P_{B}(t_{0}) = P_{\!_\parallel}(t_{0})=P_{T}\frac{\Delta\sigma_{\!_\parallel}}{\sigma_{\textrm{tot}}} ~, 
     &\;\;\;& \textrm{when} \;\;\boldsymbol{\zeta}_{T}\cdot\boldsymbol{\hat{p}} = \pm 1 ~.
\end{eqnarray}
\end{subequations}

\begin{figure}[!t]
   \centering
      \includegraphics[width=0.9\textwidth]{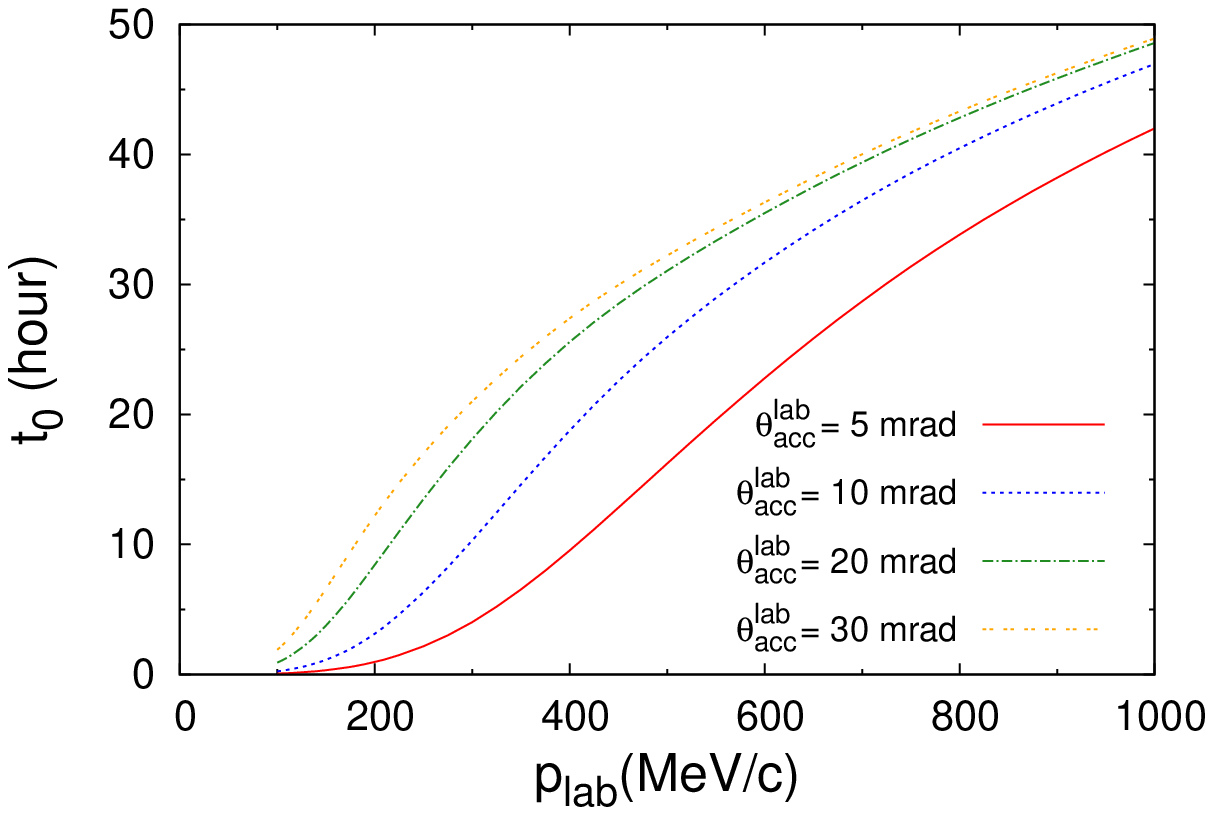}
      \vspace{-1.5em}
\caption{\label{t0}(Color online) 
The time $t_{0}$ in hours with the areal density of the target $n=10^{14}$ $\rm{cm}^{-2}$ 
and the revolution frequency of the beam $f=10^{6}$ $\rm{s}^{-1}$ as function of the antiproton
beam momentum. The solid red, dotted blue, dot-dashed green, and double-dotted orange
lines show the result for acceptance angles $\theta_{\rm{acc}}^{\rm{lab}}=5$, 10, 20, and 30
mrad in the laboratory frame, respectively.} 
\end{figure}

In Fig. \ref{t0}, the optimal time $t_{0}$  as a function of laboratory momentum is plotted 
for the typical values $n=10^{14}$ $\rm{cm}^{-2}$ and $f=10^{6}$ $\rm{s}^{-1}$ and for
acceptance angles $\theta_{\rm{acc}}^{\rm{lab}} = 5$, 10, 20, and 30 mrad in the laboratory
frame (with the assumption that $P_{T}=1$). We find that $t_{0}$ is of the order of several tens of hours
for the momenta and the acceptance angles considered.

\begin{figure}[p]
   \centering
   \includegraphics[width=0.45\textwidth]{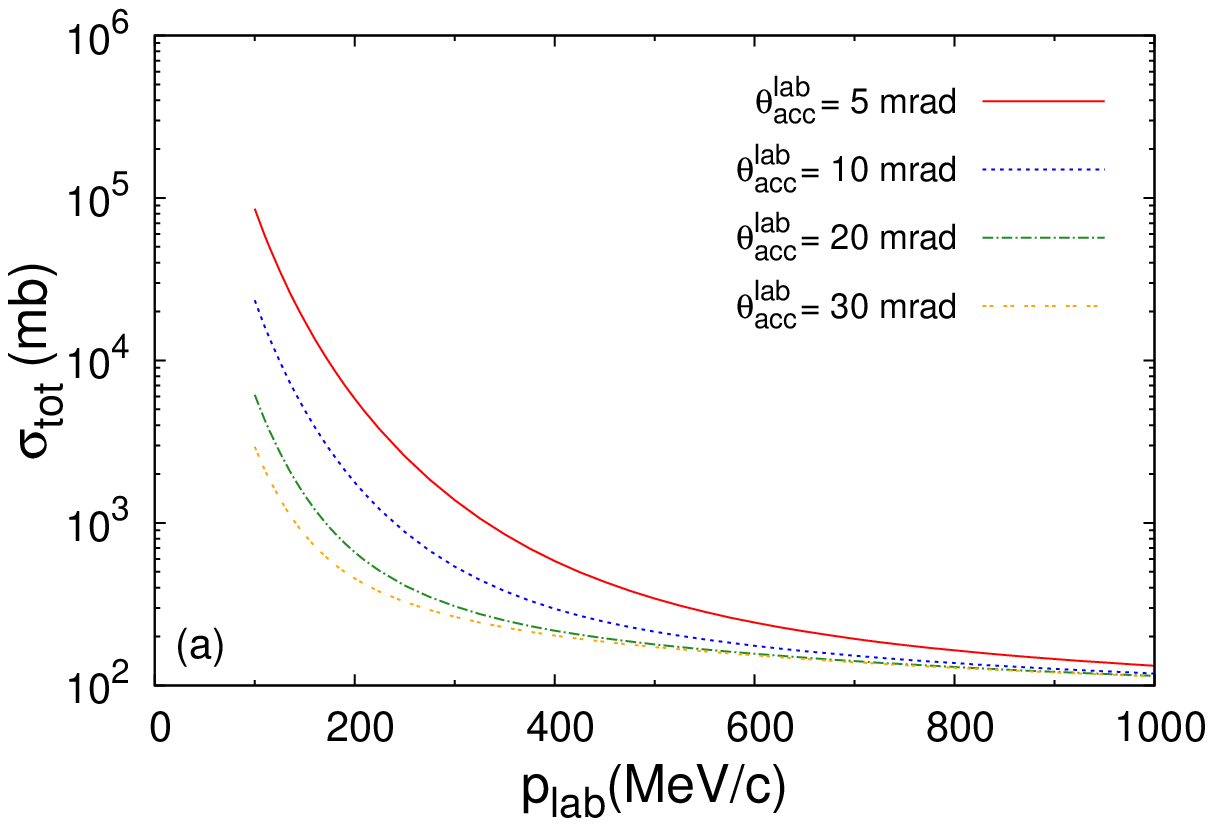} \hspace{2em}
   \includegraphics[width=0.45\textwidth]{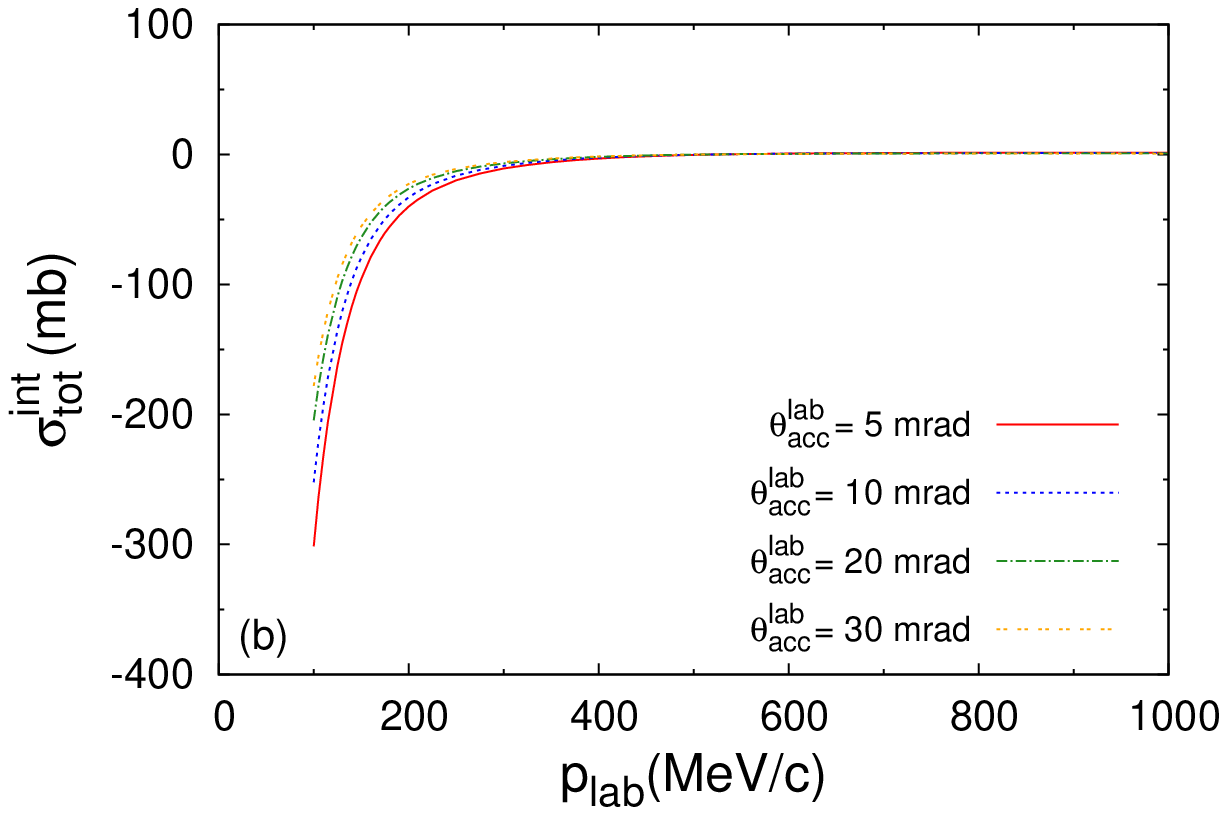}\\ \vspace{-0.9em}
   \includegraphics[width=0.45\textwidth]{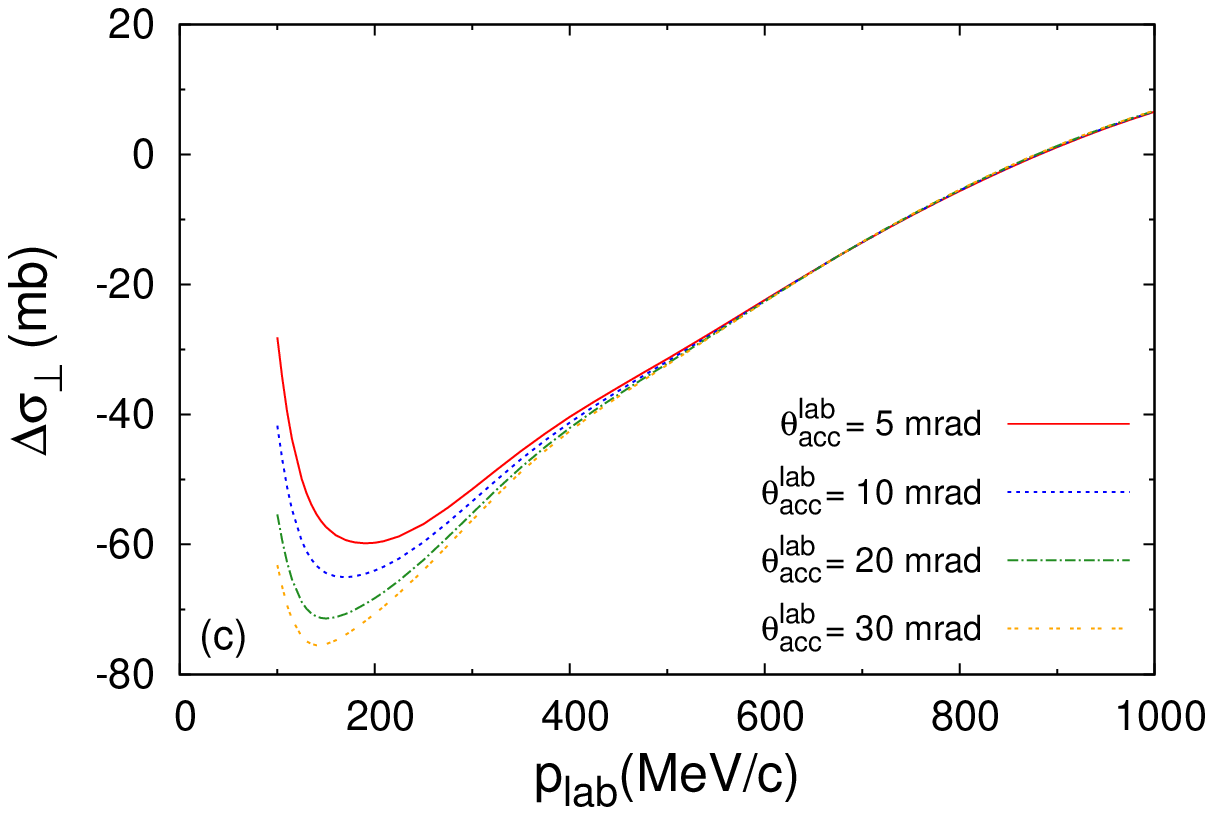} \hspace{2em}
   \includegraphics[width=0.45\textwidth]{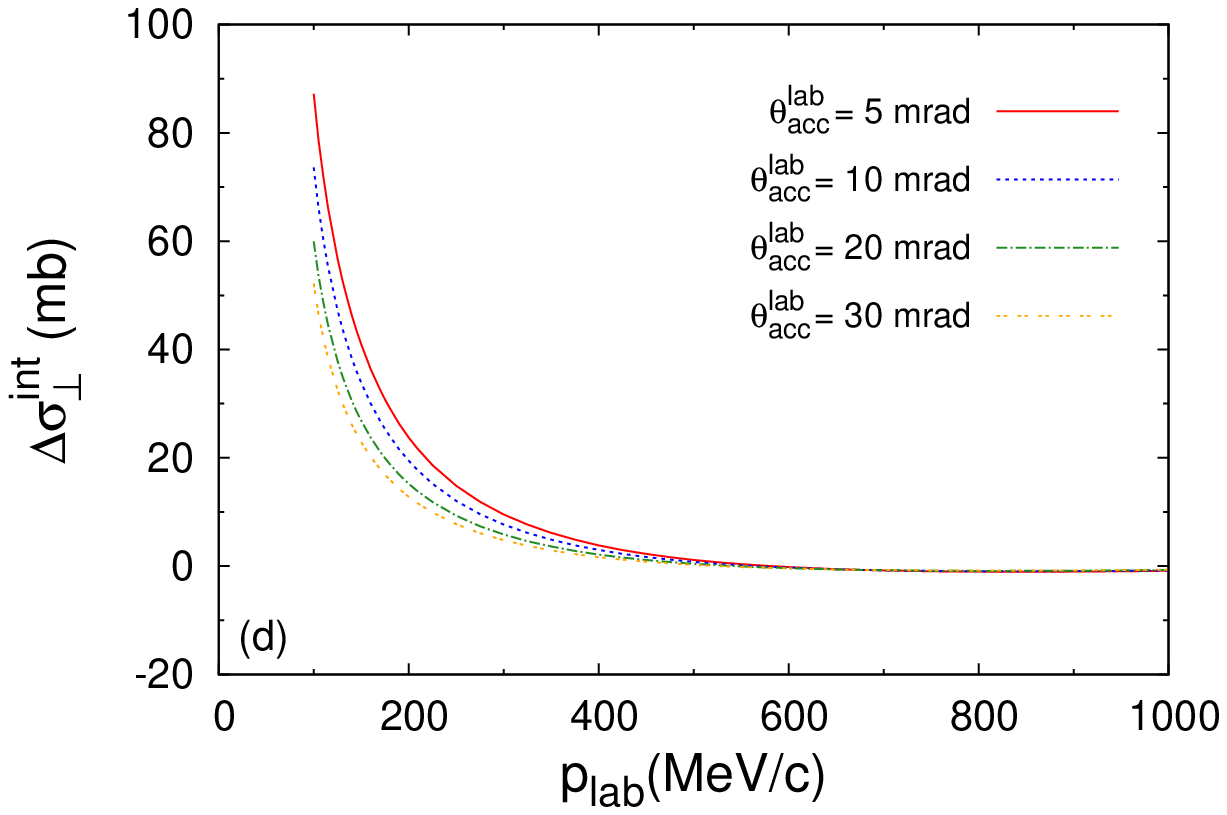}\\ \vspace{-0.9em}
   \includegraphics[width=0.45\textwidth]{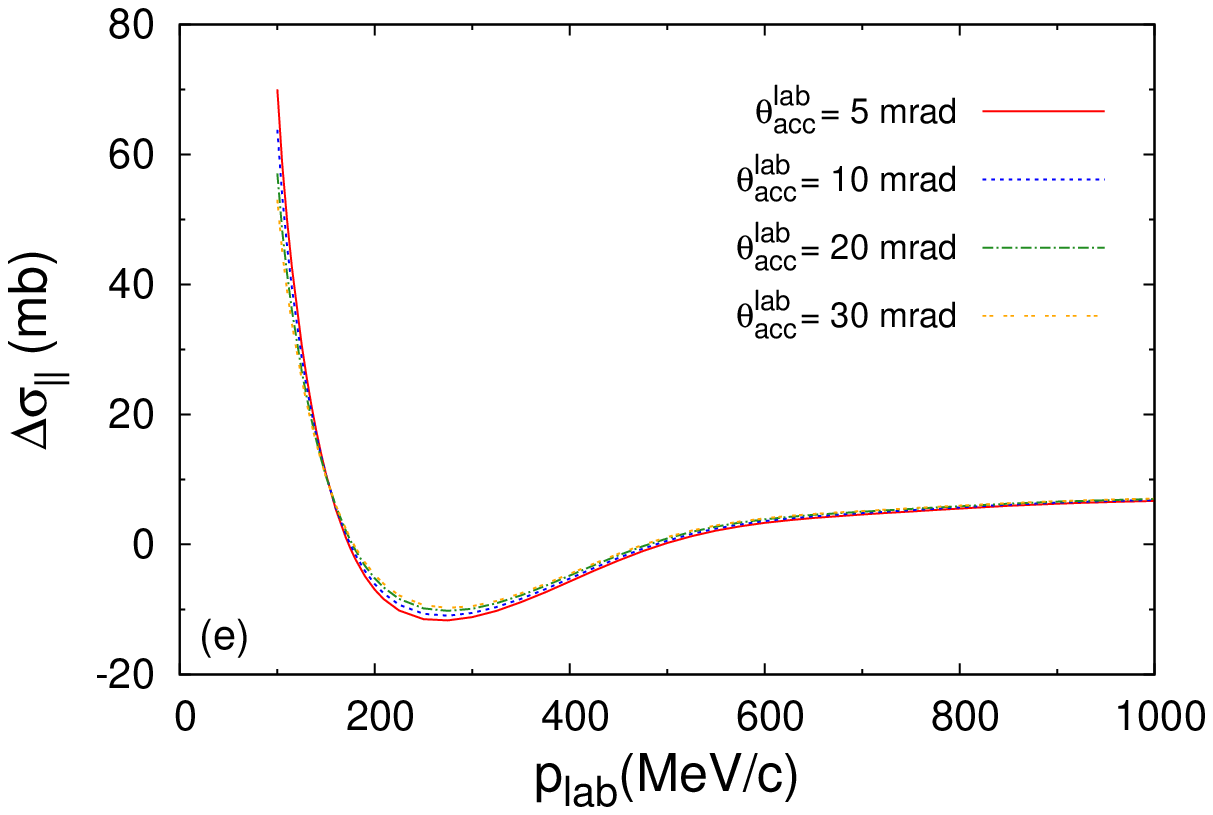} \hspace{2em}
   \includegraphics[width=0.45\textwidth]{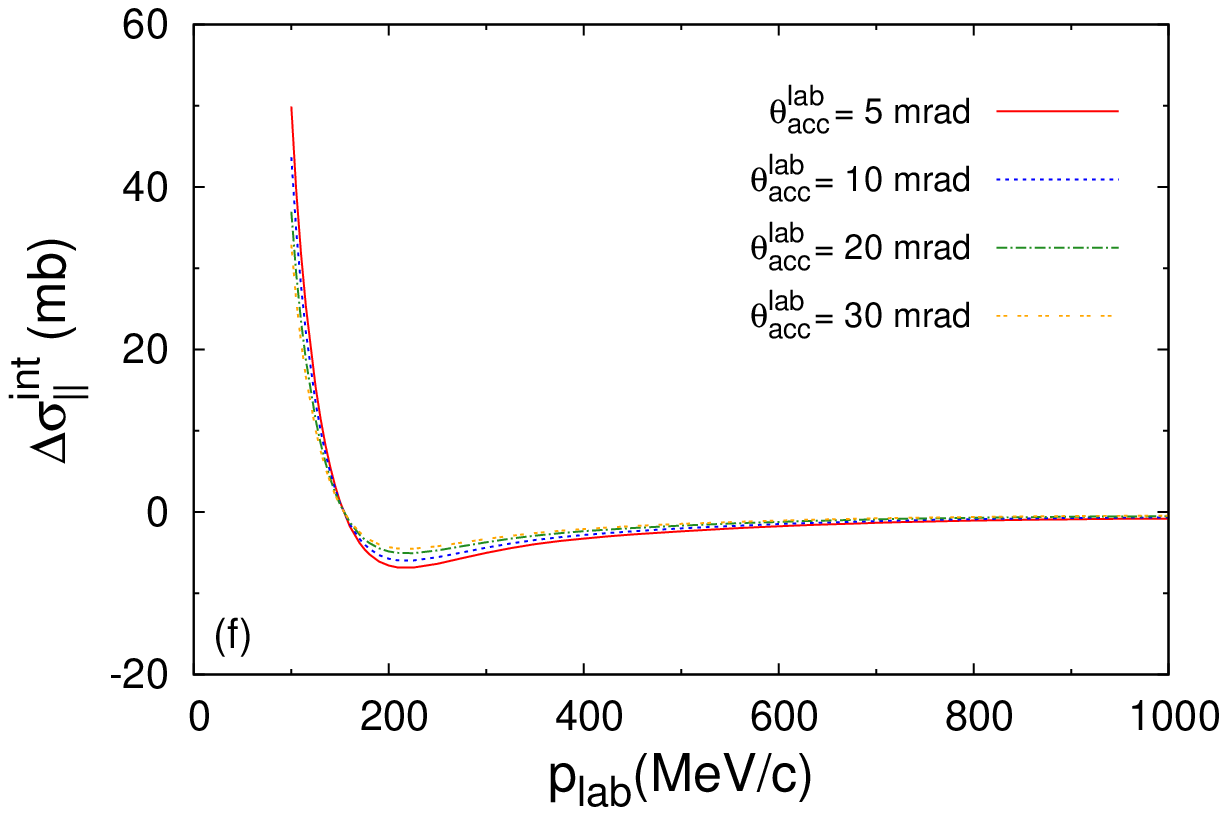}\\ \vspace{-0.9em}
   \includegraphics[width=0.45\textwidth]{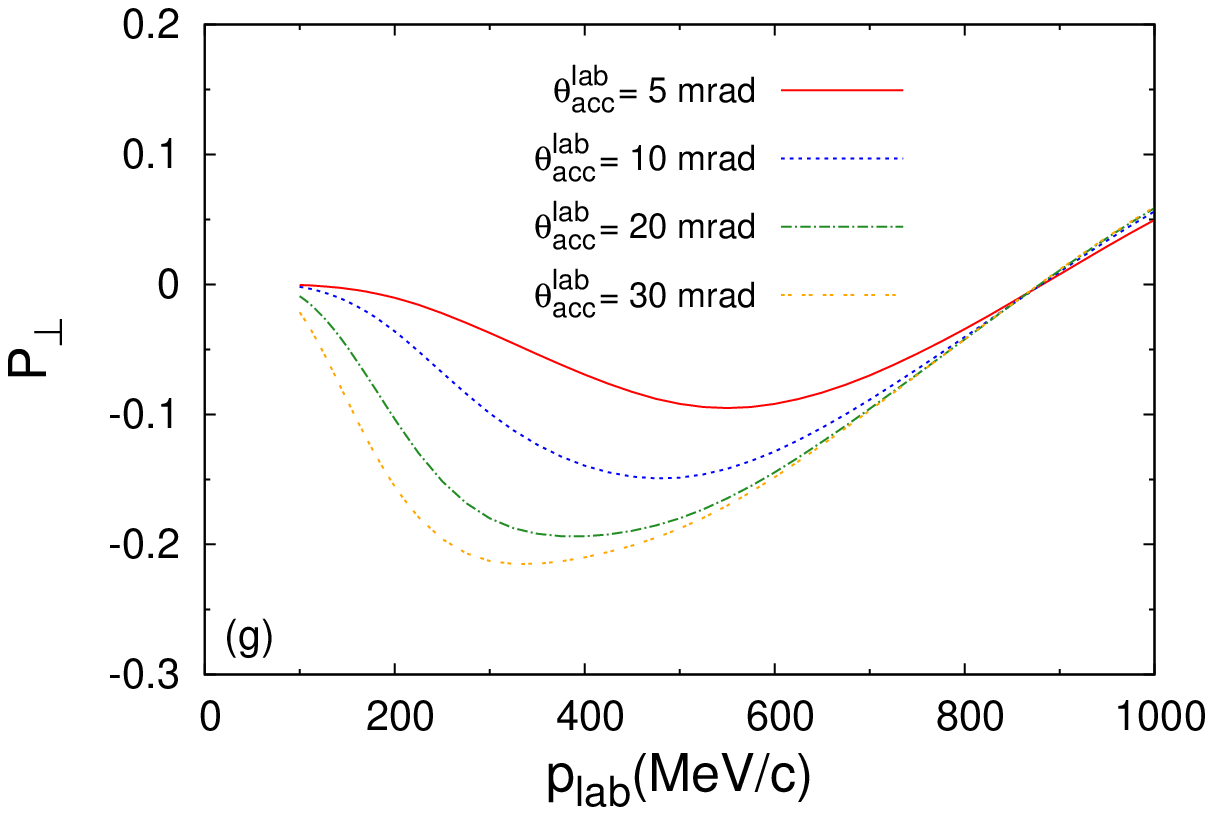} \hspace{2em}
   \includegraphics[width=0.45\textwidth]{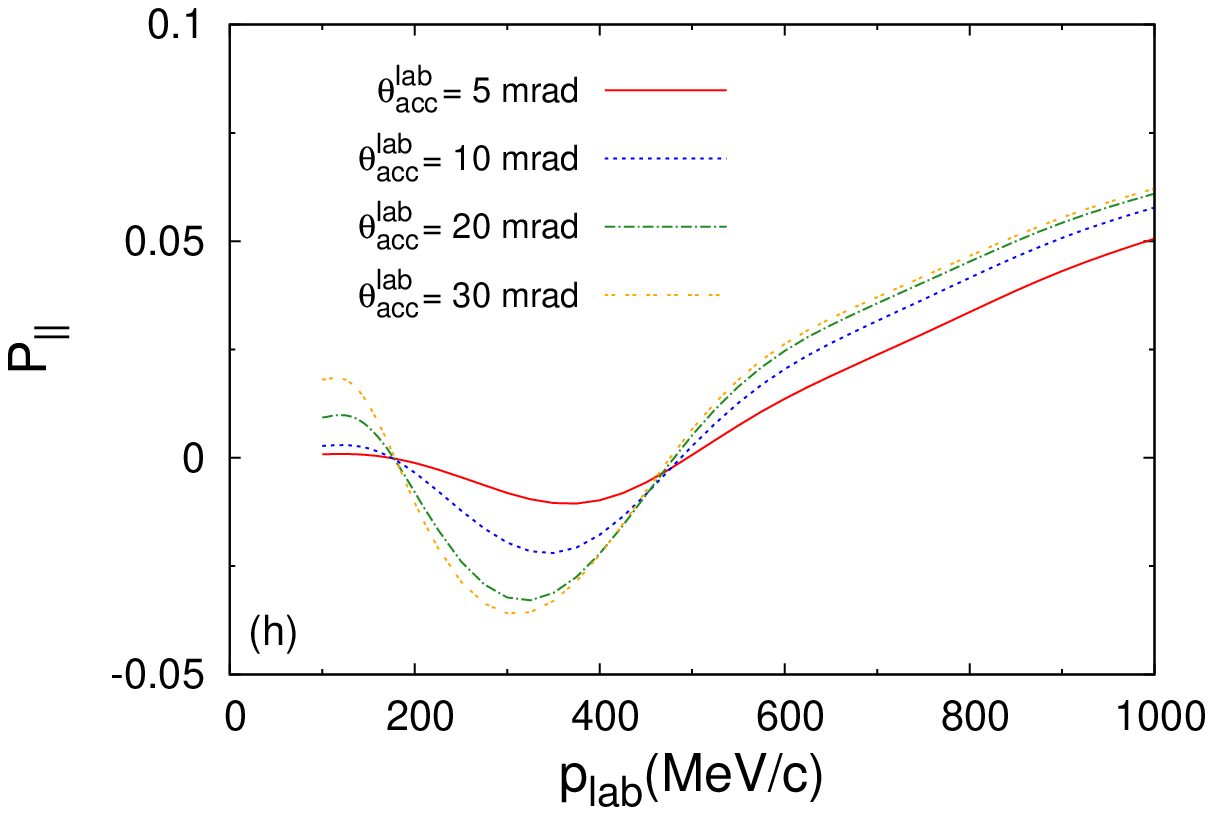}  
\caption{\label{comparison_labnoMM}(Color online) 
The total cross sections $\sigma_{\textrm{tot}}$,  the interference cross
sections $\sigma_{\textrm{tot}}^{\textrm{int}}$, the differences of the transverse and longitudinal
cross sections, $\Delta\sigma_{\!_\perp}$ and $\Delta\sigma_{\!_\parallel}$, and the
polarizations $P_{\!_\perp}$ and $P_{\!_\parallel}$ at time $t_{0}$. The solid red, dotted
blue, dot-dashed green, and double-dotted orange lines show the results for
$\theta_{\rm{acc}}^{\rm{lab}}=5$, 10, 20, and 30 mrad, respectively.} 
\end{figure}

In Fig.~\ref{comparison_labnoMM} the integrated cross sections, the integrated Coulomb-nuclear
interference cross sections, and the beam polarizations are given at time $t_{0}$ as functions of
momentum for the different acceptance angles. The interferences are important for the lower momenta.
For a target with polarization perpendicular to the direction of the incoming beam (transverse,
$P_{B}=P_{\!_\perp}$), the maximal beam polarization is around $-15\%$, depending on the
acceptance angle. At momenta around 350 MeV/$c$, the polarization can reach about
$-20\%$ for ``large'' acceptance angle $\theta_{\rm{acc}}^{\rm{lab}}=30$ mrad. For ``small''
acceptance angle $\theta_{\rm{acc}}^{\rm{lab}}=5$ mrad, the polarization is about $-10\%$
at momenta around 550 MeV/$c$. When the momentum goes up to 1000 MeV/$c$, the
polarization is around $5\%$. For a target with polarization collinear with the direction of
the incoming beam (longitudinal, $P_{B}=P_{\!_\parallel}$), the maximal beam polarization
is around $-2\%$ at low momenta and $5\%$ at the higher momenta.  At low momenta
around 300 MeV/$c$,  the polarization reaches about $-3\%$ for acceptance angle
$\theta_{\rm{acc}}^{\rm{lab}}=30$ mrad, while at momenta around 350 MeV/$c$, the
polarization reaches about $-1\%$ for acceptance angle $\theta_{\rm{acc}}^{\rm{lab}}=5$ mrad.
For momenta around 1000 MeV/$c$, the polarizations are between $5\%$ and $6\%$ when
$\theta_{\rm{acc}}^{\rm{lab}}$ varies between 5  and 30 mrad. In general, the transverse polarization
$P_{\!_\perp}$ reaches higher values than the longitudinal polarization $P_{\!_\parallel}$.
By using the error matrix of the PWA solution, we have estimated that the statistical uncertainty
in our predictions for the polarizations $P_{\!_\perp}$ and $P_{\!_\parallel}$ is less than 2\%
in the momentum range considered.

We conclude that in the momentum range considered here, a significant transverse polarization
$P_{\!_\perp}$can be achieved within a reasonable time. The signs of the polarizations imply
that the beam polarization has the same direction as the target polarization when the sign is
positive, while it has the opposite direction when the sign is negative.

\section{Summary and conclusions} \label{sec:Summary}
Based on our new energy-dependent PWA of all antiproton-proton scattering
data below 925 MeV/$c$, we presented predictions for unmeasured rank-two
spin observables, which may be tested by future experiments. Such new spin
data can improve the existing solution, provided they are (statistically)
precise enough and (systematically) accurate. Since the PWA uses
model-independent theoretical input for the long-range electromagnetic and
strong interactions and gives an optimal description of the existing database,
these predictions are robust and at present the best possible. Striking
spin effects are predicted especially for charge-exchange scattering, confirming
quantitatively some of the early, pre-LEAR findings of Ref.~\cite{Dov82}. These
effects are due to the spin dependence of the long-range one- and two-pion
exchange interactions, in particular the coherent tensor force, and the spin
dependence of the parametrized short-range interaction. In the charge-exchange
reaction the values of the polarization-transfer parameters $R_{t}'$ and $A_{t}$
are large  for the very forward angles at low energies, which suggests a way to
produce polarized antineutrons. We investigated the buildup of polarization
of a circulating antiproton beam by a polarized proton target as a function of
momentum and for several typical acceptance angles. It appears feasible to
achieve a significant transverse polarization within a reasonable time. The size
of the resulting polarization depends strongly on the momentum of the beam.

\section*{Acknowledgments}
We would like to thank D. Boer and G. Onderwater for helpful discussions.

\end{document}